\documentclass[aps,reprint,twocolumn,showpacs,superscriptaddress]{revtex4-1}
\usepackage{graphicx}
\usepackage{xcolor}
\usepackage{epstopdf}
\usepackage{dcolumn}
\usepackage{bm}
\usepackage{color}
\usepackage{subfigure}
\usepackage{amsmath}

\begin{document}


\title{Engineering topological phases in the Luttinger semimetal $\alpha$-Sn }

\author{Dongqin Zhang$^{\dagger}$}
\affiliation{National Laboratory of Solid State Microstructures and School of Physics, Nanjing University, Nanjing 210093, China}

\author{Huaiqiang Wang$^{\dagger}$}
\affiliation{National Laboratory of Solid State Microstructures and School of Physics, Nanjing University, Nanjing 210093, China}

\author{Jiawei Ruan}
\affiliation{National Laboratory of Solid State Microstructures and School of Physics, Nanjing University, Nanjing 210093, China}

\author{Ge Yao}
\affiliation{National Laboratory of Solid State Microstructures and School of Physics, Nanjing University, Nanjing 210093, China}
\affiliation{Collaborative Innovation Center of Advanced Microstructures, Nanjing University, Nanjing 210093, China}

\author{Haijun Zhang}
\email{zhanghj@nju.edu.cn}
\affiliation{National Laboratory of Solid State Microstructures and School of Physics, Nanjing University, Nanjing 210093, China}
\affiliation{Collaborative Innovation Center of Advanced Microstructures, Nanjing University, Nanjing 210093, China}






\begin{abstract}
$\alpha$-Sn is well known as a typical Luttinger semimetal with a quadratic band touching at the $\Gamma$ point. Based on the effective $k\cdot p$ analysis as well as first-principles calculations, we demonstrate that multiple topological phases with a rich diagram, including topological insulator, Dirac semimetal, and Weyl semimetal phases, can be induced and engineered in $\alpha$-Sn by external strains, magnetic fields, and circularly polarized light (CPL). Intriguingly, not only the conventional type-I Weyl nodes, but also type-II Weyl nodes and double-Weyl nodes can be generated directly from the quadratic semimetal by applying a magnetic field or CPL. Our results apply equally well to other Luttinger semimetals with similar crystal and electronic structures, and thus open an avenue for realizing and engineering multiple topological phases on a versatile platform.
\end{abstract}


\maketitle


\section{Introduction}
Recently, topological phases have been significantly extended from gapped topological insulators~\cite{Hasan2010,Qi2010}, such as quantum spin Hall states~\cite{kane2005quantum, Bernevig2006quantum, Konig2007quantum}, quantum anomalous Hall states~\cite{liu2008,Yu2010,Chang2013,weng2015quantum}, three-dimensional topological insulators~\cite{Fu2007Topological, zhang2009topological, Chen2009} and topological superconductors~\cite{Fu2008,Sau2010,Qi2010b,Alicea2010,he2017chiral}, to gapless topological semimetals, such as Weyl semimetals~\cite{Wan2011,Xu2011,yang2011, Burkov2011a, Burkov2011b, Halasz2012, Zyuzin2012, Lu2012, Das2013, Liu2014, zhang2014a, Weng2015, Xu2015a, Lv2015a,  Yang2015, Lv2015b, Xu2015b, Alidoust2015, huang2015, Xu2016, Lu2015, Ruan2016a, Ruan2016b}, Dirac semimetals~\cite{Wang2012, Young2012, Wang2013,Liu2014A, Chen2014, Neupane2014, Yang2014, Xu2015Observation, Xu2017, Huang2017, tang2016dirac, Wang2017Antiferromagetic, hua2018dirac}, and nodal-line semimetals~\cite{Yu2015Topological, Kim2015Dirac, Bian2016Drumhead, Fang2016Topological,  Yu2017Topological, Yan2017Nodal, Chen2017Topological, Li2017Type-II, Sun2017Dirac}. Unlike topological insulators, topological semimetals have stable bulk-band-touching points or closed lines in the Brillouin zone, leading to the emergence of low-energy quasiparticles, known as Weyl fermions, Dirac fermions, and nodal-line fermions, respectively. Both Dirac and Weyl semimetals manifest topologically protected surface Fermi arcs and exotic transport properties originating from the chiral anomaly, such as negative magnetoresistance and chiral magnetic effects~\cite{nielsen1983, liu2013chiral, Son2013, hosur2013, Xiong413, Huang2015Observation, Zhang2016Signatures, li2016chiral, Zyuzin2012Topological, Chang2015, vazifeh2013electromagnetic}. Dirac semimetals and nodal-line semimetals require special symmetries, such as, time-reversal symmetry (TRS), inversion symmetry (IS), and crystal symmetries~\cite{Wang2012, Yang2014, gao2016classification, tang2016dirac, Fang2016Topological, Yu2017Topological}, while Weyl semimetals can be in a stable existence without any symmetries.

Weyl nodes have been experimentally observed in several materials, for example, the TaAs family~\cite{Lv2015a, huang2015, Xu2015a, Yang2015, Lv2015b, Xu2016, Xu2015b} and WTe$_2$ family~\cite{Deng2016, wang2016gate, belopolski2016discovery, Wang2016Observation, jiang2017signature, zhang2016raman, Tamai2016}, but most of them are not ideal systems to study the intrinsic properties of Weyl semimetals because of the following considerations: (1) The Weyl nodes are mixed with topologically trivial bulk bands, so it is hard to distinguish the Weyl nodes from bulk bands in experiments, which is one important reason for obtaining qualitatively different transport measurements~\cite{arnold2016negative, klotz2016quantum, yan2017topological}. (2) All TaAs-family and WTe$_2$-family materials do not have a simple Hamiltonian to describe their electronic structure because of the low crystal symmetries and complex orbitals. Motivated by this consideration, ideal Weyl semimetals were theoretically proposed in HgTe-class materials~\cite{Ruan2016a,Ruan2016b}.  The primitive HgTe-class material is $\alpha$-Sn in which both the TRS and IS are preserved and the conduction and valence bands quadratically touch at the $\Gamma$ point. Moreover, topological phases in $\alpha$-Sn films have been confirmed by recent experiments~\cite{Barfuss2013, Ohtsubo2013, Xu2017, scholz2018, zhu2015epitaxial, liao2017superconductivity}.

Recently various band engineering techniques were developed as quite useful tools for tuning electronic structures by external perturbations, such as strains, magnetic fields, and light fields. For example, a broad range of in-plane compressive strain can induce ideal Weyl semimetals in HgTe-class materials~\cite{Ruan2016a}. A magnetic field is also an efficient route to create Weyl semimetals through breaking TRS~\cite{Hui2015Negative, Li2015Giant, Liang2015Ultrahigh, Hirschberger2016, Shekhar2016Observation, Xiong413, Cano, oh2017topological}. In addition, Floquet topological states recently attracted wide attention, and especially Floquet topological insulators and semimetals have been put forward driven by a light field~\cite{lindner2011floquet, cayssol2013floquet,Narayan2015, Hubener2017, Chan2016a, Chan2016b, Yan2016, Taguchi2016, Bomantara2016, Rui2014, Chen2016, Zhang2016, Bomantara2016generating, Zhou2016Floquet, Yan2017, Ezawa2017, Gupta2017, Wang2017Line}.

In this paper, we investigate such a typical Luttinger semimetal with a quadratic band touching in the presence of external strains, magnetic fields and circularly polarized light. It is found that multiple topological phases can be induced in $\alpha$-Sn, including Dirac semimetal, topological insulator, and Weyl semimetal phases. Intriguingly, in the Weyl semimetal phase, in addition to conventional type-I Weyl nodes, both type-II Weyl nodes~\cite{Soluyanov2015, Wang2016, Aut2016, Sun2015} and double-Weyl nodes~\cite{Xu2011, Fang2012, Huang2015New} can also be generated.

Concretely, an in-plane tensile (compressive) strain transforms $\alpha$-Sn into a topological insulator (Dirac semimetal), while a magnetic field or a circularly polarized light (CPL) incident in the [001] direction gives rise to the coexistence of single-Weyl nodes and double-Weyl nodes on the high symmetry $k_z$ axis at different energies. Upon certain perturbations, each double-Weyl node may split into two single-Weyl nodes. Further, when both an in-plane strain and a magnetic field or a CPL are simultaneously introduced, the system will display a rich phase diagram, where the number, type, and location of the Weyl nodes can be engineered. We emphasize that our results are general and applicable to other Luttinger semimetals, which provides an efficient way to engineer rich topological phases in such materials~\cite{yao2017pr, Ghorashi2018}.

This paper is organized as follows. In Sec. II, we first show the crystal structure and corresponding electronic structure of $\alpha$-Sn by first-principles calculations, and then present the unperturbed Luttinger Hamiltonian for the $k\cdot p$ model analysis. Strain effects are investigated in Sec. III. In Sec. IV, we study the Weyl nodes induced by a magnetic field in the absence and presence of in-plane strains, respectively. In Sec. V, we consider applying an off-resonant CPL to $\alpha$-Sn and derive the photo-induced Floquet Weyl and double-Weyl nodes. Strain effects will also be discussed there. In Sec. VI, we come to the discussion and conclusion.

\section{Crystal structure and electronic structure}
\begin{figure}
  \centering
  \includegraphics[scale=0.75]{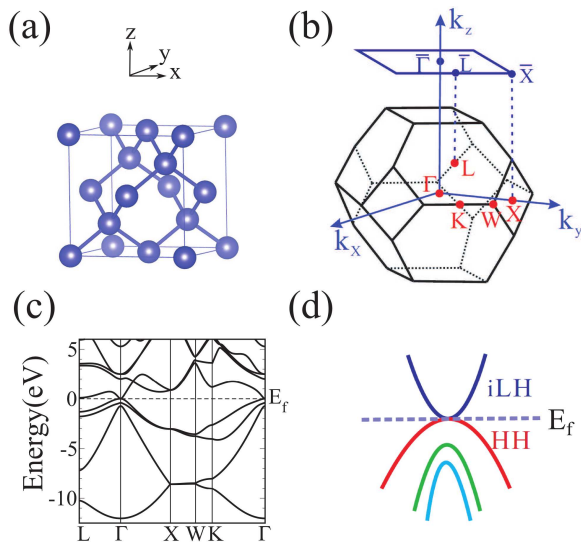}\\
  \caption{(Color online) (a) Schematic illustration of $\alpha$-Sn with a diamond crystal structure. (b) The bulk Brillouin zone (BZ) and its projection on the (001) surface of $\alpha$-Sn. (c) The band structure of $\alpha$-Sn from first-principles calculations. (d) Schematic illustration of the unoccupied inverted light-hole (iLH) and occupied heavy-hole (HH) bands of $\alpha$-Sn. }
  \label{fig.1}
\end{figure}

Usually, the group-IV compounds with the diamond structure, such as C, Si, and Ge, are insulators, where the bottom of the conduction bands comes from the $s$ orbital and the top of the valence bands comes from $p$ orbitals. There is a full energy gap between conduction and valence bands. The top of the valence bands is composed of light-hole and heavy-hole bands which are touching at the $\Gamma$ point protected by the cubic symmetry. Differently, $\alpha$-Sn is a well-known negative-gap semimetal, because the $s$ orbital bands go below the $p$ orbital bands to become occupied and the light-hole bands go up to become unoccupied with an inverted curvature, schematically shown in Fig. 1d. Comparing with C, Si, and Ge, $\alpha$-Sn has a band inversion which makes it topologically nontrivial \cite{scholz2018}.

In order to obtain band structures, first-principles calculations are performed within density functional theory (DFT) as implemented in  the Vienna {\it ab initio} simulation package (VASP)~\cite{kresse1999}. The band structure is investigated with the nonlocal Heyd-Scuseria-Ernzerhof (HSE) hybrid functional~\cite{Heyd2003}. The cutoff energy for the plane wave expansion is 500 eV and a $k$-point mesh of $16\times 16\times16$ is used for the bulk self-consistent calculations. The spin-orbit coupling (SOC) effect is taken into account self-consistently. The calculated band structure of $\alpha$-Sn is shown in Fig.~1c, and the quadratic band touching between the inverted light-hole (iLH) conduction and heavy-hole (HH) valence bands is clearly seen and protected by the cubic symmetry. In addition, due to the coexistence of TRS and IS, all the bands are doubly degenerate, so the quadratic touching point at the $\Gamma$ point has the fourfold degeneracy, marked by $\Gamma_{8}^{+}$, with the ordered basis $|J,m_j\rangle=|\frac{3}{2},\frac{3}{2}\rangle,|\frac{3}{2},\frac{1}{2}\rangle,|\frac{3}{2},-\frac{1}{2}\rangle,
|\frac{3}{2},-\frac{3}{2}\rangle$.

In contrast to other group-IV compounds (C, Si, and Ge) with the diamond structure, $\Gamma_{8}^{+}$ bands from $p$ orbitals and $\Gamma_{7}^{-}$ bands dominated by $s$ orbitals are inverted in $\alpha$-Sn, which is crucial for the emergence of topological phases. Previous researches mainly focus on topological insulator and Dirac semimetal phases of $\alpha$-Sn~\cite{Barfuss2013, Xu2013Large, Xu2017, Huang2017}, but a thorough investigation of its rich topological phases is still lacking. In what follows, we will use the effective $k\cdot p$ model analysis as well as first-principles calculations to show that $\alpha$-Sn is a perfect platform to demonstrate rich topological states ranging from topological insulators to topological semimetals with the simplest electronic structures under strains, Zeeman fields, and light fields.

The $\Gamma_{8}^{+}$ states are the only states near the Fermi level, and all other regions in the BZ are fully gapped. As a starting point, we choose the $\Gamma_{8}^{+}$ state for the $k\cdot p$ model Hamiltonian. The four low-energy bands around the $\Gamma$ point can be effectively described by the Luttinger Hamiltonian~\cite{Luttinger1956}:
\begin{equation}
\begin{split}
\label{eqn.1}
H_{\mathrm{L}}(\mathbf{k})=&\frac{\hbar^{2}}{m}\Big[(\gamma_{1}+\frac{5}{2}\gamma_{2})\frac{k^{2}}{2}-\gamma_{2}\sum_{i=x,y,z}k_{i}^{2}J_{i}^{2}\\
&-\frac{\gamma_{3}}{2}(\{k_x,k_y\}\{J_x,J_y\}+\mathrm{c.p.})\Big],
\end{split}
\end{equation}
where $\gamma_{i}(i=1,2,3)$ are dimensionless parameters, `` $\{\}$'' denotes an anticommutator, ``c.p.'' means cyclic permutations, and the angular momentum operators are given by
\begin{equation}
\begin{split}
\label{eqn.2}
         J_z&=\left(
        \begin{array}{cccc}
          \frac{3}{2} & 0 & 0 &0 \\
         0 & \frac{1}{2}& 0 & 0 \\
          0 &0 & -\frac{1}{2} & 0 \\
          0 & 0 & 0 & -\frac{3}{2} \\
        \end{array}
      \right),
  J_x=\left(
        \begin{array}{cccc}
          0 & \frac{\sqrt{3}}{2} & 0 &0 \\
          \frac{\sqrt{3}}{2} & 0 & 1 & 0 \\
          0 & 1 & 0 & \frac{\sqrt{3}}{2} \\
          0 & 0 & \frac{\sqrt{3}}{2} & 0 \\
        \end{array}
      \right),\\
  J_y=&\left(
        \begin{array}{cccc}
          0 & -\frac{\sqrt{3}i}{2} & 0 &0 \\
         \frac{\sqrt{3}i}{2} & 0 & -i & 0 \\
          0 & i & 0 & -\frac{\sqrt{3}i}{2} \\
          0 & 0 & \frac{\sqrt{3}i}{2} & 0 \\
        \end{array}
      \right).
\end{split}
\end{equation}
By fitting the first-principles band structure of $\alpha$-Sn, the parameters can be determined as $\gamma_{1}=14.97$, $\gamma_{2}=10.61$, $\gamma_{3}=8.52$~\cite{Lawaetz1971}. For notational brevity, henceforth, we absorb the factor $\frac{\hbar^{2}}{m}$ into $\gamma_{i}$. The Luttinger Hamiltonian will be used as the unperturbed Hamiltonian throughout the following sections. In addition, symmetry operations of $\alpha$-Sn can be expressed as $4\times 4$ matrices acting on the ordered $|J,j_z\rangle$ basis. For later references, we list some of them as
\begin{equation}
\begin{split}
C_{2x}=&e^{-i\pi J_{x}}=i\sigma_{x}\tau_{x},\quad
C_{2y}=e^{-i\pi J_{y}}=-i\sigma_{x}\tau_{y},\\
C_{2z}=&e^{-i\pi J_{z}}=i\sigma_{0}\tau_{z},\quad
T=e^{-i\pi J_{y}}K=-i\sigma_{x}\tau_{y}K,\\
C_{4z}I=&e^{-i\frac{\pi}{2} J_{z}}\times (-1)=\mathrm{Diag}[e^{i\frac{\pi}{4}},e^{i\frac{3\pi}{4}},e^{i\frac{5\pi}{4}},e^{i\frac{7\pi}{4}}].
\end{split}
\end{equation}
Here, $K$ denotes the complex conjugate, and the Pauli matrices $\sigma_{i}$ and $\tau_{i} \quad(i=0,x,y,z)$ are used only to simplify the expression of the $4\times 4$ matrices, but not for explicit physical meanings such as spin or orbital.

\section{Strain effects: tensile and compressive strains}
Strains serve as a useful tool to engineer electronic structures in condensed matter systems, such as the strain-induced ideal Weyl points in HgTe-class materials~\cite{Ruan2016a}. Considering the similar inverted band structure between $\alpha$-Sn and HgTe, it seems natural and interesting to investigate various strain effects in $\alpha$-Sn.  In this section, we consider applying an in-plane biaxial strain to $\alpha$-Sn, which reduces the original cubic symmetry $O_{h}^{7}$ to a tetragonal symmetry $D_{4h}^{19}$~\cite{Huang2017}. Depending on the type of the strains, say, compressive or tensile, different topological phases are expected to emerge correspondingly, as we explicitly show below.
\begin{figure*}
  \centering
  \includegraphics[scale=1.1]{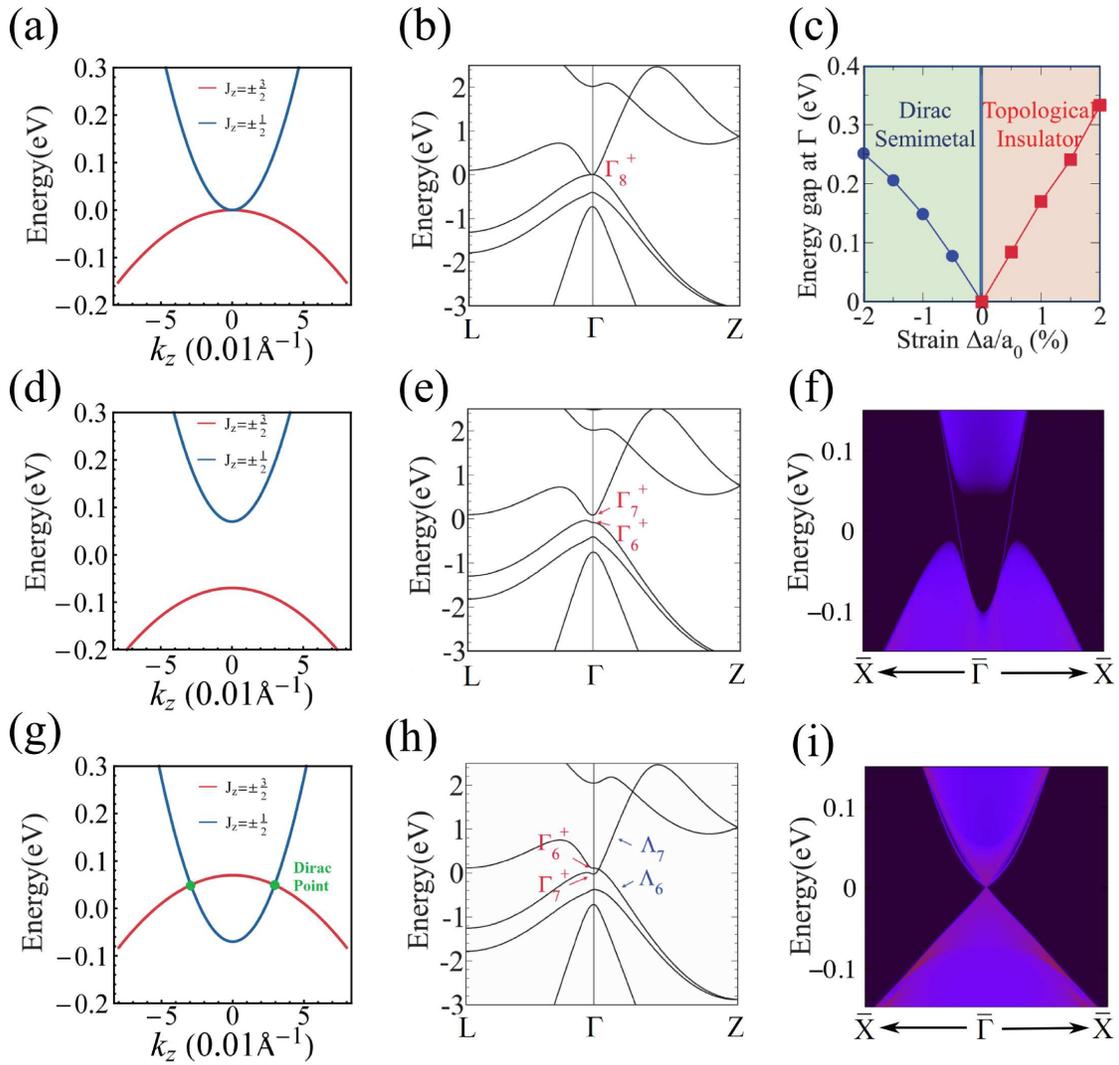}\\
  \caption{(Color online) The band structure of $\alpha$-Sn obtained from the effective $k\cdot p$ model analysis: (a) without strain, (d) under $1\%$ tensile strain and (g) under $1\%$ compressive strain. The band structure of $\alpha$-Sn from first-principles calculations: (b) without strain, (e) under $1\%$ tensile strain and (h) under $1\%$ compressive strain. LDOS on the (001) surface for $\alpha$-Sn (f) under $1\%$ tensile strain and (i) under  $1\%$ compressive  strain. (c) The complete phase diagram of $\alpha$-Sn as a function of an in-plane strain.}
  \label{fig.2}
\end{figure*}

For simplicity and without loss of generality, we consider a strain applied in the $xy$ plane, which changes the in-plane lattice constant to $a=(1+\delta)a_{0}$, with $\delta>0$ for the tensile strain and $\delta<0$ for the compressive strain. It can be treated as a perturbation described by the following Hamiltonian:
\begin{equation}
\label{}
H_{\mathrm{strain}}=\epsilon(J_{z}^{2}-\frac{5}{4}),
\end{equation}
where $\epsilon$ is determined by the strength of the strain and $\epsilon<0$ ($\epsilon>0$) for tensile (compressive) strain \cite{Ruan2016a}. Based on the reduced crystal symmetry, we naturally concentrate on the $k_{z}$ axis, where the total Hamiltonian becomes:
\begin{equation}
\label{}
H(k_{z})=(\frac{\gamma_{1}}{2}+\frac{5}{4}\gamma_{2})k_{z}^{2}-(\gamma_{2}k_{z}^{2}-\epsilon) J_{z}^{2}-\frac{5}{4}\epsilon.
\end{equation}
Since $[J_{z},H(k_{z})]=0$, $J_{z}$ is a good quantum number and thus all the bands can be labeled by their $J_{z}$ eigenvalues. Without strain, $J_{z}=\pm\frac{3}{2}$ $\big(J_{z}=\pm\frac{1}{2}\big)$ bands are doubly degenerate with a downward (upward) parabolic dispersion $E=(\frac{\gamma_{1}}{2}-\gamma_{2})k_{z}^{2}$ $\Big(E=(\frac{\gamma_{1}}{2}+\gamma_{2})k_{z}^{2}\Big)$, forming a four-band quadratic touching at the $\Gamma$ point, as shown by the band structures in Figs.~2(a) and 2(b). Consequently, unstrained $\alpha$-Sn lies in the semimetal phase with a quadratic touch.

Under a tensile strain with $\epsilon<0$, the doubly degenerate $J_{z}=\pm\frac{3}{2}$ ($\Gamma_{7}^{+}$) bands and $J_{z}=\pm\frac{1}{2}$ ($\Gamma_{6}^{+}$) bands are pushed down and up, respectively, generating a full energy gap between conduction bands and valence bands in the whole BZ, as can be seen from the band structures in Figs. 2(d) and 2(e) under $1\%$ tensile strain with $\epsilon=-0.07$eV. The topological property of this insulator phase can be characterized by the $Z_{2}$ index~\cite{Fu2007}. Thanks to the preserved IS, its $Z_{2}$ index can be simply obtained through the product of parities at time-reversal invariant momenta~\cite{Fu2007}. A detailed calculation yields a nontrivial $Z_{2}=1$, as expected, because of the inversion of $\Gamma_{8}^{+}$ and $\Gamma_{7}^{-}$ bands with opposite parities in unstrained $\alpha$-Sn. This indicates that the in-plane tensile strain changes $\alpha$-Sn into a three-dimensional topological insulator~\cite{Barfuss2013}.

In contrast, under a compressive strain with $\epsilon>0$, the doubly degenerate $J_{z}=\pm\frac{3}{2}$ ($\Gamma_{7}^{+}$) bands and $J_{z}=\pm\frac{1}{2}$ ($\Gamma_{6}^{+}$) bands are pushed up and down, respectively, which inevitably results in two four-band crossing points at $\mathbf{k_{D}^{\pm}}=(0,0,\pm k_{D})$ with $k_{D}=\sqrt{\frac{\epsilon}{\gamma_{2}}}$, as shown in Figs. 2(g) and 2(h) under $-1\%$ compressive strain with $\epsilon=0.07$eV. These two crossings are unavoidable since the $\Lambda_{6}$ and $\Lambda_{7}$ bands belong to different two-dimensional irreducible representations and the hybridization between them is strictly forbidden. Moreover, by expanding the Hamiltonian around $\mathbf{k_{D}^{\pm}}$ to linear order of relative momentum $q$, we obtain:
\begin{equation}
\label{}
H(\mathbf{k_{D}^{\pm}}+\mathbf{q})=\pm\Big\{ cq_{z}-\sigma_{z}[v_{z}q_{z}\tau_{z}+v_{\perp}(q_{x}\tau_{x}+q_{y}\tau_{y})]\Big\},
\end{equation}
where a constant shift term has been dropped and $c=\gamma_{1}k_{D}$, $v_{z}=2\gamma_{2}k_{D}$, $v_{\perp}=\sqrt{3}\gamma_{3}k_{D}$, and $\sigma_{i}$ and $\tau_{y}\quad (i=x,y,z)$ are Pauli matrices. This low-energy effective Hamiltonian describes a Dirac fermion consisting of a pair of Weyl fermions with opposite chiralities. The two Dirac points at $\mathbf{k_{D}^{\pm}}$ are related to each other by TRS or IS. As a result, a stable 3D Dirac semimetal phase is induced in $\alpha$-Sn by a compressive in-plane strain. Similar analysis shows that a strain along the (111) direction can also transform $\alpha$-Sn into a Dirac semimetal~\cite{Huang2017}, which has already been experimentally realized~\cite{Xu2017}.

As a further evidence, we have constructed maximally localized Wannier functions on the basis of first-principles calculations~\cite{Marzari1997, Souza2001, Zhang2009a}, to show the existence of topological surface states for strained $\alpha$-Sn. On the surface, H atoms are used to remove the non-physical dangling bonds. In Figs.~2(f) and 2(i), the local density of states (LDOS) on the (001) surface are plotted for the topological insulator phase and Dirac semimetal phase, respectively. We can see that there is a single Dirac cone state within the energy band gap at the $\Gamma$ point for the topological insulator phase in Fig.~2(f), and that there is a gapless linear-dispersion bulk state clearly indicating the Dirac semimetal phase in Fig.~2(i), where the two bulk Dirac nodes on the $k_{z}$ axis are projected to the same surface momentum $\overline{\Gamma}$, in obvious contrast to the unstrained $\alpha$-Sn with quadratic-dispersion bulk states.

Based on the above analysis, we now present the complete phase diagram of $\alpha$-Sn as a function of the applied in-plane strain in Fig. 2(c), where a tensile (compressive) strain changes $\alpha$-Sn from a zero-gap semiconductor to a topological insulator (Dirac semimetal).

\section{Weyl semimetals induced by magnetic fields}
It is well known that Dirac points may split into pairs of Weyl points when either TRS or IS is broken~\cite{Wan2011,Xu2011,yang2011, Burkov2011a, Burkov2011b, Halasz2012, Zyuzin2012, Lu2012, Das2013, Liu2014, zhang2014a, Weng2015, Xu2015a, Lv2015a,  Yang2015, Lv2015b, Xu2015b, Alidoust2015, huang2015, Xu2016, Lu2015, Ruan2016a, Ruan2016b}. So if we break TRS by an external magnetic field, Weyl semimetal phases are expected to emerge from the Dirac semimetal phase in compressively strained $\alpha$-Sn. Interestingly, based on the $k\cdot p$ model analysis, we find that Weyl points can be induced even directly from unstrained $\alpha$-Sn in the trivial semimetal phase and from tensile-strained $\alpha$-Sn in the topological insulator phase. In this section, we investigate such Weyl semimetal phases in both unstrained and strained $\alpha$-Sn under an external magnetic field.

For a small magnetic field $\mathbf{B}$, we can neglect orbital Landau level effects and only consider the Zeeman effect. Since the energy scale of the atomic spin-orbit coupling is much larger than those of the perturbations introduced by external fields, the total angular momentum eigenstates still act as a useful basis. Therefore, the Zeeman coupling of the Luttinger semimetal $\alpha$-Sn to $\mathbf{B}$ can simply be effectively described by~\cite{Luttinger1956}
\begin{equation}
\label{}
H_{z}=\frac{e\hbar}{m}(\kappa \mathbf{B}\cdot \mathbf{J}+q\mathbf{B}\cdot \mathbf{J}^{3}),
\end{equation}
where the dimensionless parameters $\kappa=11.84$ and $q=-0.30$ are effective Lande g-factors for $\alpha$-Sn~\cite{Lawaetz1971}. Since $\kappa\gg q$, the $q$ term can be safely omitted in the following calculations. For convenience, we absorb the factor $\frac{e\hbar}{m}$ into $\kappa$.

Generally speaking, the number and locations of the induced Weyl points depend on the direction of the magnetic field. For simplicity and clarity, we will only consider applying the magnetic field along some high-symmetry axes of unstrained and strained $\alpha$-Sn, respectively. Before the following analysis, we make an estimation of the range of the magnitude of $\mathbf{B}$, where the negligence of the orbital Landau effects remains valid. For the Landau level effect to be manifest, a rough criterion $B\gg\frac{1}{\mu}$ ($\mu$ is electron mobility) must be satisfied, which results from the condition that an electron should be able to complete at least a few orbits before losing its momentum due to scattering~\cite{Datta}. However, considering the anomalously high mobility of $\alpha$-Sn ($\sim10^{5}\mathrm{cm}^{2}\mathrm{V}^{-1}\mathrm{s}^{-1}$)~\cite{Ewald1959Electronic}, the above condition amounts to $B\gg\sim 0.1$ T. Therefore, to neglect the orbital effect in $\alpha$-Sn, the magnetic field should not be much greater than $0.1$ T. For other Luttinger semimetals with smaller electron mobilities or larger Lande $g$ factors, the range of the magnetic field could be larger. It is worth mentioning that the physics will become quite different when Landau levels are formed in a quite strong magnetic field, as shown in Refs.  \cite{Kim2017, Zhang2017Magnetic}.

\subsection{Unstrained $\alpha$-Sn}
Firstly, we apply the magnetic field along the [001] direction to unstrained $\alpha$-Sn, and the Hamiltonian on the $k_{z}$ axis becomes:
\begin{equation}
 \begin{split}
 \label{}
 H(k_{z})=&(\frac{1}{2}\gamma_{1}+\frac{5}{4}\gamma_{2})k_{z}^{2}-\gamma_{2}k_{z}^{2}J_{z}^{2}+\kappa BJ_{z},
 \end{split}
 \end{equation}
where the last term represents the Zeeman coupling. Since $J_{z}$ still commutes with $H(k_{z})$, the energy bands can be labeled by the $J_{z}$ eigenvalue, with
\begin{equation}
\begin{split}
\label{}
E_{\pm\frac{3}{2}}=&\frac{\gamma_{1}}{2}k_{z}^{2}-\gamma_{2}k_{z}^{2}\pm\frac{3}{2}\kappa B,\\
E_{\pm\frac{1}{2}}=&\frac{\gamma_{1}}{2}k_{z}^{2}+\gamma_{2}k_{z}^{2}\pm\frac{1}{2}\kappa B.
\end{split}
\end{equation}
The Zeeman splitting between the two downward dispersive $J_{z}=\pm\frac{3}{2}$ bands is larger than those between upward dispersive $J_{z}=\pm\frac{1}{2}$ bands. Consequently, the most pushed up $J_{z}=\frac{3}{2}$ band will inevitably intersect with $J_{z}=\pm\frac{1}{2}$ bands, leading to four two-band crossings, as illustrated in Fig. 3(a) with $B=1$ T. Two crossing points reside between $J_{z}=\frac{3}{2}$ and $\frac{1}{2}$ bands at $\mathbf{k^{\pm}_{sw}}=(0,0,\pm k_{sw})$ with $k_{sw}=\sqrt{\frac{\kappa B}{2\gamma_{2}}}$, and the other two between $J_{z}=\frac{3}{2}$ and $-\frac{1}{2}$ bands at $\mathbf{k^{\pm}_{dw}}=(0,0,\pm k_{dw})$ with $k_{dw}=\sqrt{\frac{\kappa B}{\gamma_{2}}}$.

The two crossings between $\frac{3}{2}$ and $\frac{1}{2}$ bands are stablized by $C_{2z}$ symmetry, because $\frac{3}{2}$ and $\frac{1}{2}$ bands have $C_{2z}$ eigenvalues $i$ and $-i$, respectively, and the hybridization between them is prohibited. Moreover, the low-energy effective two-band Hamiltonian around $\mathbf{k^{\pm}_{sw}}$ to $q$ order is derived as
\begin{equation}
\begin{split}
\label{}
H_{\frac{3}{2},\frac{1}{2}}(\mathbf{\mathbf{k_{sw}^{\pm}+q}})
=&\pm\Big[\gamma_{1}k_{sw}q_{z}-2\gamma_{2}k_{sw}q_{z}\sigma_{z}\\
&-\sqrt{3}\gamma_{3}k_{sw}(q_{x}\sigma_{x}+q_{y}\sigma_{y})\Big],
\end{split}
\end{equation}
where the constant shift term has been dropped and the Pauli matrices $\sigma_{i}$ act in the $(\frac{3}{2},\frac{1}{2})$ subspace. This Hamiltonian describes a pair of single-Weyl points with opposite chiralities $\chi_{\mathbf{\mathbf{k_{sw}^{\pm}}}}=\mp1$ (see the Appendix), which feature linear dispersions in all three directions. The two Weyl points are related to each other through IS, and act as a monopole-antimonopole pair of Berry curvature in momentum space, as shown in Fig.~3(e).

The band crossings between $\frac{3}{2}$ and $-\frac{1}{2}$ are more interesting. The low-energy effective Hamiltonian around $\mathbf{k^{\pm}_{dw}}$ to $q^{2}$ order is obtained as
\begin{equation}
\begin{split}
\label{}
H_{\frac{3}{2},-\frac{1}{2}}(\mathbf{\mathbf{k^{\pm}_{dw}+q}})
=&\pm\gamma_{1}k_{dw}q_{z}-\frac{\sqrt{3}\gamma_{2}}{2}(q_{x}^{2}-q_{y}^{2})\sigma_{x}\\
&-\sqrt{3}\gamma_{3}q_{x}q_{y}\sigma_{y}\mp2\gamma_{2}k_{dw}q_{z}\sigma_{z},
\end{split}
\end{equation}
where $k_{dw}\gg q$ is assumed and the Pauli matrices $\sigma_{i}$ act in the $(\frac{3}{2},-\frac{1}{2})$ subspace. This Hamiltonian describes a pair of double-Weyl points with chiralities $\chi_{\mathbf{\mathbf{k_{dw}^{\pm}}}}=\mp2$ (see the Appendix), whose energy dispersion is linear in the $k_{z}$ direction and quadratic in the $k_{x}$ and $k_{y}$ directions~\cite{Xu2011, Fang2012, Huang2015New}. If we introduce a perturbation such as a linear inversion-breaking term:
\begin{equation}
H_{I}=\alpha\big[k_{x}\{J_{x},J_{y}^{2}-J_{z}^{2}\}+\mathrm{c.p.}\big],
\end{equation}
then each double-Weyl point will split into two single-Weyl points, as illustrated by Figs. 3(c) and 3(d). Our reason for choosing this perturbation lies in that upon introducing $H_{I}$, the original diamond-like symmetry of $\alpha$-Sn is reduced to the zinc-blende-like symmetry of HgTe-class materials, and our analysis can be directly applied to these materials. The single-Weyl points split from double-Weyl points are symmetry-protected, as proved below by a similar argument to that in Ref.~\cite{Cano}.

In the gapped $k_{z}=0$ plane, the $\frac{3}{2}$ band exists above the $-\frac{1}{2}$ band, so at the $\Gamma$ point, the higher band and the lower band have $C_{4z}I$ ($=\mathrm{diag}[e^{i\frac{\pi}{4}},e^{i\frac{3\pi}{4}},e^{i\frac{5\pi}{4}},e^{i\frac{7\pi}{4}}]$) eigenvalues $e^{i\frac{\pi}{4}}$ and $e^{i\frac{5\pi}{4}}$, respectively. While in the gapped $k_{z}=\pi$ ($=\infty$) plane, the $\frac{3}{2}$ band rests below the $-\frac{1}{2}$ band, and thus at the $(0,0,\pi)$ point, $C_{4z}I$ eigenvalues of the higher and lower bands are inverted to that at $\Gamma$. As a result, going from $k_{z}=0$ to $k_{z}=\pi$ planes, the $C_{4z}I$ eigenvalue of the lower band has changed from $e^{i\frac{5\pi}{4}}$ to $e^{i\frac{\pi}{4}}$. According to Ref.~\cite{Fang2012}, the change of these eigenvalues is related to the difference of the Chern number between $k_{z}=0$ and $k_{z}=\pi$ planes:
\begin{equation}
\frac{e^{i\frac{\pi}{4}}}{e^{i\frac{5\pi}{4}}}=-1=i^{C_{k_{z}=0}-C_{k_{z}=\pi}}.
\end{equation}
This implies that two (mod four) single-Weyl points must exist between $k_{z}=0$ and $k_{z}=\pi$ planes for the $\frac{3}{2}$ and $-\frac{1}{2}$ bands. If we suppose one of them is located at $(k_{xw},k_{yw},k_{zw})$, then $C_{2z}$ symmetry will lead to another Weyl point of the same chirality at $(-k_{xw},-k_{yw},k_{zw})$. Moreover, although TRS, $C_{2x}$ and $C_{2y}$ symmetries are separately broken by $\mathbf{B}$ in the [001] direction, their combinations $C_{2x}T$ and $C_{2y}T$ remain as special symmetries, since the $T$ operation flips $\mathbf{B}$ and $C_{2x}$ (or $C_{2y}$) flips it back. Thus, unless $k_{xw}=0$ or $k_{yw}=0$, $C_{2x}T$ and $C_{2y}T$ symmetries will generate two additional Weyl points with the same chirality at $(-k_{xw},k_{yw},k_{zw})$ and $(k_{xw},-k_{yw},k_{zw})$, which will produce altogether four Weyl points, and obviously contradicts the Chern number argument. Therefore, the only way is that two single-Weyl points of the same chirality should exist in the $k_{x}=0$ or $k_{y}=0$ plane between $k_{z}=0$ and $k_{z}=\pi$ planes. For example, in the presence of the linear inversion-breaking term $H_{\mathrm{I}}$, with $\alpha=0.2$, two Weyl points can be found in the $k_{x}=0$ plane in the $k_{z}>0$ region, which is schematically shown in Fig. 3(d), and further verified by the Berry curvature configuration in Fig. 3(f). Without the perturbation $H_{I}$, they coincide with each other on the positive $k_{z}$ axis and constitute a double-Weyl point in Fig. 3(c).

A similar argument can be applied to the two single-Weyl points between $k_{z}=0$ and $k_{z}=-\pi$ planes, which are related to those in the $k_{z}>0$ region by $C_{4z}I$ symmetry, as shown by the Weyl points in the $k_{y}=0$ plane in the $k_{z}<0$ region in Fig. 3(d). However, they have opposite chiralities, because the inversion operation inverses chirality. They constitute the other double-Weyl point in the negative $k_{z}$ axis, as shown in Fig. 3(c), which forms a charge-two monopole-antimonopole pair with that on the positive $k_z$ axis.

It is worth mentioning that the single-Weyl nodes on the $k_z$ axis are pinned at the same energy by the $C_{4z}I$ symmetry connecting them. In addition, the remaining four single-Weyl nodes in the $k_{y}=0$ and $k_{x}=0$ plane also reside at the same energy, since they are related to each other by the $C_{2z}$ or $C_{4}I$ symmetry, or a combination of them.
\begin{figure}
  \centering
  \includegraphics[scale=0.75]{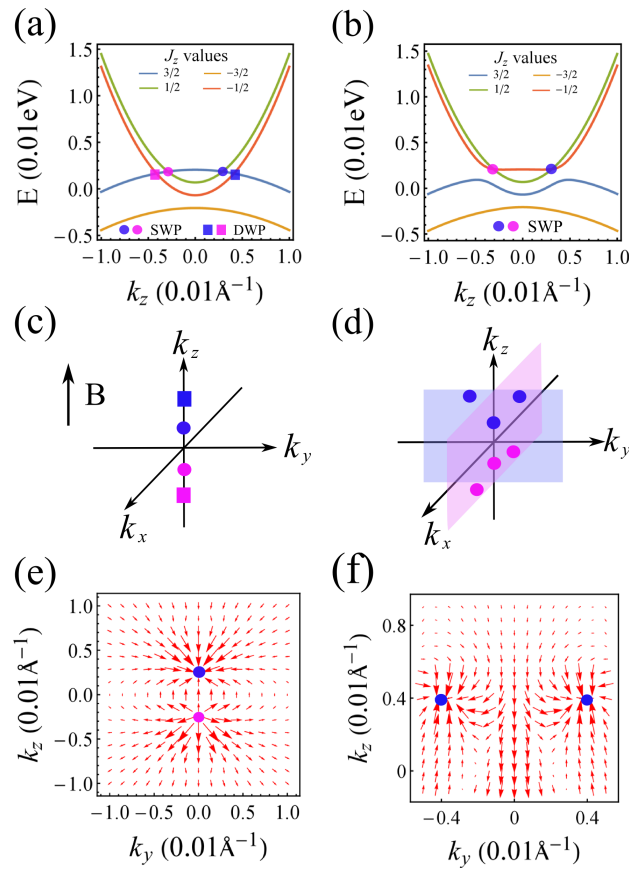}\\
  \caption{(Color online) (a) Band structure and (c) locations of Weyl points on the $k_{z}$ axis of $\alpha$-Sn under a magnetic field $B=1$ T in the [001] direction, where single-Weyl (double-Weyl) nodes are represented by circles (squares), and different colors stand for different chiralities. (b) Band structure and (d) locations of Weyl points in the presence of $H_{I}$ with $\alpha=0.2$. (e) Berry curvature configuration of the pair of single-Weyl points between $J_{z}=\frac{3}{2}$ and $J_{z}=\frac{1}{2}$ bands at $(0,0,\pm\sqrt{\frac{\kappa B}{2\gamma_{2}}}\approx\pm0.0029$\AA$^{-1})$. (f) Berry curvature configuration of the two single-Weyl points in the $k_{x}=0$ plane in the $k_{z}>0$ region.}
  \label{fig.3}
\end{figure}

So under a magnetic field $\mathbf{B}$ in the [001] direction, two Weyl points and two double-Weyl points are generated for unstrained $\alpha$-Sn. In the presence of some additional perturbations, each double-Weyl point splits into two Weyl points in the $k_{x}=0$ or $k_{y}=0$ plane, thus leading to altogether six Weyl points in $\alpha$-Sn. When $\mathbf{B}$ is applied along the [100] or [010] direction, identical analysis can be carried out due to the $C_{3,111}$ symmetry of the crystal.

\subsection{Strained $\alpha$-Sn}
When $\mathbf{B}$ is applied to strained $\alpha$-Sn, the number of generated Weyl points will depend on both the direction of $\mathbf{B}$ and the relative magnitude between Zeeman-induced and strain-induced band splittings. Moreover, in consideration of the broken $C_{3,111}$ symmetry due to the in-plane strain, $\alpha$-Sn should exhibit different phases under the magnetic field in the $[001]$ and $[100]$ directions. In this section, we will first consider the $[001]$ case and then the $[100]$ case.

\subsubsection{Magnetic field $\mathbf{B}$ in the $[001]$ direction}
\begin{figure*}[htbp]
  \centering
  \includegraphics[scale=0.92]{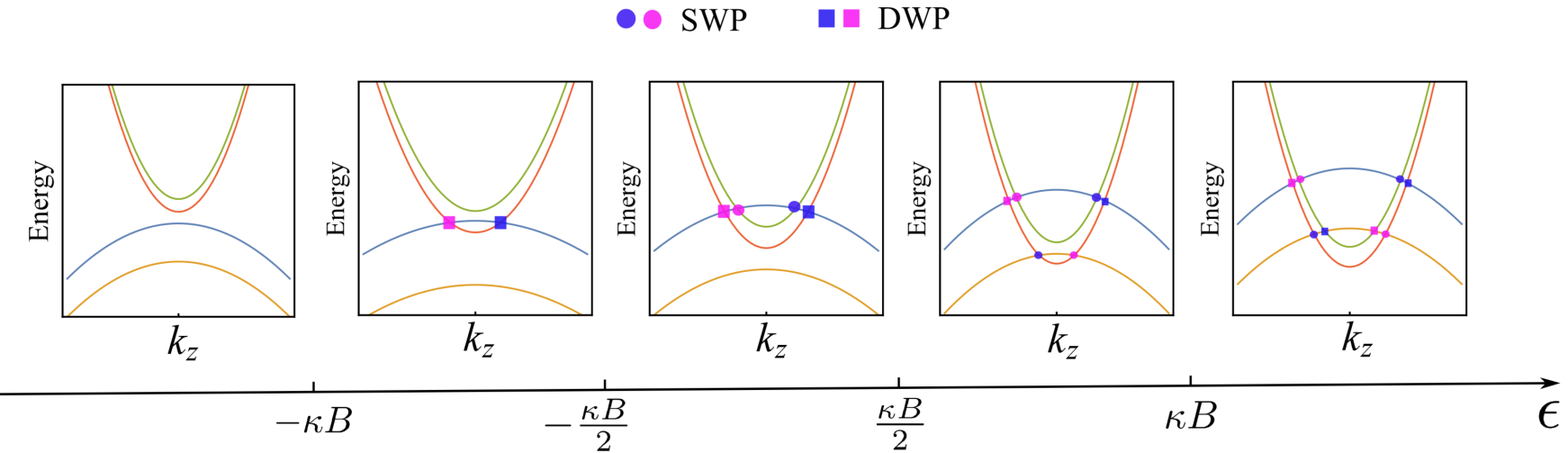}\\
  \caption{(Color online) $B$-$\epsilon$ phase diagram of Weyl points generated in strained $\alpha$-Sn under the magnetic field $\mathbf{B}$ in the [001] direction, where single-Weyl (double-Weyl) nodes are represented by circles (squares), and different colors stand for different chiralities. }
  \label{fig.4}
\end{figure*}
When $\mathbf{B}$ is applied along the [001] direction to strained $\alpha$-Sn, the Hamiltonian on the $k_{z}$ axis can be written as
\begin{equation}
\begin{split}
\label{}
H(k_{z})=&(\frac{\gamma_{1}}{2}+\frac{5\gamma_{2}}{4})k_{z}^{2}-\frac{5\epsilon}{4}
-(\gamma_{2}k_{z}^{2}-\epsilon)J_{z}^{2}+\kappa BJ_{z}.
\end{split}
\end{equation}
$J_{z}$ is still a good quantum number, and the energy spectra in the $J_{z}$ basis are simply given by
\begin{equation}
\begin{split}
\label{}
E_{\pm\frac{3}{2}}=&\frac{\gamma_{1}}{2}k_{z}^{2}-(\gamma_{2}k_{z}^{2}-\epsilon)\pm\frac{3}{2}\kappa B,\\
E_{\pm\frac{1}{2}}=&\frac{\gamma_{1}}{2}k_{z}^{2}+(\gamma_{2}k_{z}^{2}-\epsilon)\pm\frac{1}{2}\kappa B.
\end{split}
\end{equation}
Depending on the relation between strain-induced and Zeeman-induced band splittings, 0, 2, 4, 6 or 8 two-band crossings are found, as demonstrated by the evolution of band structures in Fig.~4. After the explicit derivation of the low-energy $k\cdot p$ effective Hamiltonian, as listed in Table.~I, these band crossings are found to be single-Weyl or double-Weyl points. The single-Weyl points reside between $\frac{3}{2}$ and $\frac{1}{2}$ bands or between $-\frac{1}{2}$ and $-\frac{3}{2}$ bands. They are protected by the intact $C_{2z}$ symmetry, since the crossing bands have different $C_{2z}$ eigenvalues. In contrast, double-Weyl points occur between $\frac{3}{2}$ and $-\frac{1}{2}$ bands or between $\frac{1}{2}$ and $-\frac{3}{2}$ bands, and they may split into single-Weyl points when extra perturbations are introduced.
\begin{table*}[htbp]
\centering
\caption{Properties of induced Weyl points in the coexistence of strain and magnetic field $\mathbf{B}$ in the [001] direction.}
\begin{tabular} {|c|c|c|c|c|}
  \hline
  Bands \& conditions & Location  & $H_{\mathrm{eff}}$& $\chi$ & Type\\
  \hline
   ($\frac{3}{2}$,$\frac{1}{2}$), $\epsilon>-\frac{\kappa B}{2}$&$(0,0,\pm k_{s1}=\pm\sqrt{\frac{\kappa B+2\epsilon}{2\gamma_{2}}})$  & $
\pm k_{s1}\Big[\gamma_{1}q_{z}-2\gamma_{2}q_{z}\sigma_{z}-\sqrt{3}\gamma_{3}(q_{x}\sigma_{x}+q_{y}\sigma_{y})\Big]$&$\mp 1$&single  \\
\hline
   ($\frac{3}{2}$,$\frac{-1}{2}$), $\epsilon>-\kappa B$& $(0,0,\pm k_{d1}=\pm\sqrt{\frac{\kappa B+\epsilon}{\gamma_{2}}})$ &$\pm\gamma_{1}k_{d1}q_{z}-\frac{\sqrt{3}\gamma_{2}}{2}(q_{x}^{2}-q_{y}^{2})\sigma_{x}-\sqrt{3}\gamma_{3}q_{x}q_{y}\sigma_{y}\mp2\gamma_{2}k_{d1}q_{z}\sigma_{z}$ &$\mp 2$& double  \\\hline
   ($\frac{1}{2}$, $\frac{-3}{2}$), $\epsilon>\kappa B$& $(0,0,\pm k_{d2}=\pm\sqrt{\frac{\epsilon-\kappa B}{\gamma_{2}}})$ &$\pm\gamma_{1}k_{d2}q_{z}-\frac{\sqrt{3}\gamma_{2}}{2}(q_{x}^{2}-q_{y}^{2})\sigma_{x}-\sqrt{3}\gamma_{3}q_{x}q_{y}\sigma_{y}\pm2\gamma_{2}k_{d2}q_{z}\sigma_{z}$  & $\pm2$ & double \\\hline
   ($\frac{-1}{2}$, $\frac{-3}{2}$),  $\epsilon>\frac{\kappa B}{2}$& $(0,0,\pm k_{s2}=\pm\sqrt{\frac{2\epsilon-\kappa B}{2\gamma_{2}}})$ &$\pm k_{s2}\Big[\gamma_{1}q_{z}+2\gamma_{2}q_{z}\sigma_{z}+\sqrt{3}\gamma_{3}(q_{x}\sigma_{x}+q_{y}\sigma_{y})\Big]$&$\pm1$ &single  \\
  \hline
\end{tabular}
\end{table*}

\subsubsection{Magnetic field $\mathbf{B}$ in the $[100]$ direction}
\begin{figure}[htbp]
  \centering
  \includegraphics[scale=0.65]{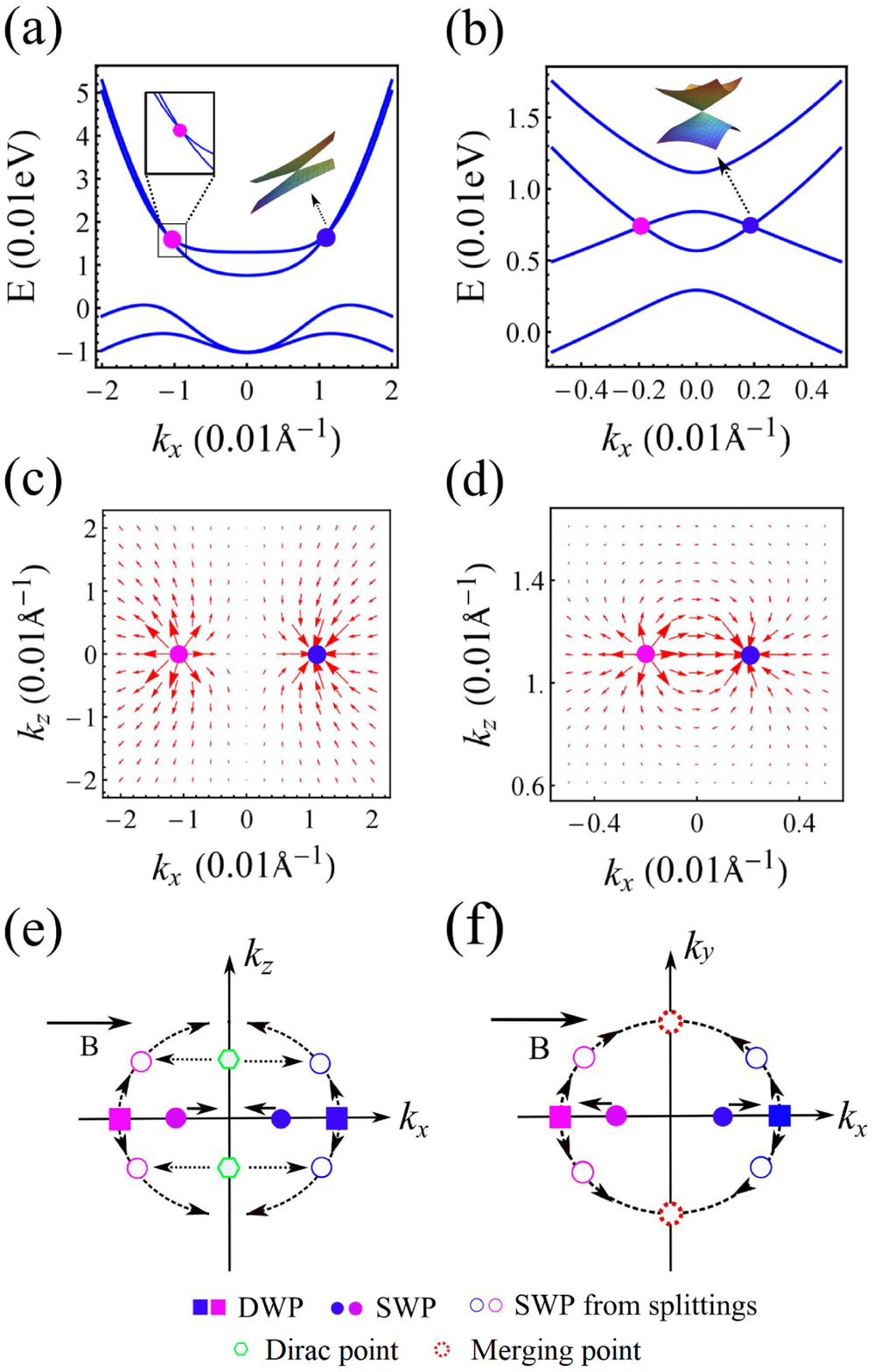}\\
  \caption{With the application of a magnetic field $\mathbf{B}=2$ T in the [100] direction. (a) Band structure on the $k_{x}$ axis under a tensile strain $\epsilon=-0.01$eV, where a pair of Weyl points are found between the third and fourth band. Inset: dispersion around the single-Weyl point. (b) Band structure at $k_{z}=k_{D}$ under a compressive strain $\epsilon=0.01$eV, where a pair of Weyl points exist between the second and third bands, which results from the splitting of the Dirac point in the $k_{z}>0$ region. Inset: dispersion around the single-Weyl point. (c) and (d) Berry curvature configurations of the Weyl points in (a) and (b), respectively. (e) and (f) Illustrations of the evolutions of single-Weyl (solid circles) and double-Weyl (solid squares) points with the increasing strength of a compressive and a tensile strain, respectively.}
  \label{fig.5}
\end{figure}

When $\mathbf{B}$ is applied along the [100] direction to strained $\alpha$-Sn, things become different due to the broken $C_{3,111}$ symmetry. For a better demonstration of the underlying physics, we first consider the Zeeman effect and then include the in-plane strain effect. According to above discussion on the Zeeman effect in unstrained $\alpha$-Sn, the application of magnetic field $\mathbf{B}$ along the [100] direction should give rise to a pair of single-Weyl points and a pair of double-Weyl points on the $k_{x}$ axis.

The inclusion of a strain in the $xy$ plane preserves $C_{2x}$ symmetry, so the single-Weyl points should still stay in the $k_x$ axis since they are protected by $C_{2x}$ symmetry. Depending on the direction of the strain (compressive or tensile), they move towards opposite directions on the $k_{x}$ axis, as we show in detail below. On the $k_{x}$ axis, the total Hamiltonian is given by
\begin{equation}
\begin{split}
\label{}
H(k_{x})=(\frac{\gamma_{1}}{2}+\frac{5}{4}\gamma_{2})k_{x}^{2}-\gamma_{2}k_{x}^{2}J_{x}^{2}+\epsilon(J_{z}^{2}-\frac{5}{4})+\kappa BJ_{x}.\\
\end{split}
\end{equation}
None of angular momentum operators $J_{x,y,z}$ commute with $H(k_{x})$, so we can no longer use their eigenvalues to label each band. Instead, we label them by their energies at the $\Gamma$ point as $E_{1}(\Gamma)<E_{2}(\Gamma)<E_{3}(\Gamma)<E_{4}(\Gamma)$. Band crossings are found only between $E_{3}$ and $E_{4}$ bands, which are located at $\mathbf{k_{w1}^{\pm}}=(\pm k_{x1},0,0)$, with $k_{x1}$ given as
\begin{equation}
\label{}
k_{x1}=\sqrt{\frac{\sqrt{\kappa^{2}B^{2}+\epsilon^{2}}-\epsilon}{2\gamma_{2}}},
\end{equation}
as concretely illustrated in Fig.~5(a) with $B=2$ T and $\epsilon=-0.01$ eV. These crossing points always exist regardless of the value of $\epsilon$. It is also verified that $E_{3}$ and $E_{4}$ bands have $C_{2x}(=i\sigma_{x}\tau_{x})$ eigenvalues $-i$ and $i$, respectively, and thus band hybridization between them is forbidden. Instead of analytically deriving the complex low-energy effective Hamiltonian around each crossing point, we numerically plot the energy dispersion around $\mathbf{k_{w1}^{+}}$ in the inset of Fig.~5(a) as well as the Berry curvature in Fig. 5(c), which indeed corresponds to a pair of Weyl fermions. Note that it has a large tilt in the $k_{z}$ direction and is a type-II Weyl fermion~\cite{Soluyanov2015, Wang2016, Aut2016, Tamai2016, Deng2016, Sun2015}. Moreover, according to Eq. (17), a compressive (tensile) strain with $\epsilon>0$ ($\epsilon<0$) pushes the pair of single-Weyl points towards (away from) each other on the $k_{x}$ axis, as illustrated in Fig. 5(e) (5(f)).

We continue to study the fate of the double-Weyl points on the $k_{x}$ axis after including strain effects. Under a compressive strain, each double-Weyl point is found to split apart into two Weyl points in the $k_{y}=0$ plane, and by increasing $\epsilon$, they evolve along the path denoted by oriented dashed lines in Fig. 5(e), where pairs of Weyl points approach each other but never merge. These Weyl points are protected by the combined $C_{2y}T$ symmetry as proved below. In the $C_{2y}T$ invariant $k_{y}=0$ plane, the Hamiltonian satisfies:
\begin{equation}
\label{}
\big(C_{2y}T\big)H(k_{x},0,k_{z})\big(C_{2y}T\big)^{-1}=H(k_{x},0,k_{z}).
 \end{equation}
In the $J_{z}$ basis, $C_{2y}=i\sigma_{x}\tau_{y}$ and $T=i\sigma_{x}\tau_{y}K$, where $K$ means complex conjugate. Their product yields $C_{2y}T=K$, and the Eq. (17) can be simplified as
\begin{equation}
\label{}
H(k_{x},0,k_{z})^{*}=H(k_{x},0,k_{z}).
\end{equation}
For a generic two-band Hamiltonian expanded by Pauli matrices, $H(k_{x},0,k_{z})=\sum_{i=0,x,y,z} d_{i}\sigma_{i}$, where $d_{i}$ are functions of $k_{x}$ and $k_{z}$, the above condition restricts that $d_{y}=0$. Occurrence of band crossings requires $d_{x}=d_{z}=0$, which can be satisfied by tuning the two parameters $k_{x}$ and $k_{z}$ in the $k_{y}=0$ plane. Once they are formed, small perturbations can only shift them within the plane but can not destroy them unless they meet each other and annihilate in pairs.

When the compressive strain is considered before the Zeeman effect, the Weyl points would have a clear physical picture induced from the Dirac points along the direction of magnetic field $\mathbf{B}$. At $k_{z}=k_{D}$, the total Hamiltonian is obtained as:
\begin{equation}
\begin{split}
\label{}
H(k_{x})=&\kappa BJ_{x}+(\frac{\gamma_{1}}{2}+\frac{5}{4}\gamma_{2})k_{x}^{2}-\gamma_{2}k_{x}^{2}J_{x}^{2}+\frac{\gamma_{1}\epsilon}{2\gamma_{2}}\\
&-\gamma_{3}k_{x}\sqrt{\frac{\epsilon}{\gamma_{2}}}\big\{J_{z},J_{x}\big\}.
\end{split}
\end{equation}
$J_{x,y,z}$ are not good quantum numbers and the four bands are labeled by their energies at the $\Gamma$ point as $E_{1}(\Gamma)<E_{2}(\Gamma)<E_{3}(\Gamma)<E_{4}(\Gamma)$. It is found that only $E_{2}$ and $E_{3}$ bands intersect with each other at $\mathbf{k_{w2}^{\pm}}=(\pm k_{x2},0,k_{D})$, where
\begin{equation}
\label{}
k_{x2}=\sqrt{\frac{\sqrt{4\gamma_{3}\epsilon^{2}+\gamma_{2}^{4}\kappa^{2}B^{2}}-2\gamma_{3}^{2}\epsilon}{\gamma_{2}^{3}}},
\end{equation}
This expression analytically gives the location of the Weyl points in the $k_{y}=0$ plane in the $k_{z}>0$ region, as verified by the concrete band structure in Fig. 5(b) with $B=2$ T and $\epsilon=0.01$ eV. Analogously, the splitting of the Dirac point at $\mathbf{k_{D}^{-}}$ generates the other two Weyl points in the $k_{z}<0$ region. We have also numerically calculated the band structure around $\mathbf{k_{w2}^{+}}$ in the inset of Fig.~5(b) and the corresponding Berry curvature configuration in Fig. 5(d) as a further proof. Equation (21) also conforms to the finding that these Weyl points never meet each other and annihilate. This can be explained by the notion that once a compressive in-plane strain is imposed, regardless of its value, Dirac points always emerge on the $k_{z}$ axis, and a nonzero magnetic field $\mathbf{B}$ along $[100]$ direction inevitably splits it and gives rise the above Weyl points.

However, under a tensile strain, each double-Weyl point splits into two Weyl points in the $k_{z}=0$ plane which evolves with increasing strength of the strain along the dashed path in Fig. 5(f) until pairs of Weyl points meet and annihilate with each other on the $k_{y}$ axis. These Weyl points are protected by the combined $C_{2z}T$ symmetry. Since $C_{2z}T(=i\sigma_{x}\tau_{x}K)$ is an antiunitary operator that squares to one, it can be represented by $K$ through basis transformation. Then a similar argument can be made on the stability of a generic two-band crossing in the $k_{z}=0$ plane. In addition, the disappearance of these Weyl points under a large tensile strain is due to the fact that such a strain generates a full gap between the lower two bands and upper two bands (or $E2$ and $E3$), which survives under the small Zeeman splitting. Thus no gapless points such as Weyl points can be found between the second and third band.

Finally, we sweep through the whole BZ, and no more Weyl points are found. So, when the magnetic field $\mathbf{B}$ is applied along the [100] direction to strained $\alpha$-Sn, two Weyl points are always induced on the $k_{x}$ axis, which are protected by $C_{2x}$ symmetry. For a compressively strained $\alpha$-Sn in the Dirac semimetal phase, four additional Weyl points are always generated on the $k_{y}=0$ plane, which originate from the splittings of Dirac points and are protected by the combined $C_{2y}T$ symmetry. In contrast, for a tensile-strained $\alpha$-Sn in the topological insulator phase, four additional Weyl points can be found on the $k_{z}=0$ plane only under a small tensile strain, which are protected by $C_{2z}T$ symmetry but will annihilate in pairs and totally vanish beyond a critical tensile strain strength.

\section{Weyl semimetals induced by a circularly-polarized light}
Strictly speaking, for $\alpha$-Sn with such an anomalously high mobility, which is quite easy to enter the quantum limit, we should also take orbital Landau level effects into consideration even with a relatively low magnetic field. This poses much difficulty in analyzing the energy spectrum of $\alpha$-Sn. Differently, an off-resonant circularly polarized light (CPL) is an alternative TRS-breaking scheme to generate Weyl points, and it has been widely used in photoinduced Floquet topological semimetal phases~\cite{Narayan2015, Hubener2017, Chan2016a, Chan2016b, Yan2016, Taguchi2016, Bomantara2016, Rui2014, Chen2016, Zhang2016, Bomantara2016generating, Zhou2016Floquet, Yan2017, Ezawa2017, Gupta2017, Wang2017Line}. For simplicity, in this section, we investigate the case where the applied CPL propagates along the [001] direction. Since the CPL breaks TRS, it is expected that Weyl points may split from Dirac points in compressively strained $\alpha$-Sn with the Dirac semimetal phase~\cite{Chan2016a, Chan2016b, Hubener2017}. However, based on the $k\cdot p$ model analysis under the off-resonant approximation, we find that Weyl points can be generated from not only compressively strained $\alpha$-Sn, but also unstrained and tensile-strained $\alpha$-Sn.

For a CPL incident along the [001] direction, as shown in Fig. 6(a), the vector potential $\mathbf{A}$ is given as
\begin{equation}
\label{eqn.23}
\mathbf{A}=A(\cos{\omega t},\eta\sin{\omega t},0),
\end{equation}
where $\eta=+1(-1)$ denotes right(left) CPL, $A$ represents the amplitude of the vector potential, and $\omega$ is the frequency of the CPL. The time-dependant periodic Hamiltonian $H(\mathbf{k},t)$ can be obtained by taking the Peierls substitution $\mathbf{k}\rightarrow \mathbf{k}+e\mathbf{A}$ (henceforth, $e$ is absorbed into $A$ for simplicity). For calculational convenience, we introduce the following five $4\times 4$ $\Gamma$ matrices:
\begin{equation}
\begin{split}
\label{eqn.3}
\Gamma_{1}=&\frac{1}{\sqrt{3}}(J_{x}^{2}-J_{y}^{2})=\sigma_{x}\mathrm{I},\\
\Gamma_{2}=&\frac{1}{3}(2J_{z}^{2}-J_{x}^{2}-J_{y}^{2})=\sigma_{z}\tau_{z},\\
\Gamma_{3}=&\frac{1}{\sqrt{3}}\{J_{x},J_{y}\}=\sigma_{y}\mathrm{I},\\
\Gamma_{4}=&\frac{1}{\sqrt{3}}\{J_{z},J_{x}\}=\sigma_{z}\tau_{x},\\
\Gamma_{5}=&\frac{1}{\sqrt{3}}\{J_{y},J_{z}\}=\sigma_{z}\tau_{y},\\
\end{split}
\end{equation}
where $\sigma_{i}$ and $\tau_{i}$ are conventional Pauli matrices, and $\Gamma_{1...5}$ satisfy the Clifford algebra $\{\Gamma_{i},{\Gamma_{j}}\}=2\delta_{i,j}$. With these matrices, the Luttinger Hamiltonian can be rewritten as
\begin{equation}
\begin{split}
\label{eqn.4}
H_{\mathrm{L}}(\mathbf{k})=&\frac{\gamma_{1}}{2}k^{2}
-\frac{\sqrt{3}\gamma_{2}}{2}\big(k_{x}^{2}-k_{y}^{2}\big)\Gamma_{1}\\
&-\frac{\gamma_{2}}{2}\big(2k_{z}^{2}-k_{x}^{2}-k_{y}^{2}\big)\Gamma_{2}\\
&-\sqrt{3}\gamma_{3}\big(k_{x}k_{y}\Gamma_{3}+k_{z}k_{x}\Gamma_{4}+k_{y}k_{z}\Gamma_{5}\big).
\end{split}
\end{equation}

\begin{table*}[htbp]
\centering
\caption{Properties of Weyl points induced in $\alpha$-Sn by a strain in the $xy$ plane and circularly-polarized light incident in the [001] direction.}
\begin{tabular} {|c|c|c|c|c|}
  \hline
  Bands & Conditions & Location on the $k_{z}$ axis & Type\\
  \hline
   $\big(\frac{3}{2}$,$\frac{1}{2}\big)$& $\epsilon>-\frac{\gamma_{2}A^{2}}{2}\& \gamma_{2}>-\frac{3\eta A^{2}\gamma_{3}^{2}}{\omega}$ & $\pm\sqrt{\frac{\epsilon+(\gamma_{2}A^{2}/2)}{\gamma_{2}+(3\eta A^{2}\gamma_{3}^{2} /\omega)}} $ &single\\
\hline
$\big(\frac{3}{2}$,$-\frac{1}{2}\big)$& $\epsilon>-\gamma_{2}A^{2}\big(\frac{1}{2}-\frac{3\eta\gamma_{3}A^{2}}{8\omega}\big)$ & $\pm\sqrt{\frac{\epsilon}{\gamma_{2}}+\frac{A^{2}}{2}-\frac{3\eta\gamma_{3}A^{4}}{8\omega}}$ &double\\
\hline
   $\big(\frac{1}{2}$,$-\frac{1}{2}\big)$& always exist & $\pm\sqrt{\frac{\gamma_{2}A^{2}}{8\gamma_{3}}} $ &single \\
\hline
 $\big(\frac{1}{2}$,$-\frac{3}{2}\big)$& $\epsilon>-\gamma_{2}A^{2}\big(\frac{1}{2}+\frac{3\eta\gamma_{3}A^{2}}{8\omega}\big)$ & $\pm\sqrt{\frac{\epsilon}{\gamma_{2}}+\frac{A^{2}}{2}+\frac{3\eta\gamma_{3}A^{4}}{8\omega}} $ &double\\
\hline
$\big(-\frac{1}{2}$,$-\frac{3}{2}\big)$& $\epsilon>-\frac{\gamma_{2}A^{2}}{2}\& \gamma_{2}>\frac{3\eta A^{2}\gamma_{3}^{2}}{\omega}$ & $\pm\sqrt{\frac{\epsilon+(\gamma_{2}A^{2}/2)}{\gamma_{2}-(3\eta A^{2}\gamma_{3}^{2} /\omega)}} $ &single \\
\hline
\end{tabular}
\end{table*}

For an off-resonant light, i.e., when $\omega$ is large compared to the system's energy scale, the off-resonant approximation can be taken~\cite{Kitagawa2011, Yan2016}, and the system can be described by the following effective static Hamiltonian:
\begin{equation}
\label{eqn.24}
H_{\mathrm{eff}}(\mathbf{k})=H_{0}(\mathbf{k})+\Delta H_{0}+\sum_{n\geq1}\frac{[H_{-n},H_{n}]}{n\omega}+O(\frac{1}{\omega^{2}})
\end{equation}
where $H_{n}=\frac{1}{T}\int^{T}_{0}H(t)e^{i n\omega t}dt$ is the Fourier component in the frequency space. $\Delta H_{0}=\frac{A^{2}}{2}(\gamma_{1}+\gamma_{2}\Gamma_{2})$ describes the correction to the zeroth-order Hamiltonian in the presence of a CPL. The $\frac{[H_{-n},H_{n}]}{n\omega}$ term stems from virtual photon absorption and emission processes, which belongs to a second-order perturbation and can be derived as

\begin{equation}
\begin{split}
\label{}
&\frac{[H_{-1},H_{1}]}{\omega}=\frac{\eta A^{2}}{\omega}\Big\{-2\sqrt{3}\gamma_{2}^{2}k_{x}k_{y}\Gamma_{12}\\
&+3\gamma_{2}\gamma_{3}\Big[(k_{x}^{2}+k_{y}^{2})\Gamma_{13}+k_{y}k_{z}\Gamma_{14}+k_{z}k_{x}\Gamma_{15}\Big]\\
&-\sqrt{3}\gamma_{2}\gamma_{3}\Big[(-k_{x}^{2}+k_{y}^{2})\Gamma_{23}-k_{y}k_{z}\Gamma_{24}+k_{z}k_{x}\Gamma_{25}\Big]\\
&-3\gamma_{3}^{2}k_{z}(k_{x}\Gamma_{34}-k_{y}\Gamma_{35}-k_{z}\Gamma_{45})\Big\},
\end{split}
\end{equation}

\begin{equation}
\begin{split}
\label{}
\frac{[H_{-2},H_{2}]}{2\omega}=&\frac{3\eta A^{4}\gamma_{2}\gamma_{3}}{8\omega}\Gamma_{13},
\end{split}
\end{equation}
and
\begin{equation}
\label{}
\frac{[H_{-n},H_{n}]}{n\omega}=0, (n>2)
\end{equation}
where $\Gamma_{ab}\equiv-\frac{1}{2i}[\Gamma_{a},\Gamma_{b}]$. Based on the reduced crystal symmetry, we focus on the $k_{z}$ axis with the following effective Hamiltonian:
\begin{equation}
\begin{split}
\label{eqn.26}
H_{\mathrm{eff}}(k_{z})=&\frac{\gamma_{1}}{2}\big(k_{z}^{2}+A^{2}\big)-\big(\gamma_{2}k_{z}^{2}-\epsilon-\frac{\gamma_{2}}{2}A^{2}\big)\Gamma_{2}\\
&+\frac{3\eta A^{2}}{\omega}\gamma_{3}^{2} k_{z}^{2}\Gamma_{45}+\frac{3\eta A^{4}}{8\omega}\gamma_{2}\gamma_{3}\Gamma_{13},
\end{split}
\end{equation}
where an in-plane strain term is included. Here, $\Gamma_{2}=\sigma_{z}\tau_{z}$, $\Gamma_{45}=-\sigma_{0}\tau_{z}$, and $\Gamma_{13}=-\sigma_{z}\tau_{0}$, so $H_{\mathrm{eff}}(k_{z})$ is diagonal in the $J_{z}$ basis and each band can still be labeled by its $J_{z}$ eigenvalue. The $\frac{\gamma_{2}}{2}A^{2}\Gamma_{2}$  term plays the role of an effective compressive strain, which pushes up (down) the down-dispersive $J_{z}=\pm\frac{3}{2}$ (the up-dispersive $J_{z}=\pm\frac{1}{2}$) bands. The $\Gamma_{45}$ term modifies the parabolic dispersion of each band, where the coefficient before $k_{z}^{2}$ is changed by $-\frac{3}{\omega}\eta A^{2}\gamma_{3}^{2}$ $\big(+\frac{3}{\omega}\eta A^{2}\gamma_{3}^{2}\big)$ for the $J_{z}=\frac{3}{2}$ and $-\frac{1}{2}$ $\big(J_{z}=-\frac{3}{2}$ and $\frac{1}{2}\big)$ bands. The $\Gamma_{13}$ term acts as a Zeeman-like term which splits both $J_{z}=\pm\frac{3}{2}$ bands and $J_{z}=\pm\frac{1}{2}$ bands.
\begin{figure}[htbp]
  \centering
  \includegraphics[scale=0.9]{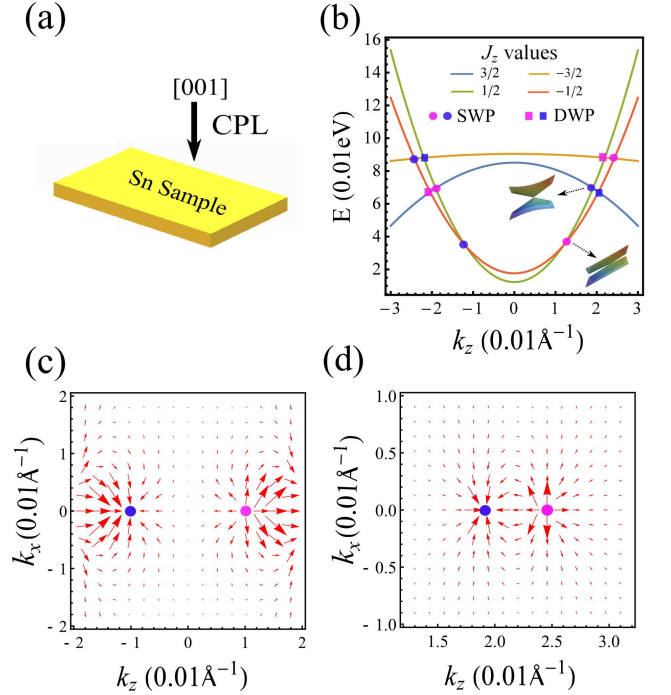}\\
  \caption{(Color online) (a) Schematic illustration of applying a CPL to $\alpha$-Sn. (b) Band structure of unstrained $\alpha$-Sn under a right CPL ($\eta=1$) with $A=0.03$ \AA$^{-1}$ and $\omega=0.6$ eV. Insets: dispersions of the Weyl points between $\pm\frac{1}{2}$ bands and between $\frac{3}{2}$ and $\frac{1}{2}$ bands, respectively. (c) and (d) Berry curvature configurations of the single-Weyl points between $\pm\frac{1}{2}$ bands and between the second and third bands, respectively.}
  \label{fig.6}
\end{figure}
Depending on $\epsilon$, $A$ and $\omega$, different numbers of Weyl points can be generated, whose detailed information is listed in Table.II.  All the single-Weyl points are protected by $C_{2z}$ symmetry because the corresponding crossing bands have different $C_{2z}$ eigenvalues. In contrast, similar to the discussions in previous sections, the double-Weyl points may split into single-Weyl points when introducing extra perturbations. It should be emphasized that there always exists a pair of single-Weyl points between the $J_{z}=\pm\frac{1}{2}$ bands irrespective of the values of $\epsilon$ or $\omega$, whose location only depends on $A$.

As a concrete example, in Fig. 6(b), we present the band structure of unstrained $\alpha$-Sn under a right CPL ($\eta=1$) with $A=0.03$ \AA$^{-1}$ and $\omega=0.6$ eV ($\sim8\times10^{14}$Hz), where three pairs of single-Weyl points and two pairs of double-Weyl points can be found on the $k_{z}$ axis, which agrees with Table.II. We also plot the low-energy dispersions of the single crossing points between the $J_{z}=\pm\frac{1}{2}$ bands and between the $J_{z}=\frac{3}{2}$ and $\frac{1}{2}$ bands in the inset of Fig. 6(b) as well as their Berry curvature configurations in Figs. 6(c) and 6(d), respectively. By decreasing $A$, the pair of Weyl points between $J_{z}=\pm\frac{1}{2}$ bands move towards each other on the $k_{z}$ axis until they finally annihilate at the $\Gamma$ point when $A=0$, while for other pairs of Weyl points, by decreasing $A$, although they also move towards each other, whether they annihilate or not depends on the direction of strain. For a tensile strain, they will meet and annihilate in pairs at the $\Gamma$ point for a nonzero critical $A$, while for a compressive strain, they will not meet until $A=0$, where they overlap with each other and constitute Dirac points, thus recovering the Dirac semimetal phase.

So tunable Weyl semimetal phases can be induced from (un)strained $\alpha$-Sn under a CPL propagating in the [001] direction, where the number and locations of generated Weyl points can be easily manipulated by both the strain and the CPL. Compared to previous studies on photo-induced Weyl semimetal phases from Dirac materials, our proposal does not require a Dirac semimetal phase as a prerequisite, and instead both type-I and type-II Weyl points can be induced directly from $\alpha$-Sn.

\section{Conclusion}
In summary, through the effective $k\cdot p$ model analysis and first-principles calculations, we have shown that multiple topological phases can be induced in $\alpha$-Sn in the presence of external strains, magnetic fields, and circularly polarized light. An in-plane biaxial tensile (compressive) strain changes $\alpha$-Sn into a topological insulator (Dirac semimetal). A magnetic field or circularly polarized light alone can drive $\alpha$-Sn into a Weyl semimetal phase, where type-I Weyl nodes, type-II Weyl nodes, and double-Weyl nodes can all be generated. Further, we present a rich phase diagram with tunable number, type, and locations of Weyl nodes by considering an in-plane strain and a magnetic field or circularly polarized light simultaneously. Our results can also be applied to other Luttinger semimetals. Our findings suggest the Luttinger materials as a versatile platform to realize and engineer various topological phases.

\textit{Note added.} When preparing this manuscript, we became aware of a recent paper~\cite{Ghorashi2018} which also investigated an irradiated three-dimensional Luttinger semimetal and found Floquet Weyl and double-Weyl nodes.

\appendix
\section{Calculation of the monopole charge for both single-Weyl and double-Weyl nodes}
In this section, we explicitly derive the monopole charge of the Weyl nodes in the main text. Both a single-Weyl node and a double-Weyl node can be described by a two-band Hamiltonian:
\begin{equation}
\label{A1}
 h(\mathbf{k})=\sum_{i}d_{i}(\mathbf{k})\sigma_{i},
\end{equation}
where $\sigma_{i=x,y,z}$ are Pauli matrices. The dispersions of the conduction and valence bands are given by $\pm d$, with $d=\sqrt{d_{x}^{2}+d_{y}^{2}+d_{z}^{2}}$. The three-dimensional Berry curvature of the valence bands can be written as $\mathbf{\Omega}=(\Omega_{x},\Omega_{y},\Omega_{z})=(F_{yz},F_{zx},F_{xy})$, with ~\cite{Bernevig2013Topological}
\begin{equation}
\label{A2}
F_{ij}=\frac{1}{2d^{3}}\epsilon_{abc}d_{a}\partial_{i}d_{b}\partial_{j}d_{c}.
\end{equation}
Then the monopole charge $\chi$ of each node can be obtained by integrating the Berry curvature over a sphere that encloses the node
\begin{equation}
\label{A3}
\chi=\frac{1}{2\pi}\int_{\Sigma}d\mathbf{S}\cdot\mathbf{\Omega}.
\end{equation}

Concretely, we first consider the single-Weyl nodes in our paper, which takes the form
\begin{equation}
\label{A4}
h_{\mathrm{s}}(\mathbf{k})=v_{x}k_{x}\sigma_{x}+v_{y}k_{y}\sigma_{y}+v_{z}k_{z}\sigma_{z}.
\end{equation}
The dispersions $E_{S}=\pm\sqrt{\sum_{i}v_{i}^{2}k_{i}^{2}}$ are linear in all three directions. By Eq. (A2), we get
\begin{equation}
\label{A5}
\mathbf{\Omega_{\mathrm{s}}}=\frac{v_{x}v_{y}v_{z}}{2d^{3}}\Big(k_{x},k_{y},k_{z}\Big)=\mathbf{sgn}[v_{x}v_{y}v_{z}]\frac{\mathbf{\widetilde{k}}}{2\widetilde{k}^{3}},
\end{equation}
where $\mathbf{\widetilde{k}}\equiv\big(|v_{x}|k_{x},|v_{y}|k_{y},|v_{z}|k_{z}\big)$. Via Eq. (A3), we obtain
\begin{equation}
\label{A6}
\chi_{\mathrm{s}}=\mathbf{sgn}[v_{x}v_{y}v_{z}].
\end{equation}
Based on Eq. (A6) The chirality of each single-Weyl node in the main text can now be readily obtained. For example, for the single-Weyl node in Eq. (9), $v_{x}(\mathbf{\mathbf{k_{sw}^{\pm}}})=v_{y}(\mathbf{\mathbf{k_{sw}^{\pm}}})= \mp\sqrt{3}\gamma_{3}k_{\mathrm{sw}}$, and $v_{z}(\mathbf{\mathbf{k_{sw}^{\pm}}})=\mp2\gamma_{2}k_{\mathrm{sw}}$, and thus $\chi_{\mathbf{\mathbf{k_{sw}^{\pm}}}}=\mp1$.\\

We go on to study the double-Weyl nodes in our paper with the form
\begin{equation}
\label{A7}
h_{\mathrm{D}}(\mathbf{k})=v_{x}(k_{x}^{2}-k_{y}^{2})\sigma_{x}+2v_{y}k_{x}k_{y}\sigma_{y}+v_{z}k_{z}\sigma_{z}.
\end{equation}
The dispersions are now given by $E=\pm\sqrt{v_{z}^{2}k_{z}^{2}+v_{x}^{2}(k_{x}^{2}-k_{y}^{2})^{2}+4v_{y}^{2}k_{x}^{2}k_{y}^{2}}$, which take a linear form in the $k_{z}$ direction and a quadratic form in both the $k_{x}$ and $k_{y}$ directions. Similarly, through Eq. (A2), we obtain
\begin{equation}
\label{A8}
\mathbf{\Omega_{\mathrm{d}}}=\frac{v_{x}v_{y}v_{z}(k_{x}^{2}+k_{y}^{2})}{d^{3}}\Big(k_{x},k_{y},2k_{z}\Big).
\end{equation}
It can be seen from Eq. (A8) that whether $\mathbf{\Omega_{d}}$ points inwards or outwards depends on $\mathbf{sgn}[v_{x}v_{y}v_{z}]$. In order to derive the monopole charge of each double-Weyl node, instead of choosing a sphere, we now choose a closed surface composed of two infinitely large $k_{x}-k_{y}$ planes at $k_{z}=0^{+}$ and $0^{-}$, respectively, and two infinitesimal side surfaces. By calculating the Berry curvature flux threading this closed surface, we obtain
\begin{equation}
\label{A9}
\chi_{\mathrm{d}}=\frac{1}{2\pi}\int dk_{x} dk_{y}(\Omega_{z}^{+}-\Omega_{z}^{-})=2\mathbf{sgn}[v_{x}v_{y}v_{z}].
\end{equation}
This expression can also be understood from the fact that the monopole charge of the node equals the change of the Chern number of the two-dimensional slice when it passes through the node. Now the chirality of the double-Weyl node in this paper can be determined by Eq. (A9). Take the double-Weyl node in Eq. (10) for instance, the velocities are given by $v_{x}(\mathbf{\mathbf{k_{dw}^{\pm}}})=-\frac{\sqrt{3}}{2}\gamma_{2}$, $v_{y}(\mathbf{\mathbf{k_{dw}^{\pm}}})= -\frac{\sqrt{3}}{2}\gamma_{3}$, and $v_{z}(\mathbf{\mathbf{k_{dw}^{\pm}}})=\mp2\gamma_{2} k_{\mathrm{dw}}$, leading to $\chi_{\mathbf{\mathbf{k_{dw}^{\pm}}}}=\mp2$.

In addition, the calculations get simplified for an isotropic Luttinger semimetal with $\gamma_{2}=\gamma_{3}$. In this case the velocities in the $k_{x}-k_{y}$ plane are isotropic as $v_{x}=v_{y}=v_{\bot}$, and thus both the single-Weyl and double-Weyl nodes can be written in the compact form ~\cite{lu2017quantum}:
\begin{equation}
\label{A10}
H=\left[
    \begin{array}{cc}
      v_{z}k_{z} & v_{\bot}(k_{-})^{N} \\
      v_{\bot}(k_{+})^{N} & -v_{z}k_{z} \\
    \end{array}
  \right],
\end{equation}
where $k_{\pm}=k_{x}\pm ik_{y}$, and $N=1$ $(N=2)$ describes a single-Weyl (double-Weyl) node. The dispersions of the conduction and valence bands are given by $\pm E$ with $E=\sqrt{(v_{z}k_{z})^{2}+\big(v_{\bot}(k_{\bot})^{N}\big)^{2}}$.  The eigenstate of the valence band is obtained as
\begin{equation}
\label{A11}
|u_{-}\rangle=\left[ \begin{array}{c} -\sin(\frac{\theta}{2})\\ e^{iN\varphi}\cos(\frac{\theta}{2}) \\ \end{array}  \right],
\end{equation}
where $\cos\theta=\frac{v_{z}q_{z}}{E}$ and $\tan\varphi=\frac{q_{y}}{q_{x}}$. For an arbitrary two-dimensional sphere enclosing the node, the Berry connection is defined as $\mathbf{A}=(A_{\theta}, A_{\varphi})$, with~\cite{Xiao2010, Bernevig2013Topological}
\begin{equation}
\begin{split}
A_{\theta}=& i\langle u_{-}|\partial_{\theta}|u_{-}\rangle=0 \\
A_{\varphi}=& i\langle u_{-}|\partial_{\varphi}|u_{-}\rangle=-N\cos^{2}(\frac{\theta}{2}). \\
\end{split}
\end{equation}
The Berry curvature can then be determined as
\begin{equation}
F_{\theta\varphi}=\partial_{\theta}A_{\varphi}-\partial_{\varphi}A_{\theta}=\frac{N\sin\theta}{2}.
\end{equation}
Finally, by integrating the Berry curvature over the two-dimensional sphere, the monopole charge is given by
\begin{equation}
\chi=\frac{\mathrm{sgn}(v_{z})}{2\pi}\int^{2\pi}_{0}d\varphi\int^{\pi}_{0}d\theta F_{\theta\varphi}=\mathbf{sgn}(v_{z})N,
\end{equation}
where $\mathbf{sgn}(v_{z})$ comes from opposite integration directions of $\theta$ for positive and negative $v_{z}$ values. Now the chirality of each Weyl node can be simply determined as $\mathbf{sgn}(v_{z})N$.

\begin{acknowledgments}
We thank L.B. Shao and M. N. Chen for helpful discussions. This work was supported by the National Natural Science Foundation of China (No. 11674165), the Fok Ying-Tong Education Foundation of China (Grant No. 161006) and the Fundamental Research Funds for the Central Universities (No. 020414380038).

Dongqin Zhang and Huai-Qiang Wang equally contributed to this work.
\end{acknowledgments}

\bibliography{reference}

\begin{thebibliography}{139}%
\makeatletter
\providecommand \@ifxundefined [1]{%
 \@ifx{#1\undefined}
}%
\providecommand \@ifnum [1]{%
 \ifnum #1\expandafter \@firstoftwo
 \else \expandafter \@secondoftwo
 \fi
}%
\providecommand \@ifx [1]{%
 \ifx #1\expandafter \@firstoftwo
 \else \expandafter \@secondoftwo
 \fi
}%
\providecommand \natexlab [1]{#1}%
\providecommand \enquote  [1]{``#1''}%
\providecommand \bibnamefont  [1]{#1}%
\providecommand \bibfnamefont [1]{#1}%
\providecommand \citenamefont [1]{#1}%
\providecommand \href@noop [0]{\@secondoftwo}%
\providecommand \href [0]{\begingroup \@sanitize@url \@href}%
\providecommand \@href[1]{\@@startlink{#1}\@@href}%
\providecommand \@@href[1]{\endgroup#1\@@endlink}%
\providecommand \@sanitize@url [0]{\catcode `\\12\catcode `\$12\catcode
  `\&12\catcode `\#12\catcode `\^12\catcode `\_12\catcode `\%12\relax}%
\providecommand \@@startlink[1]{}%
\providecommand \@@endlink[0]{}%
\providecommand \url  [0]{\begingroup\@sanitize@url \@url }%
\providecommand \@url [1]{\endgroup\@href {#1}{\urlprefix }}%
\providecommand \urlprefix  [0]{URL }%
\providecommand \Eprint [0]{\href }%
\providecommand \doibase [0]{http://dx.doi.org/}%
\providecommand \selectlanguage [0]{\@gobble}%
\providecommand \bibinfo  [0]{\@secondoftwo}%
\providecommand \bibfield  [0]{\@secondoftwo}%
\providecommand \translation [1]{[#1]}%
\providecommand \BibitemOpen [0]{}%
\providecommand \bibitemStop [0]{}%
\providecommand \bibitemNoStop [0]{.\EOS\space}%
\providecommand \EOS [0]{\spacefactor3000\relax}%
\providecommand \BibitemShut  [1]{\csname bibitem#1\endcsname}%
\let\auto@bib@innerbib\@empty
\bibitem [{\citenamefont {Hasan}\ and\ \citenamefont {Kane}(2010)}]{Hasan2010}%
  \BibitemOpen
  \bibfield  {author} {\bibinfo {author} {\bibfnamefont {M.~Z.}\ \bibnamefont
  {Hasan}}\ and\ \bibinfo {author} {\bibfnamefont {C.~L.}\ \bibnamefont
  {Kane}},\ }\href {\doibase 10.1103/RevModPhys.82.3045} {\bibfield  {journal}
  {\bibinfo  {journal} {Rev. Mod. Phys.}\ }\textbf {\bibinfo {volume} {82}},\
  \bibinfo {pages} {3045} (\bibinfo {year} {2010})}\BibitemShut {NoStop}%
\bibitem [{\citenamefont {Qi}\ and\ \citenamefont {Zhang}(2011)}]{Qi2010}%
  \BibitemOpen
  \bibfield  {author} {\bibinfo {author} {\bibfnamefont {X.-L.}\ \bibnamefont
  {Qi}}\ and\ \bibinfo {author} {\bibfnamefont {S.-C.}\ \bibnamefont {Zhang}},\
  }\href {\doibase 10.1103/RevModPhys.83.1057} {\bibfield  {journal} {\bibinfo
  {journal} {Rev. Mod. Phys.}\ }\textbf {\bibinfo {volume} {83}},\ \bibinfo
  {pages} {1057} (\bibinfo {year} {2011})}\BibitemShut {NoStop}%
\bibitem [{\citenamefont {Kane}\ and\ \citenamefont
  {Mele}(2005)}]{kane2005quantum}%
  \BibitemOpen
  \bibfield  {author} {\bibinfo {author} {\bibfnamefont {C.~L.}\ \bibnamefont
  {Kane}}\ and\ \bibinfo {author} {\bibfnamefont {E.~J.}\ \bibnamefont
  {Mele}},\ }\href@noop {} {\bibfield  {journal} {\bibinfo  {journal} {Phys.
  Rev. Lett.}\ }\textbf {\bibinfo {volume} {95}},\ \bibinfo {pages} {226801}
  (\bibinfo {year} {2005})}\BibitemShut {NoStop}%
\bibitem [{\citenamefont {Bernevig}\ \emph {et~al.}(2006)\citenamefont
  {Bernevig}, \citenamefont {Hughes},\ and\ \citenamefont
  {Zhang}}]{Bernevig2006quantum}%
  \BibitemOpen
  \bibfield  {author} {\bibinfo {author} {\bibfnamefont {B.~A.}\ \bibnamefont
  {Bernevig}}, \bibinfo {author} {\bibfnamefont {T.~L.}\ \bibnamefont
  {Hughes}}, \ and\ \bibinfo {author} {\bibfnamefont {S.-C.}\ \bibnamefont
  {Zhang}},\ }\href@noop {} {\bibfield  {journal} {\bibinfo  {journal}
  {Science}\ }\textbf {\bibinfo {volume} {314}},\ \bibinfo {pages} {1757}
  (\bibinfo {year} {2006})}\BibitemShut {NoStop}%
\bibitem [{\citenamefont {K{\"o}nig}\ \emph {et~al.}(2007)\citenamefont
  {K{\"o}nig}, \citenamefont {Wiedmann}, \citenamefont {Br{\"u}ne},
  \citenamefont {Roth}, \citenamefont {Buhmann}, \citenamefont {Molenkamp},
  \citenamefont {Qi},\ and\ \citenamefont {Zhang}}]{Konig2007quantum}%
  \BibitemOpen
  \bibfield  {author} {\bibinfo {author} {\bibfnamefont {M.}~\bibnamefont
  {K{\"o}nig}}, \bibinfo {author} {\bibfnamefont {S.}~\bibnamefont {Wiedmann}},
  \bibinfo {author} {\bibfnamefont {C.}~\bibnamefont {Br{\"u}ne}}, \bibinfo
  {author} {\bibfnamefont {A.}~\bibnamefont {Roth}}, \bibinfo {author}
  {\bibfnamefont {H.}~\bibnamefont {Buhmann}}, \bibinfo {author} {\bibfnamefont
  {L.~W.}\ \bibnamefont {Molenkamp}}, \bibinfo {author} {\bibfnamefont {X.-L.}\
  \bibnamefont {Qi}}, \ and\ \bibinfo {author} {\bibfnamefont {S.-C.}\
  \bibnamefont {Zhang}},\ }\href@noop {} {\bibfield  {journal} {\bibinfo
  {journal} {Science}\ }\textbf {\bibinfo {volume} {318}},\ \bibinfo {pages}
  {766} (\bibinfo {year} {2007})}\BibitemShut {NoStop}%
\bibitem [{\citenamefont {Liu}\ \emph {et~al.}(2008)\citenamefont {Liu},
  \citenamefont {Qi}, \citenamefont {Dai}, \citenamefont {Fang},\ and\
  \citenamefont {Zhang}}]{liu2008}%
  \BibitemOpen
  \bibfield  {author} {\bibinfo {author} {\bibfnamefont {C.-X.}\ \bibnamefont
  {Liu}}, \bibinfo {author} {\bibfnamefont {X.-L.}\ \bibnamefont {Qi}},
  \bibinfo {author} {\bibfnamefont {X.}~\bibnamefont {Dai}}, \bibinfo {author}
  {\bibfnamefont {Z.}~\bibnamefont {Fang}}, \ and\ \bibinfo {author}
  {\bibfnamefont {S.-C.}\ \bibnamefont {Zhang}},\ }\href {\doibase
  10.1103/PhysRevLett.101.146802} {\bibfield  {journal} {\bibinfo  {journal}
  {Phys. Rev. Lett.}\ }\textbf {\bibinfo {volume} {101}},\ \bibinfo {pages}
  {146802} (\bibinfo {year} {2008})}\BibitemShut {NoStop}%
\bibitem [{\citenamefont {Yu}\ \emph {et~al.}(2010)\citenamefont {Yu},
  \citenamefont {Zhang}, \citenamefont {Zhang}, \citenamefont {Zhang},
  \citenamefont {Dai},\ and\ \citenamefont {Fang}}]{Yu2010}%
  \BibitemOpen
  \bibfield  {author} {\bibinfo {author} {\bibfnamefont {R.}~\bibnamefont
  {Yu}}, \bibinfo {author} {\bibfnamefont {W.}~\bibnamefont {Zhang}}, \bibinfo
  {author} {\bibfnamefont {H.-J.}\ \bibnamefont {Zhang}}, \bibinfo {author}
  {\bibfnamefont {S.-C.}\ \bibnamefont {Zhang}}, \bibinfo {author}
  {\bibfnamefont {X.}~\bibnamefont {Dai}}, \ and\ \bibinfo {author}
  {\bibfnamefont {Z.}~\bibnamefont {Fang}},\ }\href {\doibase
  10.1126/science.1187485} {\bibfield  {journal} {\bibinfo  {journal}
  {Science}\ }\textbf {\bibinfo {volume} {329}},\ \bibinfo {pages} {61}
  (\bibinfo {year} {2010})}\BibitemShut {NoStop}%
\bibitem [{\citenamefont {Chang}\ \emph {et~al.}(2013)\citenamefont {Chang},
  \citenamefont {Zhang}, \citenamefont {Feng}, \citenamefont {Shen},
  \citenamefont {Zhang}, \citenamefont {Guo}, \citenamefont {Li}, \citenamefont
  {Ou}, \citenamefont {Wei}, \citenamefont {Wang}, \citenamefont {Ji},
  \citenamefont {Feng}, \citenamefont {Ji}, \citenamefont {Chen}, \citenamefont
  {Jia}, \citenamefont {Dai}, \citenamefont {Fang}, \citenamefont {Zhang},
  \citenamefont {He}, \citenamefont {Wang}, \citenamefont {Lu}, \citenamefont
  {Ma},\ and\ \citenamefont {Xue}}]{Chang2013}%
  \BibitemOpen
  \bibfield  {author} {\bibinfo {author} {\bibfnamefont {C.-Z.}\ \bibnamefont
  {Chang}}, \bibinfo {author} {\bibfnamefont {J.}~\bibnamefont {Zhang}},
  \bibinfo {author} {\bibfnamefont {X.}~\bibnamefont {Feng}}, \bibinfo {author}
  {\bibfnamefont {J.}~\bibnamefont {Shen}}, \bibinfo {author} {\bibfnamefont
  {Z.}~\bibnamefont {Zhang}}, \bibinfo {author} {\bibfnamefont
  {M.}~\bibnamefont {Guo}}, \bibinfo {author} {\bibfnamefont {K.}~\bibnamefont
  {Li}}, \bibinfo {author} {\bibfnamefont {Y.}~\bibnamefont {Ou}}, \bibinfo
  {author} {\bibfnamefont {P.}~\bibnamefont {Wei}}, \bibinfo {author}
  {\bibfnamefont {L.-L.}\ \bibnamefont {Wang}}, \bibinfo {author}
  {\bibfnamefont {Z.-Q.}\ \bibnamefont {Ji}}, \bibinfo {author} {\bibfnamefont
  {Y.}~\bibnamefont {Feng}}, \bibinfo {author} {\bibfnamefont {S.}~\bibnamefont
  {Ji}}, \bibinfo {author} {\bibfnamefont {X.}~\bibnamefont {Chen}}, \bibinfo
  {author} {\bibfnamefont {J.}~\bibnamefont {Jia}}, \bibinfo {author}
  {\bibfnamefont {X.}~\bibnamefont {Dai}}, \bibinfo {author} {\bibfnamefont
  {Z.}~\bibnamefont {Fang}}, \bibinfo {author} {\bibfnamefont {S.-C.}\
  \bibnamefont {Zhang}}, \bibinfo {author} {\bibfnamefont {K.}~\bibnamefont
  {He}}, \bibinfo {author} {\bibfnamefont {Y.}~\bibnamefont {Wang}}, \bibinfo
  {author} {\bibfnamefont {L.}~\bibnamefont {Lu}}, \bibinfo {author}
  {\bibfnamefont {X.-C.}\ \bibnamefont {Ma}}, \ and\ \bibinfo {author}
  {\bibfnamefont {Q.-K.}\ \bibnamefont {Xue}},\ }\href {\doibase
  10.1126/science.1234414} {\bibfield  {journal} {\bibinfo  {journal}
  {Science}\ }\textbf {\bibinfo {volume} {340}},\ \bibinfo {pages} {167}
  (\bibinfo {year} {2013})}\BibitemShut {NoStop}%
\bibitem [{\citenamefont {Weng}\ \emph
  {et~al.}(2015{\natexlab{a}})\citenamefont {Weng}, \citenamefont {Yu},
  \citenamefont {Hu}, \citenamefont {Dai},\ and\ \citenamefont
  {Fang}}]{weng2015quantum}%
  \BibitemOpen
  \bibfield  {author} {\bibinfo {author} {\bibfnamefont {H.}~\bibnamefont
  {Weng}}, \bibinfo {author} {\bibfnamefont {R.}~\bibnamefont {Yu}}, \bibinfo
  {author} {\bibfnamefont {X.}~\bibnamefont {Hu}}, \bibinfo {author}
  {\bibfnamefont {X.}~\bibnamefont {Dai}}, \ and\ \bibinfo {author}
  {\bibfnamefont {Z.}~\bibnamefont {Fang}},\ }\href@noop {} {\bibfield
  {journal} {\bibinfo  {journal} {Adv. Phys.}\ }\textbf {\bibinfo {volume}
  {64}},\ \bibinfo {pages} {227} (\bibinfo {year}
  {2015}{\natexlab{a}})}\BibitemShut {NoStop}%
\bibitem [{\citenamefont {Fu}\ \emph {et~al.}(2007)\citenamefont {Fu},
  \citenamefont {Kane},\ and\ \citenamefont {Mele}}]{Fu2007Topological}%
  \BibitemOpen
  \bibfield  {author} {\bibinfo {author} {\bibfnamefont {L.}~\bibnamefont
  {Fu}}, \bibinfo {author} {\bibfnamefont {C.~L.}\ \bibnamefont {Kane}}, \ and\
  \bibinfo {author} {\bibfnamefont {E.~J.}\ \bibnamefont {Mele}},\ }\href
  {\doibase 10.1103/PhysRevLett.98.106803} {\bibfield  {journal} {\bibinfo
  {journal} {Phys. Rev. Lett.}\ }\textbf {\bibinfo {volume} {98}},\ \bibinfo
  {pages} {106803} (\bibinfo {year} {2007})}\BibitemShut {NoStop}%
\bibitem [{\citenamefont {Zhang}\ \emph
  {et~al.}(2009{\natexlab{a}})\citenamefont {Zhang}, \citenamefont {Liu},
  \citenamefont {Qi}, \citenamefont {Dai}, \citenamefont {Fang},\ and\
  \citenamefont {Zhang}}]{zhang2009topological}%
  \BibitemOpen
  \bibfield  {author} {\bibinfo {author} {\bibfnamefont {H.}~\bibnamefont
  {Zhang}}, \bibinfo {author} {\bibfnamefont {C.-X.}\ \bibnamefont {Liu}},
  \bibinfo {author} {\bibfnamefont {X.-L.}\ \bibnamefont {Qi}}, \bibinfo
  {author} {\bibfnamefont {X.}~\bibnamefont {Dai}}, \bibinfo {author}
  {\bibfnamefont {Z.}~\bibnamefont {Fang}}, \ and\ \bibinfo {author}
  {\bibfnamefont {S.-C.}\ \bibnamefont {Zhang}},\ }\href@noop {} {\bibfield
  {journal} {\bibinfo  {journal} {Nat. Phys.}\ }\textbf {\bibinfo {volume}
  {5}},\ \bibinfo {pages} {438} (\bibinfo {year}
  {2009}{\natexlab{a}})}\BibitemShut {NoStop}%
\bibitem [{\citenamefont {Chen}\ \emph {et~al.}(2009)\citenamefont {Chen},
  \citenamefont {Analytis}, \citenamefont {Chu}, \citenamefont {Liu},
  \citenamefont {Mo}, \citenamefont {Qi}, \citenamefont {Zhang}, \citenamefont
  {Lu}, \citenamefont {Dai}, \citenamefont {Fang}, \citenamefont {Zhang},
  \citenamefont {Fisher}, \citenamefont {Hussain},\ and\ \citenamefont
  {Shen}}]{Chen2009}%
  \BibitemOpen
  \bibfield  {author} {\bibinfo {author} {\bibfnamefont {Y.~L.}\ \bibnamefont
  {Chen}}, \bibinfo {author} {\bibfnamefont {J.~G.}\ \bibnamefont {Analytis}},
  \bibinfo {author} {\bibfnamefont {J.-H.}\ \bibnamefont {Chu}}, \bibinfo
  {author} {\bibfnamefont {Z.~K.}\ \bibnamefont {Liu}}, \bibinfo {author}
  {\bibfnamefont {S.-K.}\ \bibnamefont {Mo}}, \bibinfo {author} {\bibfnamefont
  {X.~L.}\ \bibnamefont {Qi}}, \bibinfo {author} {\bibfnamefont {H.~J.}\
  \bibnamefont {Zhang}}, \bibinfo {author} {\bibfnamefont {D.~H.}\ \bibnamefont
  {Lu}}, \bibinfo {author} {\bibfnamefont {X.}~\bibnamefont {Dai}}, \bibinfo
  {author} {\bibfnamefont {Z.}~\bibnamefont {Fang}}, \bibinfo {author}
  {\bibfnamefont {S.~C.}\ \bibnamefont {Zhang}}, \bibinfo {author}
  {\bibfnamefont {I.~R.}\ \bibnamefont {Fisher}}, \bibinfo {author}
  {\bibfnamefont {Z.}~\bibnamefont {Hussain}}, \ and\ \bibinfo {author}
  {\bibfnamefont {Z.-X.}\ \bibnamefont {Shen}},\ }\href {\doibase
  10.1126/science.1173034} {\bibfield  {journal} {\bibinfo  {journal}
  {Science}\ }\textbf {\bibinfo {volume} {325}},\ \bibinfo {pages} {178}
  (\bibinfo {year} {2009})}\BibitemShut {NoStop}%
\bibitem [{\citenamefont {Fu}\ and\ \citenamefont {Kane}(2008)}]{Fu2008}%
  \BibitemOpen
  \bibfield  {author} {\bibinfo {author} {\bibfnamefont {L.}~\bibnamefont
  {Fu}}\ and\ \bibinfo {author} {\bibfnamefont {C.~L.}\ \bibnamefont {Kane}},\
  }\href {\doibase 10.1103/PhysRevLett.100.096407} {\bibfield  {journal}
  {\bibinfo  {journal} {Phys. Rev. Lett.}\ }\textbf {\bibinfo {volume} {100}},\
  \bibinfo {pages} {096407} (\bibinfo {year} {2008})}\BibitemShut {NoStop}%
\bibitem [{\citenamefont {Sau}\ \emph {et~al.}(2010)\citenamefont {Sau},
  \citenamefont {Lutchyn}, \citenamefont {Tewari},\ and\ \citenamefont
  {Das~Sarma}}]{Sau2010}%
  \BibitemOpen
  \bibfield  {author} {\bibinfo {author} {\bibfnamefont {J.~D.}\ \bibnamefont
  {Sau}}, \bibinfo {author} {\bibfnamefont {R.~M.}\ \bibnamefont {Lutchyn}},
  \bibinfo {author} {\bibfnamefont {S.}~\bibnamefont {Tewari}}, \ and\ \bibinfo
  {author} {\bibfnamefont {S.}~\bibnamefont {Das~Sarma}},\ }\href {\doibase
  10.1103/PhysRevLett.104.040502} {\bibfield  {journal} {\bibinfo  {journal}
  {Phys. Rev. Lett.}\ }\textbf {\bibinfo {volume} {104}},\ \bibinfo {pages}
  {040502} (\bibinfo {year} {2010})}\BibitemShut {NoStop}%
\bibitem [{\citenamefont {Qi}\ \emph {et~al.}(2010)\citenamefont {Qi},
  \citenamefont {Hughes},\ and\ \citenamefont {Zhang}}]{Qi2010b}%
  \BibitemOpen
  \bibfield  {author} {\bibinfo {author} {\bibfnamefont {X.-L.}\ \bibnamefont
  {Qi}}, \bibinfo {author} {\bibfnamefont {T.~L.}\ \bibnamefont {Hughes}}, \
  and\ \bibinfo {author} {\bibfnamefont {S.-C.}\ \bibnamefont {Zhang}},\ }\href
  {\doibase 10.1103/PhysRevB.82.184516} {\bibfield  {journal} {\bibinfo
  {journal} {Phys. Rev. B}\ }\textbf {\bibinfo {volume} {82}},\ \bibinfo
  {pages} {184516} (\bibinfo {year} {2010})}\BibitemShut {NoStop}%
\bibitem [{\citenamefont {Alicea}(2010)}]{Alicea2010}%
  \BibitemOpen
  \bibfield  {author} {\bibinfo {author} {\bibfnamefont {J.}~\bibnamefont
  {Alicea}},\ }\href {\doibase 10.1103/PhysRevB.81.125318} {\bibfield
  {journal} {\bibinfo  {journal} {Phys. Rev. B}\ }\textbf {\bibinfo {volume}
  {81}},\ \bibinfo {pages} {125318} (\bibinfo {year} {2010})}\BibitemShut
  {NoStop}%
\bibitem [{\citenamefont {He}\ \emph {et~al.}(2017)\citenamefont {He},
  \citenamefont {Pan}, \citenamefont {Stern}, \citenamefont {Burks},
  \citenamefont {Che}, \citenamefont {Yin}, \citenamefont {Wang}, \citenamefont
  {Lian}, \citenamefont {Zhou}, \citenamefont {Choi}, \citenamefont {Murata},
  \citenamefont {Kou}, \citenamefont {Chen}, \citenamefont {Nie}, \citenamefont
  {Shao}, \citenamefont {Fan}, \citenamefont {Zhang}, \citenamefont {Liu},
  \citenamefont {Xia},\ and\ \citenamefont {Wang}}]{he2017chiral}%
  \BibitemOpen
  \bibfield  {author} {\bibinfo {author} {\bibfnamefont {Q.~L.}\ \bibnamefont
  {He}}, \bibinfo {author} {\bibfnamefont {L.}~\bibnamefont {Pan}}, \bibinfo
  {author} {\bibfnamefont {A.~L.}\ \bibnamefont {Stern}}, \bibinfo {author}
  {\bibfnamefont {E.~C.}\ \bibnamefont {Burks}}, \bibinfo {author}
  {\bibfnamefont {X.}~\bibnamefont {Che}}, \bibinfo {author} {\bibfnamefont
  {G.}~\bibnamefont {Yin}}, \bibinfo {author} {\bibfnamefont {J.}~\bibnamefont
  {Wang}}, \bibinfo {author} {\bibfnamefont {B.}~\bibnamefont {Lian}}, \bibinfo
  {author} {\bibfnamefont {Q.}~\bibnamefont {Zhou}}, \bibinfo {author}
  {\bibfnamefont {E.~S.}\ \bibnamefont {Choi}}, \bibinfo {author}
  {\bibfnamefont {K.}~\bibnamefont {Murata}}, \bibinfo {author} {\bibfnamefont
  {X.}~\bibnamefont {Kou}}, \bibinfo {author} {\bibfnamefont {Z.}~\bibnamefont
  {Chen}}, \bibinfo {author} {\bibfnamefont {T.}~\bibnamefont {Nie}}, \bibinfo
  {author} {\bibfnamefont {Q.}~\bibnamefont {Shao}}, \bibinfo {author}
  {\bibfnamefont {Y.}~\bibnamefont {Fan}}, \bibinfo {author} {\bibfnamefont
  {S.-C.}\ \bibnamefont {Zhang}}, \bibinfo {author} {\bibfnamefont
  {K.}~\bibnamefont {Liu}}, \bibinfo {author} {\bibfnamefont {J.}~\bibnamefont
  {Xia}}, \ and\ \bibinfo {author} {\bibfnamefont {K.~L.}\ \bibnamefont
  {Wang}},\ }\href {\doibase 10.1126/science.aag2792} {\bibfield  {journal}
  {\bibinfo  {journal} {Science}\ }\textbf {\bibinfo {volume} {357}},\ \bibinfo
  {pages} {294} (\bibinfo {year} {2017})}\BibitemShut {NoStop}%
\bibitem [{\citenamefont {Wan}\ \emph {et~al.}(2011)\citenamefont {Wan},
  \citenamefont {Turner}, \citenamefont {Vishwanath},\ and\ \citenamefont
  {Savrasov}}]{Wan2011}%
  \BibitemOpen
  \bibfield  {author} {\bibinfo {author} {\bibfnamefont {X.}~\bibnamefont
  {Wan}}, \bibinfo {author} {\bibfnamefont {A.~M.}\ \bibnamefont {Turner}},
  \bibinfo {author} {\bibfnamefont {A.}~\bibnamefont {Vishwanath}}, \ and\
  \bibinfo {author} {\bibfnamefont {S.~Y.}\ \bibnamefont {Savrasov}},\ }\href
  {\doibase 10.1103/PhysRevB.83.205101} {\bibfield  {journal} {\bibinfo
  {journal} {Phys. Rev. B}\ }\textbf {\bibinfo {volume} {83}},\ \bibinfo
  {pages} {205101} (\bibinfo {year} {2011})}\BibitemShut {NoStop}%
\bibitem [{\citenamefont {Xu}\ \emph {et~al.}(2011)\citenamefont {Xu},
  \citenamefont {Weng}, \citenamefont {Wang}, \citenamefont {Dai},\ and\
  \citenamefont {Fang}}]{Xu2011}%
  \BibitemOpen
  \bibfield  {author} {\bibinfo {author} {\bibfnamefont {G.}~\bibnamefont
  {Xu}}, \bibinfo {author} {\bibfnamefont {H.}~\bibnamefont {Weng}}, \bibinfo
  {author} {\bibfnamefont {Z.}~\bibnamefont {Wang}}, \bibinfo {author}
  {\bibfnamefont {X.}~\bibnamefont {Dai}}, \ and\ \bibinfo {author}
  {\bibfnamefont {Z.}~\bibnamefont {Fang}},\ }\href {\doibase
  10.1103/PhysRevLett.107.186806} {\bibfield  {journal} {\bibinfo  {journal}
  {Phys. Rev. Lett.}\ }\textbf {\bibinfo {volume} {107}},\ \bibinfo {pages}
  {186806} (\bibinfo {year} {2011})}\BibitemShut {NoStop}%
\bibitem [{\citenamefont {Yang}\ \emph {et~al.}(2011)\citenamefont {Yang},
  \citenamefont {Lu},\ and\ \citenamefont {Ran}}]{yang2011}%
  \BibitemOpen
  \bibfield  {author} {\bibinfo {author} {\bibfnamefont {K.-Y.}\ \bibnamefont
  {Yang}}, \bibinfo {author} {\bibfnamefont {Y.-M.}\ \bibnamefont {Lu}}, \ and\
  \bibinfo {author} {\bibfnamefont {Y.}~\bibnamefont {Ran}},\ }\href {\doibase
  10.1103/PhysRevB.84.075129} {\bibfield  {journal} {\bibinfo  {journal} {Phys.
  Rev. B}\ }\textbf {\bibinfo {volume} {84}},\ \bibinfo {pages} {075129}
  (\bibinfo {year} {2011})}\BibitemShut {NoStop}%
\bibitem [{\citenamefont {Burkov}\ and\ \citenamefont
  {Balents}(2011)}]{Burkov2011a}%
  \BibitemOpen
  \bibfield  {author} {\bibinfo {author} {\bibfnamefont {A.~A.}\ \bibnamefont
  {Burkov}}\ and\ \bibinfo {author} {\bibfnamefont {L.}~\bibnamefont
  {Balents}},\ }\href {\doibase 10.1103/PhysRevLett.107.127205} {\bibfield
  {journal} {\bibinfo  {journal} {Phys. Rev. Lett.}\ }\textbf {\bibinfo
  {volume} {107}},\ \bibinfo {pages} {127205} (\bibinfo {year}
  {2011})}\BibitemShut {NoStop}%
\bibitem [{\citenamefont {Burkov}\ \emph {et~al.}(2011)\citenamefont {Burkov},
  \citenamefont {Hook},\ and\ \citenamefont {Balents}}]{Burkov2011b}%
  \BibitemOpen
  \bibfield  {author} {\bibinfo {author} {\bibfnamefont {A.~A.}\ \bibnamefont
  {Burkov}}, \bibinfo {author} {\bibfnamefont {M.~D.}\ \bibnamefont {Hook}}, \
  and\ \bibinfo {author} {\bibfnamefont {L.}~\bibnamefont {Balents}},\ }\href
  {\doibase 10.1103/PhysRevB.84.235126} {\bibfield  {journal} {\bibinfo
  {journal} {Phys. Rev. B}\ }\textbf {\bibinfo {volume} {84}},\ \bibinfo
  {pages} {235126} (\bibinfo {year} {2011})}\BibitemShut {NoStop}%
\bibitem [{\citenamefont {Hal\'asz}\ and\ \citenamefont
  {Balents}(2012)}]{Halasz2012}%
  \BibitemOpen
  \bibfield  {author} {\bibinfo {author} {\bibfnamefont {G.~B.}\ \bibnamefont
  {Hal\'asz}}\ and\ \bibinfo {author} {\bibfnamefont {L.}~\bibnamefont
  {Balents}},\ }\href {\doibase 10.1103/PhysRevB.85.035103} {\bibfield
  {journal} {\bibinfo  {journal} {Phys. Rev. B}\ }\textbf {\bibinfo {volume}
  {85}},\ \bibinfo {pages} {035103} (\bibinfo {year} {2012})}\BibitemShut
  {NoStop}%
\bibitem [{\citenamefont {Zyuzin}\ \emph {et~al.}(2012)\citenamefont {Zyuzin},
  \citenamefont {Wu},\ and\ \citenamefont {Burkov}}]{Zyuzin2012}%
  \BibitemOpen
  \bibfield  {author} {\bibinfo {author} {\bibfnamefont {A.~A.}\ \bibnamefont
  {Zyuzin}}, \bibinfo {author} {\bibfnamefont {S.}~\bibnamefont {Wu}}, \ and\
  \bibinfo {author} {\bibfnamefont {A.~A.}\ \bibnamefont {Burkov}},\ }\href
  {\doibase 10.1103/PhysRevB.85.165110} {\bibfield  {journal} {\bibinfo
  {journal} {Phys. Rev. B}\ }\textbf {\bibinfo {volume} {85}},\ \bibinfo
  {pages} {165110} (\bibinfo {year} {2012})}\BibitemShut {NoStop}%
\bibitem [{\citenamefont {Lu}\ \emph {et~al.}(2012)\citenamefont {Lu},
  \citenamefont {Fu}, \citenamefont {Joannopoulos},\ and\ \citenamefont
  {Soljacic}}]{Lu2012}%
  \BibitemOpen
  \bibfield  {author} {\bibinfo {author} {\bibfnamefont {L.}~\bibnamefont
  {Lu}}, \bibinfo {author} {\bibfnamefont {L.}~\bibnamefont {Fu}}, \bibinfo
  {author} {\bibfnamefont {J.}~\bibnamefont {Joannopoulos}}, \ and\ \bibinfo
  {author} {\bibfnamefont {M.}~\bibnamefont {Soljacic}},\ }\href@noop {}
  {\bibfield  {journal} {\bibinfo  {journal} {Nat. Photon.}\ }\textbf {\bibinfo
  {volume} {7}},\ \bibinfo {pages} {294} (\bibinfo {year} {2012})}\BibitemShut
  {NoStop}%
\bibitem [{\citenamefont {Das}(2013)}]{Das2013}%
  \BibitemOpen
  \bibfield  {author} {\bibinfo {author} {\bibfnamefont {T.}~\bibnamefont
  {Das}},\ }\href {\doibase 10.1103/PhysRevB.88.035444} {\bibfield  {journal}
  {\bibinfo  {journal} {Phys. Rev. B}\ }\textbf {\bibinfo {volume} {88}},\
  \bibinfo {pages} {035444} (\bibinfo {year} {2013})}\BibitemShut {NoStop}%
\bibitem [{\citenamefont {Liu}\ and\ \citenamefont
  {Vanderbilt}(2014)}]{Liu2014}%
  \BibitemOpen
  \bibfield  {author} {\bibinfo {author} {\bibfnamefont {J.}~\bibnamefont
  {Liu}}\ and\ \bibinfo {author} {\bibfnamefont {D.}~\bibnamefont
  {Vanderbilt}},\ }\href {\doibase 10.1103/PhysRevB.90.155316} {\bibfield
  {journal} {\bibinfo  {journal} {Phys. Rev. B}\ }\textbf {\bibinfo {volume}
  {90}},\ \bibinfo {pages} {155316} (\bibinfo {year} {2014})}\BibitemShut
  {NoStop}%
\bibitem [{\citenamefont {Zhang}\ \emph {et~al.}(2014)\citenamefont {Zhang},
  \citenamefont {Wang}, \citenamefont {Xu}, \citenamefont {Xu},\ and\
  \citenamefont {Zhang}}]{zhang2014a}%
  \BibitemOpen
  \bibfield  {author} {\bibinfo {author} {\bibfnamefont {H.}~\bibnamefont
  {Zhang}}, \bibinfo {author} {\bibfnamefont {J.}~\bibnamefont {Wang}},
  \bibinfo {author} {\bibfnamefont {G.}~\bibnamefont {Xu}}, \bibinfo {author}
  {\bibfnamefont {Y.}~\bibnamefont {Xu}}, \ and\ \bibinfo {author}
  {\bibfnamefont {S.-C.}\ \bibnamefont {Zhang}},\ }\href@noop {} {\bibfield
  {journal} {\bibinfo  {journal} {Phys. Rev. Lett.}\ }\textbf {\bibinfo
  {volume} {112}},\ \bibinfo {pages} {096804} (\bibinfo {year}
  {2014})}\BibitemShut {NoStop}%
\bibitem [{\citenamefont {Weng}\ \emph
  {et~al.}(2015{\natexlab{b}})\citenamefont {Weng}, \citenamefont {Fang},
  \citenamefont {Fang}, \citenamefont {Bernevig},\ and\ \citenamefont
  {Dai}}]{Weng2015}%
  \BibitemOpen
  \bibfield  {author} {\bibinfo {author} {\bibfnamefont {H.}~\bibnamefont
  {Weng}}, \bibinfo {author} {\bibfnamefont {C.}~\bibnamefont {Fang}}, \bibinfo
  {author} {\bibfnamefont {Z.}~\bibnamefont {Fang}}, \bibinfo {author}
  {\bibfnamefont {B.~A.}\ \bibnamefont {Bernevig}}, \ and\ \bibinfo {author}
  {\bibfnamefont {X.}~\bibnamefont {Dai}},\ }\href {\doibase
  10.1103/PhysRevX.5.011029} {\bibfield  {journal} {\bibinfo  {journal} {Phys.
  Rev. X}\ }\textbf {\bibinfo {volume} {5}},\ \bibinfo {pages} {011029}
  (\bibinfo {year} {2015}{\natexlab{b}})}\BibitemShut {NoStop}%
\bibitem [{\citenamefont {Xu}\ \emph {et~al.}(2015{\natexlab{a}})\citenamefont
  {Xu}, \citenamefont {Belopolski}, \citenamefont {Alidoust}, \citenamefont
  {Neupane}, \citenamefont {Bian}, \citenamefont {Zhang}, \citenamefont
  {Sankar}, \citenamefont {Chang}, \citenamefont {Yuan}, \citenamefont {Lee},
  \citenamefont {Huang}, \citenamefont {Zheng}, \citenamefont {Ma},
  \citenamefont {Sanchez}, \citenamefont {Wang}, \citenamefont {Bansil},
  \citenamefont {Chou}, \citenamefont {Shibayev}, \citenamefont {Lin},
  \citenamefont {Jia},\ and\ \citenamefont {Hasan}}]{Xu2015a}%
  \BibitemOpen
  \bibfield  {author} {\bibinfo {author} {\bibfnamefont {S.-Y.}\ \bibnamefont
  {Xu}}, \bibinfo {author} {\bibfnamefont {I.}~\bibnamefont {Belopolski}},
  \bibinfo {author} {\bibfnamefont {N.}~\bibnamefont {Alidoust}}, \bibinfo
  {author} {\bibfnamefont {M.}~\bibnamefont {Neupane}}, \bibinfo {author}
  {\bibfnamefont {G.}~\bibnamefont {Bian}}, \bibinfo {author} {\bibfnamefont
  {C.}~\bibnamefont {Zhang}}, \bibinfo {author} {\bibfnamefont
  {R.}~\bibnamefont {Sankar}}, \bibinfo {author} {\bibfnamefont
  {G.}~\bibnamefont {Chang}}, \bibinfo {author} {\bibfnamefont
  {Z.}~\bibnamefont {Yuan}}, \bibinfo {author} {\bibfnamefont {C.-C.}\
  \bibnamefont {Lee}}, \bibinfo {author} {\bibfnamefont {S.-M.}\ \bibnamefont
  {Huang}}, \bibinfo {author} {\bibfnamefont {H.}~\bibnamefont {Zheng}},
  \bibinfo {author} {\bibfnamefont {J.}~\bibnamefont {Ma}}, \bibinfo {author}
  {\bibfnamefont {D.~S.}\ \bibnamefont {Sanchez}}, \bibinfo {author}
  {\bibfnamefont {B.}~\bibnamefont {Wang}}, \bibinfo {author} {\bibfnamefont
  {A.}~\bibnamefont {Bansil}}, \bibinfo {author} {\bibfnamefont
  {F.}~\bibnamefont {Chou}}, \bibinfo {author} {\bibfnamefont {P.~P.}\
  \bibnamefont {Shibayev}}, \bibinfo {author} {\bibfnamefont {H.}~\bibnamefont
  {Lin}}, \bibinfo {author} {\bibfnamefont {S.}~\bibnamefont {Jia}}, \ and\
  \bibinfo {author} {\bibfnamefont {M.~Z.}\ \bibnamefont {Hasan}},\ }\href
  {\doibase 10.1126/science.aaa9297} {\bibfield  {journal} {\bibinfo  {journal}
  {Science}\ }\textbf {\bibinfo {volume} {349}},\ \bibinfo {pages} {613}
  (\bibinfo {year} {2015}{\natexlab{a}})}\BibitemShut {NoStop}%
\bibitem [{\citenamefont {Lv}\ \emph {et~al.}(2015{\natexlab{a}})\citenamefont
  {Lv}, \citenamefont {Weng}, \citenamefont {Fu}, \citenamefont {Wang},
  \citenamefont {Miao}, \citenamefont {Ma}, \citenamefont {Richard},
  \citenamefont {Huang}, \citenamefont {Zhao}, \citenamefont {Chen},
  \citenamefont {Fang}, \citenamefont {Dai}, \citenamefont {Qian},\ and\
  \citenamefont {Ding}}]{Lv2015a}%
  \BibitemOpen
  \bibfield  {author} {\bibinfo {author} {\bibfnamefont {B.~Q.}\ \bibnamefont
  {Lv}}, \bibinfo {author} {\bibfnamefont {H.~M.}\ \bibnamefont {Weng}},
  \bibinfo {author} {\bibfnamefont {B.~B.}\ \bibnamefont {Fu}}, \bibinfo
  {author} {\bibfnamefont {X.~P.}\ \bibnamefont {Wang}}, \bibinfo {author}
  {\bibfnamefont {H.}~\bibnamefont {Miao}}, \bibinfo {author} {\bibfnamefont
  {J.}~\bibnamefont {Ma}}, \bibinfo {author} {\bibfnamefont {P.}~\bibnamefont
  {Richard}}, \bibinfo {author} {\bibfnamefont {X.~C.}\ \bibnamefont {Huang}},
  \bibinfo {author} {\bibfnamefont {L.~X.}\ \bibnamefont {Zhao}}, \bibinfo
  {author} {\bibfnamefont {G.~F.}\ \bibnamefont {Chen}}, \bibinfo {author}
  {\bibfnamefont {Z.}~\bibnamefont {Fang}}, \bibinfo {author} {\bibfnamefont
  {X.}~\bibnamefont {Dai}}, \bibinfo {author} {\bibfnamefont {T.}~\bibnamefont
  {Qian}}, \ and\ \bibinfo {author} {\bibfnamefont {H.}~\bibnamefont {Ding}},\
  }\href {\doibase 10.1103/PhysRevX.5.031013} {\bibfield  {journal} {\bibinfo
  {journal} {Phys. Rev. X}\ }\textbf {\bibinfo {volume} {5}},\ \bibinfo {pages}
  {031013} (\bibinfo {year} {2015}{\natexlab{a}})}\BibitemShut {NoStop}%
\bibitem [{\citenamefont {Yang}\ \emph {et~al.}(2015)\citenamefont {Yang},
  \citenamefont {Liu}, \citenamefont {Sun}, \citenamefont {Peng}, \citenamefont
  {Yang}, \citenamefont {Zhang}, \citenamefont {Zhou}, \citenamefont {Zhang},
  \citenamefont {Guo}, \citenamefont {Rahn}, \citenamefont {Prabhakaran},
  \citenamefont {Hussain}, \citenamefont {Mo}, \citenamefont {Felser},
  \citenamefont {Yan},\ and\ \citenamefont {Chen}}]{Yang2015}%
  \BibitemOpen
  \bibfield  {author} {\bibinfo {author} {\bibfnamefont {L.~X.}\ \bibnamefont
  {Yang}}, \bibinfo {author} {\bibfnamefont {Z.~K.}\ \bibnamefont {Liu}},
  \bibinfo {author} {\bibfnamefont {Y.}~\bibnamefont {Sun}}, \bibinfo {author}
  {\bibfnamefont {H.}~\bibnamefont {Peng}}, \bibinfo {author} {\bibfnamefont
  {H.~F.}\ \bibnamefont {Yang}}, \bibinfo {author} {\bibfnamefont
  {T.}~\bibnamefont {Zhang}}, \bibinfo {author} {\bibfnamefont
  {B.}~\bibnamefont {Zhou}}, \bibinfo {author} {\bibfnamefont {Y.}~\bibnamefont
  {Zhang}}, \bibinfo {author} {\bibfnamefont {Y.~F.}\ \bibnamefont {Guo}},
  \bibinfo {author} {\bibfnamefont {M.}~\bibnamefont {Rahn}}, \bibinfo {author}
  {\bibfnamefont {D.}~\bibnamefont {Prabhakaran}}, \bibinfo {author}
  {\bibfnamefont {Z.}~\bibnamefont {Hussain}}, \bibinfo {author} {\bibfnamefont
  {S.-K.}\ \bibnamefont {Mo}}, \bibinfo {author} {\bibfnamefont
  {C.}~\bibnamefont {Felser}}, \bibinfo {author} {\bibfnamefont
  {B.}~\bibnamefont {Yan}}, \ and\ \bibinfo {author} {\bibfnamefont {Y.~L.}\
  \bibnamefont {Chen}},\ }\href@noop {} {\bibfield  {journal} {\bibinfo
  {journal} {Nat. Phys.}\ }\textbf {\bibinfo {volume} {11}},\ \bibinfo {pages}
  {879} (\bibinfo {year} {2015})}\BibitemShut {NoStop}%
\bibitem [{\citenamefont {Lv}\ \emph {et~al.}(2015{\natexlab{b}})\citenamefont
  {Lv}, \citenamefont {Lv}, \citenamefont {Xu}, \citenamefont {Weng},
  \citenamefont {Ma}, \citenamefont {Richard}, \citenamefont {Huang},
  \citenamefont {Zhao}, \citenamefont {Chen}, \citenamefont {Matt},
  \citenamefont {Bisti}, \citenamefont {Strocov}, \citenamefont {Mesot},
  \citenamefont {Fang}, \citenamefont {Dai}, \citenamefont {Qian},
  \citenamefont {Shi},\ and\ \citenamefont {Ding}}]{Lv2015b}%
  \BibitemOpen
  \bibfield  {author} {\bibinfo {author} {\bibfnamefont {B.~Q.}\ \bibnamefont
  {Lv}}, \bibinfo {author} {\bibfnamefont {B.~Q.}\ \bibnamefont {Lv}}, \bibinfo
  {author} {\bibfnamefont {N.}~\bibnamefont {Xu}}, \bibinfo {author}
  {\bibfnamefont {H.~M.}\ \bibnamefont {Weng}}, \bibinfo {author}
  {\bibfnamefont {J.~Z.}\ \bibnamefont {Ma}}, \bibinfo {author} {\bibfnamefont
  {P.}~\bibnamefont {Richard}}, \bibinfo {author} {\bibfnamefont {X.~C.}\
  \bibnamefont {Huang}}, \bibinfo {author} {\bibfnamefont {L.~X.}\ \bibnamefont
  {Zhao}}, \bibinfo {author} {\bibfnamefont {G.~F.}\ \bibnamefont {Chen}},
  \bibinfo {author} {\bibfnamefont {C.~E.}\ \bibnamefont {Matt}}, \bibinfo
  {author} {\bibfnamefont {F.}~\bibnamefont {Bisti}}, \bibinfo {author}
  {\bibfnamefont {V.~N.}\ \bibnamefont {Strocov}}, \bibinfo {author}
  {\bibfnamefont {J.}~\bibnamefont {Mesot}}, \bibinfo {author} {\bibfnamefont
  {Z.}~\bibnamefont {Fang}}, \bibinfo {author} {\bibfnamefont {X.}~\bibnamefont
  {Dai}}, \bibinfo {author} {\bibfnamefont {T.}~\bibnamefont {Qian}}, \bibinfo
  {author} {\bibfnamefont {M.}~\bibnamefont {Shi}}, \ and\ \bibinfo {author}
  {\bibfnamefont {H.}~\bibnamefont {Ding}},\ }\href@noop {} {\bibfield
  {journal} {\bibinfo  {journal} {Nat. Phys.}\ }\textbf {\bibinfo {volume}
  {11}},\ \bibinfo {pages} {724} (\bibinfo {year}
  {2015}{\natexlab{b}})}\BibitemShut {NoStop}%
\bibitem [{\citenamefont {Xu}\ \emph {et~al.}(2015{\natexlab{b}})\citenamefont
  {Xu}, \citenamefont {Belopolski}, \citenamefont {Sanchez}, \citenamefont
  {Zhang}, \citenamefont {Chang}, \citenamefont {Guo}, \citenamefont {Bian},
  \citenamefont {Yuan}, \citenamefont {Lu}, \citenamefont {Chang},
  \citenamefont {Shibayev}, \citenamefont {Prokopovych}, \citenamefont
  {Alidoust}, \citenamefont {Zheng}, \citenamefont {Lee}, \citenamefont
  {Huang}, \citenamefont {Sankar}, \citenamefont {Chou}, \citenamefont {Hsu},
  \citenamefont {Jeng}, \citenamefont {Bansil}, \citenamefont {Neupert},
  \citenamefont {Strocov}, \citenamefont {Lin}, \citenamefont {Jia},\ and\
  \citenamefont {Hasan}}]{Xu2015b}%
  \BibitemOpen
  \bibfield  {author} {\bibinfo {author} {\bibfnamefont {S.-Y.}\ \bibnamefont
  {Xu}}, \bibinfo {author} {\bibfnamefont {I.}~\bibnamefont {Belopolski}},
  \bibinfo {author} {\bibfnamefont {D.~S.}\ \bibnamefont {Sanchez}}, \bibinfo
  {author} {\bibfnamefont {C.}~\bibnamefont {Zhang}}, \bibinfo {author}
  {\bibfnamefont {G.}~\bibnamefont {Chang}}, \bibinfo {author} {\bibfnamefont
  {C.}~\bibnamefont {Guo}}, \bibinfo {author} {\bibfnamefont {G.}~\bibnamefont
  {Bian}}, \bibinfo {author} {\bibfnamefont {Z.}~\bibnamefont {Yuan}}, \bibinfo
  {author} {\bibfnamefont {H.}~\bibnamefont {Lu}}, \bibinfo {author}
  {\bibfnamefont {T.-R.}\ \bibnamefont {Chang}}, \bibinfo {author}
  {\bibfnamefont {P.~P.}\ \bibnamefont {Shibayev}}, \bibinfo {author}
  {\bibfnamefont {M.~L.}\ \bibnamefont {Prokopovych}}, \bibinfo {author}
  {\bibfnamefont {N.}~\bibnamefont {Alidoust}}, \bibinfo {author}
  {\bibfnamefont {H.}~\bibnamefont {Zheng}}, \bibinfo {author} {\bibfnamefont
  {C.-C.}\ \bibnamefont {Lee}}, \bibinfo {author} {\bibfnamefont {S.-M.}\
  \bibnamefont {Huang}}, \bibinfo {author} {\bibfnamefont {R.}~\bibnamefont
  {Sankar}}, \bibinfo {author} {\bibfnamefont {F.}~\bibnamefont {Chou}},
  \bibinfo {author} {\bibfnamefont {C.-H.}\ \bibnamefont {Hsu}}, \bibinfo
  {author} {\bibfnamefont {H.-T.}\ \bibnamefont {Jeng}}, \bibinfo {author}
  {\bibfnamefont {A.}~\bibnamefont {Bansil}}, \bibinfo {author} {\bibfnamefont
  {T.}~\bibnamefont {Neupert}}, \bibinfo {author} {\bibfnamefont {V.~N.}\
  \bibnamefont {Strocov}}, \bibinfo {author} {\bibfnamefont {H.}~\bibnamefont
  {Lin}}, \bibinfo {author} {\bibfnamefont {S.}~\bibnamefont {Jia}}, \ and\
  \bibinfo {author} {\bibfnamefont {M.~Z.}\ \bibnamefont {Hasan}},\ }\href
  {http://advances.sciencemag.org/content/1/10/e1501092} {\bibfield  {journal}
  {\bibinfo  {journal} {Sci. Adv.}\ }\textbf {\bibinfo {volume} {1}},\ \bibinfo
  {pages} {e1501092} (\bibinfo {year} {2015}{\natexlab{b}})}\BibitemShut
  {NoStop}%
\bibitem [{\citenamefont {Alidoust}\ \emph {et~al.}(2015)\citenamefont
  {Alidoust}, \citenamefont {Xu}, \citenamefont {Belopolski}, \citenamefont
  {Bian}, \citenamefont {Zheng}, \citenamefont {Sanchez}, \citenamefont
  {Neupert}, \citenamefont {Hasan}, \citenamefont {Yuan},\ and\ \citenamefont
  {Zhang}}]{Alidoust2015}%
  \BibitemOpen
  \bibfield  {author} {\bibinfo {author} {\bibfnamefont {N.}~\bibnamefont
  {Alidoust}}, \bibinfo {author} {\bibfnamefont {S.~Y.}\ \bibnamefont {Xu}},
  \bibinfo {author} {\bibfnamefont {I.}~\bibnamefont {Belopolski}}, \bibinfo
  {author} {\bibfnamefont {G.}~\bibnamefont {Bian}}, \bibinfo {author}
  {\bibfnamefont {H.}~\bibnamefont {Zheng}}, \bibinfo {author} {\bibfnamefont
  {D.~S.}\ \bibnamefont {Sanchez}}, \bibinfo {author} {\bibfnamefont
  {T.}~\bibnamefont {Neupert}}, \bibinfo {author} {\bibfnamefont {M.~Z.}\
  \bibnamefont {Hasan}}, \bibinfo {author} {\bibfnamefont {Z.}~\bibnamefont
  {Yuan}}, \ and\ \bibinfo {author} {\bibfnamefont {C.}~\bibnamefont {Zhang}},\
  }\href@noop {} {\bibfield  {journal} {\bibinfo  {journal} {Nat. Phys.}\
  }\textbf {\bibinfo {volume} {11}},\ \bibinfo {pages} {748} (\bibinfo {year}
  {2015})}\BibitemShut {NoStop}%
\bibitem [{\citenamefont {Huang}\ \emph
  {et~al.}(2015{\natexlab{a}})\citenamefont {Huang}, \citenamefont {Xu},
  \citenamefont {Belopolski}, \citenamefont {Lee}, \citenamefont {Chang},
  \citenamefont {Wang}, \citenamefont {Alidoust}, \citenamefont {Bian},
  \citenamefont {Neupane}, \citenamefont {Zhang}, \citenamefont {Jia},
  \citenamefont {Bansil}, \citenamefont {Lin},\ and\ \citenamefont
  {Hasan}}]{huang2015}%
  \BibitemOpen
  \bibfield  {author} {\bibinfo {author} {\bibfnamefont {S.-M.}\ \bibnamefont
  {Huang}}, \bibinfo {author} {\bibfnamefont {S.-Y.}\ \bibnamefont {Xu}},
  \bibinfo {author} {\bibfnamefont {I.}~\bibnamefont {Belopolski}}, \bibinfo
  {author} {\bibfnamefont {C.-C.}\ \bibnamefont {Lee}}, \bibinfo {author}
  {\bibfnamefont {G.}~\bibnamefont {Chang}}, \bibinfo {author} {\bibfnamefont
  {B.}~\bibnamefont {Wang}}, \bibinfo {author} {\bibfnamefont {N.}~\bibnamefont
  {Alidoust}}, \bibinfo {author} {\bibfnamefont {G.}~\bibnamefont {Bian}},
  \bibinfo {author} {\bibfnamefont {M.}~\bibnamefont {Neupane}}, \bibinfo
  {author} {\bibfnamefont {C.}~\bibnamefont {Zhang}}, \bibinfo {author}
  {\bibfnamefont {S.}~\bibnamefont {Jia}}, \bibinfo {author} {\bibfnamefont
  {A.}~\bibnamefont {Bansil}}, \bibinfo {author} {\bibfnamefont
  {H.}~\bibnamefont {Lin}}, \ and\ \bibinfo {author} {\bibfnamefont {M.~Z.}\
  \bibnamefont {Hasan}},\ }\href@noop {} {\bibfield  {journal} {\bibinfo
  {journal} {Nat. Commun.}\ }\textbf {\bibinfo {volume} {6}},\ \bibinfo {pages}
  {7373} (\bibinfo {year} {2015}{\natexlab{a}})}\BibitemShut {NoStop}%
\bibitem [{\citenamefont {Xu}\ \emph {et~al.}(2016)\citenamefont {Xu},
  \citenamefont {Weng}, \citenamefont {Lv}, \citenamefont {Matt}, \citenamefont
  {Park}, \citenamefont {Bisti}, \citenamefont {Strocov}, \citenamefont
  {Gawryluk}, \citenamefont {Pomjakushina},\ and\ \citenamefont
  {Conder}}]{Xu2016}%
  \BibitemOpen
  \bibfield  {author} {\bibinfo {author} {\bibfnamefont {N.}~\bibnamefont
  {Xu}}, \bibinfo {author} {\bibfnamefont {H.~M.}\ \bibnamefont {Weng}},
  \bibinfo {author} {\bibfnamefont {B.~Q.}\ \bibnamefont {Lv}}, \bibinfo
  {author} {\bibfnamefont {C.~E.}\ \bibnamefont {Matt}}, \bibinfo {author}
  {\bibfnamefont {J.}~\bibnamefont {Park}}, \bibinfo {author} {\bibfnamefont
  {F.}~\bibnamefont {Bisti}}, \bibinfo {author} {\bibfnamefont {V.~N.}\
  \bibnamefont {Strocov}}, \bibinfo {author} {\bibfnamefont {D.}~\bibnamefont
  {Gawryluk}}, \bibinfo {author} {\bibfnamefont {E.}~\bibnamefont
  {Pomjakushina}}, \ and\ \bibinfo {author} {\bibfnamefont {K.}~\bibnamefont
  {Conder}},\ }\href@noop {} {\bibfield  {journal} {\bibinfo  {journal} {Nat.
  Commun.}\ }\textbf {\bibinfo {volume} {7}},\ \bibinfo {pages} {11006}
  (\bibinfo {year} {2016})}\BibitemShut {NoStop}%
\bibitem [{\citenamefont {Lu}\ \emph {et~al.}(2015)\citenamefont {Lu},
  \citenamefont {Wang}, \citenamefont {Ye}, \citenamefont {Ran}, \citenamefont
  {Fu}, \citenamefont {Joannopoulos},\ and\ \citenamefont {Solja{\v
  c}i{\'c}}}]{Lu2015}%
  \BibitemOpen
  \bibfield  {author} {\bibinfo {author} {\bibfnamefont {L.}~\bibnamefont
  {Lu}}, \bibinfo {author} {\bibfnamefont {Z.}~\bibnamefont {Wang}}, \bibinfo
  {author} {\bibfnamefont {D.}~\bibnamefont {Ye}}, \bibinfo {author}
  {\bibfnamefont {L.}~\bibnamefont {Ran}}, \bibinfo {author} {\bibfnamefont
  {L.}~\bibnamefont {Fu}}, \bibinfo {author} {\bibfnamefont {J.~D.}\
  \bibnamefont {Joannopoulos}}, \ and\ \bibinfo {author} {\bibfnamefont
  {M.}~\bibnamefont {Solja{\v c}i{\'c}}},\ }\href {\doibase
  10.1126/science.aaa9273} {\bibfield  {journal} {\bibinfo  {journal}
  {Science}\ }\textbf {\bibinfo {volume} {349}},\ \bibinfo {pages} {622}
  (\bibinfo {year} {2015})}\BibitemShut {NoStop}%
\bibitem [{\citenamefont {Ruan}\ \emph
  {et~al.}(2016{\natexlab{a}})\citenamefont {Ruan}, \citenamefont {Jian},
  \citenamefont {Yao}, \citenamefont {Zhang}, \citenamefont {Zhang},\ and\
  \citenamefont {Xing}}]{Ruan2016a}%
  \BibitemOpen
  \bibfield  {author} {\bibinfo {author} {\bibfnamefont {J.}~\bibnamefont
  {Ruan}}, \bibinfo {author} {\bibfnamefont {S.~K.}\ \bibnamefont {Jian}},
  \bibinfo {author} {\bibfnamefont {H.}~\bibnamefont {Yao}}, \bibinfo {author}
  {\bibfnamefont {H.}~\bibnamefont {Zhang}}, \bibinfo {author} {\bibfnamefont
  {S.~C.}\ \bibnamefont {Zhang}}, \ and\ \bibinfo {author} {\bibfnamefont
  {D.}~\bibnamefont {Xing}},\ }\href@noop {} {\bibfield  {journal} {\bibinfo
  {journal} {Nat. Commun.}\ }\textbf {\bibinfo {volume} {7}},\ \bibinfo {pages}
  {11136} (\bibinfo {year} {2016}{\natexlab{a}})}\BibitemShut {NoStop}%
\bibitem [{\citenamefont {Ruan}\ \emph
  {et~al.}(2016{\natexlab{b}})\citenamefont {Ruan}, \citenamefont {Jian},
  \citenamefont {Zhang}, \citenamefont {Yao}, \citenamefont {Zhang},
  \citenamefont {Zhang},\ and\ \citenamefont {Xing}}]{Ruan2016b}%
  \BibitemOpen
  \bibfield  {author} {\bibinfo {author} {\bibfnamefont {J.}~\bibnamefont
  {Ruan}}, \bibinfo {author} {\bibfnamefont {S.-K.}\ \bibnamefont {Jian}},
  \bibinfo {author} {\bibfnamefont {D.}~\bibnamefont {Zhang}}, \bibinfo
  {author} {\bibfnamefont {H.}~\bibnamefont {Yao}}, \bibinfo {author}
  {\bibfnamefont {H.}~\bibnamefont {Zhang}}, \bibinfo {author} {\bibfnamefont
  {S.-C.}\ \bibnamefont {Zhang}}, \ and\ \bibinfo {author} {\bibfnamefont
  {D.}~\bibnamefont {Xing}},\ }\href {\doibase 10.1103/PhysRevLett.116.226801}
  {\bibfield  {journal} {\bibinfo  {journal} {Phys. Rev. Lett.}\ }\textbf
  {\bibinfo {volume} {116}},\ \bibinfo {pages} {226801} (\bibinfo {year}
  {2016}{\natexlab{b}})}\BibitemShut {NoStop}%
\bibitem [{\citenamefont {Wang}\ \emph {et~al.}(2012)\citenamefont {Wang},
  \citenamefont {Sun}, \citenamefont {Chen}, \citenamefont {Franchini},
  \citenamefont {Xu}, \citenamefont {Weng}, \citenamefont {Dai},\ and\
  \citenamefont {Fang}}]{Wang2012}%
  \BibitemOpen
  \bibfield  {author} {\bibinfo {author} {\bibfnamefont {Z.}~\bibnamefont
  {Wang}}, \bibinfo {author} {\bibfnamefont {Y.}~\bibnamefont {Sun}}, \bibinfo
  {author} {\bibfnamefont {X.-Q.}\ \bibnamefont {Chen}}, \bibinfo {author}
  {\bibfnamefont {C.}~\bibnamefont {Franchini}}, \bibinfo {author}
  {\bibfnamefont {G.}~\bibnamefont {Xu}}, \bibinfo {author} {\bibfnamefont
  {H.}~\bibnamefont {Weng}}, \bibinfo {author} {\bibfnamefont {X.}~\bibnamefont
  {Dai}}, \ and\ \bibinfo {author} {\bibfnamefont {Z.}~\bibnamefont {Fang}},\
  }\href {\doibase 10.1103/PhysRevB.85.195320} {\bibfield  {journal} {\bibinfo
  {journal} {Phys. Rev. B}\ }\textbf {\bibinfo {volume} {85}},\ \bibinfo
  {pages} {195320} (\bibinfo {year} {2012})}\BibitemShut {NoStop}%
\bibitem [{\citenamefont {Young}\ \emph {et~al.}(2012)\citenamefont {Young},
  \citenamefont {Zaheer}, \citenamefont {Teo}, \citenamefont {Kane},
  \citenamefont {Mele},\ and\ \citenamefont {Rappe}}]{Young2012}%
  \BibitemOpen
  \bibfield  {author} {\bibinfo {author} {\bibfnamefont {S.~M.}\ \bibnamefont
  {Young}}, \bibinfo {author} {\bibfnamefont {S.}~\bibnamefont {Zaheer}},
  \bibinfo {author} {\bibfnamefont {J.~C.~Y.}\ \bibnamefont {Teo}}, \bibinfo
  {author} {\bibfnamefont {C.~L.}\ \bibnamefont {Kane}}, \bibinfo {author}
  {\bibfnamefont {E.~J.}\ \bibnamefont {Mele}}, \ and\ \bibinfo {author}
  {\bibfnamefont {A.~M.}\ \bibnamefont {Rappe}},\ }\href {\doibase
  10.1103/PhysRevLett.108.140405} {\bibfield  {journal} {\bibinfo  {journal}
  {Phys. Rev. Lett.}\ }\textbf {\bibinfo {volume} {108}},\ \bibinfo {pages}
  {140405} (\bibinfo {year} {2012})}\BibitemShut {NoStop}%
\bibitem [{\citenamefont {Wang}\ \emph {et~al.}(2013)\citenamefont {Wang},
  \citenamefont {Weng}, \citenamefont {Wu}, \citenamefont {Dai},\ and\
  \citenamefont {Fang}}]{Wang2013}%
  \BibitemOpen
  \bibfield  {author} {\bibinfo {author} {\bibfnamefont {Z.}~\bibnamefont
  {Wang}}, \bibinfo {author} {\bibfnamefont {H.}~\bibnamefont {Weng}}, \bibinfo
  {author} {\bibfnamefont {Q.}~\bibnamefont {Wu}}, \bibinfo {author}
  {\bibfnamefont {X.}~\bibnamefont {Dai}}, \ and\ \bibinfo {author}
  {\bibfnamefont {Z.}~\bibnamefont {Fang}},\ }\href {\doibase
  10.1103/PhysRevB.88.125427} {\bibfield  {journal} {\bibinfo  {journal} {Phys.
  Rev. B}\ }\textbf {\bibinfo {volume} {88}},\ \bibinfo {pages} {125427}
  (\bibinfo {year} {2013})}\BibitemShut {NoStop}%
\bibitem [{\citenamefont {Liu}\ \emph {et~al.}(2014)\citenamefont {Liu},
  \citenamefont {Jiang}, \citenamefont {Zhou}, \citenamefont {Wang},
  \citenamefont {Zhang}, \citenamefont {Weng}, \citenamefont {Prabhakaran},
  \citenamefont {Mo}, \citenamefont {Peng},\ and\ \citenamefont
  {Dudin}}]{Liu2014A}%
  \BibitemOpen
  \bibfield  {author} {\bibinfo {author} {\bibfnamefont {Z.~K.}\ \bibnamefont
  {Liu}}, \bibinfo {author} {\bibfnamefont {J.}~\bibnamefont {Jiang}}, \bibinfo
  {author} {\bibfnamefont {B.}~\bibnamefont {Zhou}}, \bibinfo {author}
  {\bibfnamefont {Z.~J.}\ \bibnamefont {Wang}}, \bibinfo {author}
  {\bibfnamefont {Y.}~\bibnamefont {Zhang}}, \bibinfo {author} {\bibfnamefont
  {H.~M.}\ \bibnamefont {Weng}}, \bibinfo {author} {\bibfnamefont
  {D.}~\bibnamefont {Prabhakaran}}, \bibinfo {author} {\bibfnamefont {S.~K.}\
  \bibnamefont {Mo}}, \bibinfo {author} {\bibfnamefont {H.}~\bibnamefont
  {Peng}}, \ and\ \bibinfo {author} {\bibfnamefont {P.}~\bibnamefont {Dudin}},\
  }\href@noop {} {\bibfield  {journal} {\bibinfo  {journal} {Nat. Mater.}\
  }\textbf {\bibinfo {volume} {13}},\ \bibinfo {pages} {677} (\bibinfo {year}
  {2014})}\BibitemShut {NoStop}%
\bibitem [{\citenamefont {Chen}\ \emph {et~al.}(2014)\citenamefont {Chen},
  \citenamefont {Liu}, \citenamefont {Zhou}, \citenamefont {Zhang},
  \citenamefont {Wang}, \citenamefont {Weng}, \citenamefont {Prabhakran},
  \citenamefont {Mo}, \citenamefont {Shen},\ and\ \citenamefont
  {Fang}}]{Chen2014}%
  \BibitemOpen
  \bibfield  {author} {\bibinfo {author} {\bibfnamefont {Y.~L.}\ \bibnamefont
  {Chen}}, \bibinfo {author} {\bibfnamefont {Z.~K.}\ \bibnamefont {Liu}},
  \bibinfo {author} {\bibfnamefont {B.}~\bibnamefont {Zhou}}, \bibinfo {author}
  {\bibfnamefont {Y.}~\bibnamefont {Zhang}}, \bibinfo {author} {\bibfnamefont
  {Z.~J.}\ \bibnamefont {Wang}}, \bibinfo {author} {\bibfnamefont {H.~M.}\
  \bibnamefont {Weng}}, \bibinfo {author} {\bibfnamefont {D.}~\bibnamefont
  {Prabhakran}}, \bibinfo {author} {\bibfnamefont {S.~K.}\ \bibnamefont {Mo}},
  \bibinfo {author} {\bibfnamefont {Z.~X.}\ \bibnamefont {Shen}}, \ and\
  \bibinfo {author} {\bibfnamefont {Z.}~\bibnamefont {Fang}},\ }\href@noop {}
  {\bibfield  {journal} {\bibinfo  {journal} {Science}\ }\textbf {\bibinfo
  {volume} {343}},\ \bibinfo {pages} {864} (\bibinfo {year}
  {2014})}\BibitemShut {NoStop}%
\bibitem [{\citenamefont {Neupane}\ \emph {et~al.}(2014)\citenamefont
  {Neupane}, \citenamefont {Xu}, \citenamefont {Sankar}, \citenamefont
  {Alidoust}, \citenamefont {Bian}, \citenamefont {Liu}, \citenamefont
  {Belopolski}, \citenamefont {Chang}, \citenamefont {Jeng},\ and\
  \citenamefont {Lin}}]{Neupane2014}%
  \BibitemOpen
  \bibfield  {author} {\bibinfo {author} {\bibfnamefont {M.}~\bibnamefont
  {Neupane}}, \bibinfo {author} {\bibfnamefont {S.~Y.}\ \bibnamefont {Xu}},
  \bibinfo {author} {\bibfnamefont {R.}~\bibnamefont {Sankar}}, \bibinfo
  {author} {\bibfnamefont {N.}~\bibnamefont {Alidoust}}, \bibinfo {author}
  {\bibfnamefont {G.}~\bibnamefont {Bian}}, \bibinfo {author} {\bibfnamefont
  {C.}~\bibnamefont {Liu}}, \bibinfo {author} {\bibfnamefont {I.}~\bibnamefont
  {Belopolski}}, \bibinfo {author} {\bibfnamefont {T.~R.}\ \bibnamefont
  {Chang}}, \bibinfo {author} {\bibfnamefont {H.~T.}\ \bibnamefont {Jeng}}, \
  and\ \bibinfo {author} {\bibfnamefont {H.}~\bibnamefont {Lin}},\ }\href@noop
  {} {\bibfield  {journal} {\bibinfo  {journal} {Nat. Commun.}\ }\textbf
  {\bibinfo {volume} {5}},\ \bibinfo {pages} {3786} (\bibinfo {year}
  {2014})}\BibitemShut {NoStop}%
\bibitem [{\citenamefont {Yang}\ and\ \citenamefont
  {Nagaosa}(2014)}]{Yang2014}%
  \BibitemOpen
  \bibfield  {author} {\bibinfo {author} {\bibfnamefont {B.~J.}\ \bibnamefont
  {Yang}}\ and\ \bibinfo {author} {\bibfnamefont {N.}~\bibnamefont {Nagaosa}},\
  }\href@noop {} {\bibfield  {journal} {\bibinfo  {journal} {Nat. Commun.}\
  }\textbf {\bibinfo {volume} {5}},\ \bibinfo {pages} {4898} (\bibinfo {year}
  {2014})}\BibitemShut {NoStop}%
\bibitem [{\citenamefont {Xu}\ \emph {et~al.}(2015{\natexlab{c}})\citenamefont
  {Xu}, \citenamefont {Liu}, \citenamefont {Kushwaha}, \citenamefont {Sankar},
  \citenamefont {Krizan}, \citenamefont {Belopolski}, \citenamefont {Neupane},
  \citenamefont {Bian}, \citenamefont {Alidoust},\ and\ \citenamefont
  {Chang}}]{Xu2015Observation}%
  \BibitemOpen
  \bibfield  {author} {\bibinfo {author} {\bibfnamefont {S.~Y.}\ \bibnamefont
  {Xu}}, \bibinfo {author} {\bibfnamefont {C.}~\bibnamefont {Liu}}, \bibinfo
  {author} {\bibfnamefont {S.~K.}\ \bibnamefont {Kushwaha}}, \bibinfo {author}
  {\bibfnamefont {R.}~\bibnamefont {Sankar}}, \bibinfo {author} {\bibfnamefont
  {J.~W.}\ \bibnamefont {Krizan}}, \bibinfo {author} {\bibfnamefont
  {I.}~\bibnamefont {Belopolski}}, \bibinfo {author} {\bibfnamefont
  {M.}~\bibnamefont {Neupane}}, \bibinfo {author} {\bibfnamefont
  {G.}~\bibnamefont {Bian}}, \bibinfo {author} {\bibfnamefont {N.}~\bibnamefont
  {Alidoust}}, \ and\ \bibinfo {author} {\bibfnamefont {T.~R.}\ \bibnamefont
  {Chang}},\ }\href@noop {} {\bibfield  {journal} {\bibinfo  {journal}
  {Science}\ }\textbf {\bibinfo {volume} {347}},\ \bibinfo {pages} {294}
  (\bibinfo {year} {2015}{\natexlab{c}})}\BibitemShut {NoStop}%
\bibitem [{\citenamefont {Xu}\ \emph {et~al.}(2017)\citenamefont {Xu},
  \citenamefont {Chan}, \citenamefont {Chen}, \citenamefont {Chen},
  \citenamefont {Wang}, \citenamefont {Dejoie}, \citenamefont {Wong},
  \citenamefont {Hlevyack}, \citenamefont {Ryu}, \citenamefont {Kee},
  \citenamefont {Tamura}, \citenamefont {Chou}, \citenamefont {Hussain},
  \citenamefont {Mo},\ and\ \citenamefont {Chiang}}]{Xu2017}%
  \BibitemOpen
  \bibfield  {author} {\bibinfo {author} {\bibfnamefont {C.-Z.}\ \bibnamefont
  {Xu}}, \bibinfo {author} {\bibfnamefont {Y.-H.}\ \bibnamefont {Chan}},
  \bibinfo {author} {\bibfnamefont {Y.}~\bibnamefont {Chen}}, \bibinfo {author}
  {\bibfnamefont {P.}~\bibnamefont {Chen}}, \bibinfo {author} {\bibfnamefont
  {X.}~\bibnamefont {Wang}}, \bibinfo {author} {\bibfnamefont {C.}~\bibnamefont
  {Dejoie}}, \bibinfo {author} {\bibfnamefont {M.-H.}\ \bibnamefont {Wong}},
  \bibinfo {author} {\bibfnamefont {J.~A.}\ \bibnamefont {Hlevyack}}, \bibinfo
  {author} {\bibfnamefont {H.}~\bibnamefont {Ryu}}, \bibinfo {author}
  {\bibfnamefont {H.-Y.}\ \bibnamefont {Kee}}, \bibinfo {author} {\bibfnamefont
  {N.}~\bibnamefont {Tamura}}, \bibinfo {author} {\bibfnamefont {M.-Y.}\
  \bibnamefont {Chou}}, \bibinfo {author} {\bibfnamefont {Z.}~\bibnamefont
  {Hussain}}, \bibinfo {author} {\bibfnamefont {S.-K.}\ \bibnamefont {Mo}}, \
  and\ \bibinfo {author} {\bibfnamefont {T.-C.}\ \bibnamefont {Chiang}},\
  }\href {\doibase 10.1103/PhysRevLett.118.146402} {\bibfield  {journal}
  {\bibinfo  {journal} {Phys. Rev. Lett.}\ }\textbf {\bibinfo {volume} {118}},\
  \bibinfo {pages} {146402} (\bibinfo {year} {2017})}\BibitemShut {NoStop}%
\bibitem [{\citenamefont {Huang}\ and\ \citenamefont {Liu}(2017)}]{Huang2017}%
  \BibitemOpen
  \bibfield  {author} {\bibinfo {author} {\bibfnamefont {H.}~\bibnamefont
  {Huang}}\ and\ \bibinfo {author} {\bibfnamefont {F.}~\bibnamefont {Liu}},\
  }\href {\doibase 10.1103/PhysRevB.95.201101} {\bibfield  {journal} {\bibinfo
  {journal} {Phys. Rev. B}\ }\textbf {\bibinfo {volume} {95}},\ \bibinfo
  {pages} {201101} (\bibinfo {year} {2017})}\BibitemShut {NoStop}%
\bibitem [{\citenamefont {Tang}\ \emph {et~al.}(2016)\citenamefont {Tang},
  \citenamefont {Zhou}, \citenamefont {Xu},\ and\ \citenamefont
  {Zhang}}]{tang2016dirac}%
  \BibitemOpen
  \bibfield  {author} {\bibinfo {author} {\bibfnamefont {P.}~\bibnamefont
  {Tang}}, \bibinfo {author} {\bibfnamefont {Q.}~\bibnamefont {Zhou}}, \bibinfo
  {author} {\bibfnamefont {G.}~\bibnamefont {Xu}}, \ and\ \bibinfo {author}
  {\bibfnamefont {S.-C.}\ \bibnamefont {Zhang}},\ }\href@noop {} {\bibfield
  {journal} {\bibinfo  {journal} {Nat. Phys.}\ }\textbf {\bibinfo {volume}
  {12}},\ \bibinfo {pages} {1100} (\bibinfo {year} {2016})}\BibitemShut
  {NoStop}%
\bibitem [{\citenamefont {Wang}(2017)}]{Wang2017Antiferromagetic}%
  \BibitemOpen
  \bibfield  {author} {\bibinfo {author} {\bibfnamefont {J.}~\bibnamefont
  {Wang}},\ }\href {\doibase 10.1103/PhysRevB.95.115138} {\bibfield  {journal}
  {\bibinfo  {journal} {Phys. Rev. B}\ }\textbf {\bibinfo {volume} {95}},\
  \bibinfo {pages} {115138} (\bibinfo {year} {2017})}\BibitemShut {NoStop}%
\bibitem [{\citenamefont {Hua}\ \emph {et~al.}(2018)\citenamefont {Hua},
  \citenamefont {Nie}, \citenamefont {Song}, \citenamefont {Yu}, \citenamefont
  {Xu},\ and\ \citenamefont {Yao}}]{hua2018dirac}%
  \BibitemOpen
  \bibfield  {author} {\bibinfo {author} {\bibfnamefont {G.}~\bibnamefont
  {Hua}}, \bibinfo {author} {\bibfnamefont {S.}~\bibnamefont {Nie}}, \bibinfo
  {author} {\bibfnamefont {Z.}~\bibnamefont {Song}}, \bibinfo {author}
  {\bibfnamefont {R.}~\bibnamefont {Yu}}, \bibinfo {author} {\bibfnamefont
  {G.}~\bibnamefont {Xu}}, \ and\ \bibinfo {author} {\bibfnamefont
  {K.}~\bibnamefont {Yao}},\ }\href@noop {} {\bibfield  {journal} {\bibinfo
  {journal} {arXiv:1801.02806}\ } (\bibinfo {year} {2018})}\BibitemShut
  {NoStop}%
\bibitem [{\citenamefont {Yu}\ \emph {et~al.}(2015)\citenamefont {Yu},
  \citenamefont {Weng}, \citenamefont {Fang}, \citenamefont {Dai},\ and\
  \citenamefont {Hu}}]{Yu2015Topological}%
  \BibitemOpen
  \bibfield  {author} {\bibinfo {author} {\bibfnamefont {R.}~\bibnamefont
  {Yu}}, \bibinfo {author} {\bibfnamefont {H.}~\bibnamefont {Weng}}, \bibinfo
  {author} {\bibfnamefont {Z.}~\bibnamefont {Fang}}, \bibinfo {author}
  {\bibfnamefont {X.}~\bibnamefont {Dai}}, \ and\ \bibinfo {author}
  {\bibfnamefont {X.}~\bibnamefont {Hu}},\ }\href {\doibase
  10.1103/PhysRevLett.115.036807} {\bibfield  {journal} {\bibinfo  {journal}
  {Phys. Rev. Lett.}\ }\textbf {\bibinfo {volume} {115}},\ \bibinfo {pages}
  {036807} (\bibinfo {year} {2015})}\BibitemShut {NoStop}%
\bibitem [{\citenamefont {Kim}\ \emph {et~al.}(2015)\citenamefont {Kim},
  \citenamefont {Wieder}, \citenamefont {Kane},\ and\ \citenamefont
  {Rappe}}]{Kim2015Dirac}%
  \BibitemOpen
  \bibfield  {author} {\bibinfo {author} {\bibfnamefont {Y.}~\bibnamefont
  {Kim}}, \bibinfo {author} {\bibfnamefont {B.~J.}\ \bibnamefont {Wieder}},
  \bibinfo {author} {\bibfnamefont {C.~L.}\ \bibnamefont {Kane}}, \ and\
  \bibinfo {author} {\bibfnamefont {A.~M.}\ \bibnamefont {Rappe}},\ }\href
  {\doibase 10.1103/PhysRevLett.115.036806} {\bibfield  {journal} {\bibinfo
  {journal} {Phys. Rev. Lett.}\ }\textbf {\bibinfo {volume} {115}},\ \bibinfo
  {pages} {036806} (\bibinfo {year} {2015})}\BibitemShut {NoStop}%
\bibitem [{\citenamefont {Bian}\ \emph {et~al.}(2016)\citenamefont {Bian},
  \citenamefont {Chang}, \citenamefont {Zheng}, \citenamefont {Velury},
  \citenamefont {Xu}, \citenamefont {Neupert}, \citenamefont {Chiu},
  \citenamefont {Huang}, \citenamefont {Sanchez}, \citenamefont {Belopolski},
  \citenamefont {Alidoust}, \citenamefont {Chen}, \citenamefont {Chang},
  \citenamefont {Bansil}, \citenamefont {Jeng}, \citenamefont {Lin},\ and\
  \citenamefont {Hasan}}]{Bian2016Drumhead}%
  \BibitemOpen
  \bibfield  {author} {\bibinfo {author} {\bibfnamefont {G.}~\bibnamefont
  {Bian}}, \bibinfo {author} {\bibfnamefont {T.-R.}\ \bibnamefont {Chang}},
  \bibinfo {author} {\bibfnamefont {H.}~\bibnamefont {Zheng}}, \bibinfo
  {author} {\bibfnamefont {S.}~\bibnamefont {Velury}}, \bibinfo {author}
  {\bibfnamefont {S.-Y.}\ \bibnamefont {Xu}}, \bibinfo {author} {\bibfnamefont
  {T.}~\bibnamefont {Neupert}}, \bibinfo {author} {\bibfnamefont {C.-K.}\
  \bibnamefont {Chiu}}, \bibinfo {author} {\bibfnamefont {S.-M.}\ \bibnamefont
  {Huang}}, \bibinfo {author} {\bibfnamefont {D.~S.}\ \bibnamefont {Sanchez}},
  \bibinfo {author} {\bibfnamefont {I.}~\bibnamefont {Belopolski}}, \bibinfo
  {author} {\bibfnamefont {N.}~\bibnamefont {Alidoust}}, \bibinfo {author}
  {\bibfnamefont {P.-J.}\ \bibnamefont {Chen}}, \bibinfo {author}
  {\bibfnamefont {G.}~\bibnamefont {Chang}}, \bibinfo {author} {\bibfnamefont
  {A.}~\bibnamefont {Bansil}}, \bibinfo {author} {\bibfnamefont {H.-T.}\
  \bibnamefont {Jeng}}, \bibinfo {author} {\bibfnamefont {H.}~\bibnamefont
  {Lin}}, \ and\ \bibinfo {author} {\bibfnamefont {M.~Z.}\ \bibnamefont
  {Hasan}},\ }\href {\doibase 10.1103/PhysRevB.93.121113} {\bibfield  {journal}
  {\bibinfo  {journal} {Phys. Rev. B}\ }\textbf {\bibinfo {volume} {93}},\
  \bibinfo {pages} {121113} (\bibinfo {year} {2016})}\BibitemShut {NoStop}%
\bibitem [{\citenamefont {Fang}\ \emph {et~al.}(2016)\citenamefont {Fang},
  \citenamefont {Weng}, \citenamefont {Dai},\ and\ \citenamefont
  {Fang}}]{Fang2016Topological}%
  \BibitemOpen
  \bibfield  {author} {\bibinfo {author} {\bibfnamefont {C.}~\bibnamefont
  {Fang}}, \bibinfo {author} {\bibfnamefont {H.}~\bibnamefont {Weng}}, \bibinfo
  {author} {\bibfnamefont {X.}~\bibnamefont {Dai}}, \ and\ \bibinfo {author}
  {\bibfnamefont {Z.}~\bibnamefont {Fang}},\ }\href@noop {} {\bibfield
  {journal} {\bibinfo  {journal} {Chin. Phys. B}\ }\textbf {\bibinfo {volume}
  {25}},\ \bibinfo {pages} {9} (\bibinfo {year} {2016})}\BibitemShut {NoStop}%
\bibitem [{\citenamefont {Yu}\ \emph {et~al.}(2017)\citenamefont {Yu},
  \citenamefont {Fang}, \citenamefont {Dai},\ and\ \citenamefont
  {Weng}}]{Yu2017Topological}%
  \BibitemOpen
  \bibfield  {author} {\bibinfo {author} {\bibfnamefont {R.}~\bibnamefont
  {Yu}}, \bibinfo {author} {\bibfnamefont {Z.}~\bibnamefont {Fang}}, \bibinfo
  {author} {\bibfnamefont {X.}~\bibnamefont {Dai}}, \ and\ \bibinfo {author}
  {\bibfnamefont {H.}~\bibnamefont {Weng}},\ }\href@noop {} {\bibfield
  {journal} {\bibinfo  {journal} {Frontiers of Physics}\ }\textbf {\bibinfo
  {volume} {12}},\ \bibinfo {pages} {127202} (\bibinfo {year}
  {2017})}\BibitemShut {NoStop}%
\bibitem [{\citenamefont {Yan}\ \emph {et~al.}(2017)\citenamefont {Yan},
  \citenamefont {Bi}, \citenamefont {Shen}, \citenamefont {Lu}, \citenamefont
  {Zhang},\ and\ \citenamefont {Wang}}]{Yan2017Nodal}%
  \BibitemOpen
  \bibfield  {author} {\bibinfo {author} {\bibfnamefont {Z.}~\bibnamefont
  {Yan}}, \bibinfo {author} {\bibfnamefont {R.}~\bibnamefont {Bi}}, \bibinfo
  {author} {\bibfnamefont {H.}~\bibnamefont {Shen}}, \bibinfo {author}
  {\bibfnamefont {L.}~\bibnamefont {Lu}}, \bibinfo {author} {\bibfnamefont
  {S.-C.}\ \bibnamefont {Zhang}}, \ and\ \bibinfo {author} {\bibfnamefont
  {Z.}~\bibnamefont {Wang}},\ }\href {\doibase 10.1103/PhysRevB.96.041103}
  {\bibfield  {journal} {\bibinfo  {journal} {Phys. Rev. B}\ }\textbf {\bibinfo
  {volume} {96}},\ \bibinfo {pages} {041103} (\bibinfo {year}
  {2017})}\BibitemShut {NoStop}%
\bibitem [{\citenamefont {Chen}\ \emph {et~al.}(2017)\citenamefont {Chen},
  \citenamefont {Lu},\ and\ \citenamefont {Hou}}]{Chen2017Topological}%
  \BibitemOpen
  \bibfield  {author} {\bibinfo {author} {\bibfnamefont {W.}~\bibnamefont
  {Chen}}, \bibinfo {author} {\bibfnamefont {H.-Z.}\ \bibnamefont {Lu}}, \ and\
  \bibinfo {author} {\bibfnamefont {J.-M.}\ \bibnamefont {Hou}},\ }\href
  {\doibase 10.1103/PhysRevB.96.041102} {\bibfield  {journal} {\bibinfo
  {journal} {Phys. Rev. B}\ }\textbf {\bibinfo {volume} {96}},\ \bibinfo
  {pages} {041102} (\bibinfo {year} {2017})}\BibitemShut {NoStop}%
\bibitem [{\citenamefont {Li}\ \emph {et~al.}(2017)\citenamefont {Li},
  \citenamefont {Yu}, \citenamefont {Liu}, \citenamefont {Guan}, \citenamefont
  {Wang}, \citenamefont {Zhang}, \citenamefont {Yao},\ and\ \citenamefont
  {Yang}}]{Li2017Type-II}%
  \BibitemOpen
  \bibfield  {author} {\bibinfo {author} {\bibfnamefont {S.}~\bibnamefont
  {Li}}, \bibinfo {author} {\bibfnamefont {Z.-M.}\ \bibnamefont {Yu}}, \bibinfo
  {author} {\bibfnamefont {Y.}~\bibnamefont {Liu}}, \bibinfo {author}
  {\bibfnamefont {S.}~\bibnamefont {Guan}}, \bibinfo {author} {\bibfnamefont
  {S.-S.}\ \bibnamefont {Wang}}, \bibinfo {author} {\bibfnamefont
  {X.}~\bibnamefont {Zhang}}, \bibinfo {author} {\bibfnamefont
  {Y.}~\bibnamefont {Yao}}, \ and\ \bibinfo {author} {\bibfnamefont {S.~A.}\
  \bibnamefont {Yang}},\ }\href {\doibase 10.1103/PhysRevB.96.081106}
  {\bibfield  {journal} {\bibinfo  {journal} {Phys. Rev. B}\ }\textbf {\bibinfo
  {volume} {96}},\ \bibinfo {pages} {081106} (\bibinfo {year}
  {2017})}\BibitemShut {NoStop}%
\bibitem [{\citenamefont {Sun}\ \emph {et~al.}(2017)\citenamefont {Sun},
  \citenamefont {Zhang}, \citenamefont {Liu}, \citenamefont {Felser},\ and\
  \citenamefont {Yan}}]{Sun2017Dirac}%
  \BibitemOpen
  \bibfield  {author} {\bibinfo {author} {\bibfnamefont {Y.}~\bibnamefont
  {Sun}}, \bibinfo {author} {\bibfnamefont {Y.}~\bibnamefont {Zhang}}, \bibinfo
  {author} {\bibfnamefont {C.-X.}\ \bibnamefont {Liu}}, \bibinfo {author}
  {\bibfnamefont {C.}~\bibnamefont {Felser}}, \ and\ \bibinfo {author}
  {\bibfnamefont {B.}~\bibnamefont {Yan}},\ }\href {\doibase
  10.1103/PhysRevB.95.235104} {\bibfield  {journal} {\bibinfo  {journal} {Phys.
  Rev. B}\ }\textbf {\bibinfo {volume} {95}},\ \bibinfo {pages} {235104}
  (\bibinfo {year} {2017})}\BibitemShut {NoStop}%
\bibitem [{\citenamefont {Nielsen}\ and\ \citenamefont
  {Ninomiya}(1981)}]{nielsen1983}%
  \BibitemOpen
  \bibfield  {author} {\bibinfo {author} {\bibfnamefont {H.}~\bibnamefont
  {Nielsen}}\ and\ \bibinfo {author} {\bibfnamefont {M.}~\bibnamefont
  {Ninomiya}},\ }\href {\doibase
  http://dx.doi.org/10.1016/0370-2693(81)91026-1} {\bibfield  {journal}
  {\bibinfo  {journal} {Phys. Lett. B}\ }\textbf {\bibinfo {volume} {105}},\
  \bibinfo {pages} {219 } (\bibinfo {year} {1981})}\BibitemShut {NoStop}%
\bibitem [{\citenamefont {Liu}\ \emph {et~al.}(2013)\citenamefont {Liu},
  \citenamefont {Ye},\ and\ \citenamefont {Qi}}]{liu2013chiral}%
  \BibitemOpen
  \bibfield  {author} {\bibinfo {author} {\bibfnamefont {C.-X.}\ \bibnamefont
  {Liu}}, \bibinfo {author} {\bibfnamefont {P.}~\bibnamefont {Ye}}, \ and\
  \bibinfo {author} {\bibfnamefont {X.-L.}\ \bibnamefont {Qi}},\ }\href@noop {}
  {\bibfield  {journal} {\bibinfo  {journal} {Phys. Rev. B}\ }\textbf {\bibinfo
  {volume} {87}},\ \bibinfo {pages} {235306} (\bibinfo {year}
  {2013})}\BibitemShut {NoStop}%
\bibitem [{\citenamefont {Son}\ and\ \citenamefont {Spivak}(2013)}]{Son2013}%
  \BibitemOpen
  \bibfield  {author} {\bibinfo {author} {\bibfnamefont {D.~T.}\ \bibnamefont
  {Son}}\ and\ \bibinfo {author} {\bibfnamefont {B.~Z.}\ \bibnamefont
  {Spivak}},\ }\href {\doibase 10.1103/PhysRevB.88.104412} {\bibfield
  {journal} {\bibinfo  {journal} {Phys. Rev. B}\ }\textbf {\bibinfo {volume}
  {88}},\ \bibinfo {pages} {104412} (\bibinfo {year} {2013})}\BibitemShut
  {NoStop}%
\bibitem [{\citenamefont {Hosur}\ and\ \citenamefont {Qi}(2013)}]{hosur2013}%
  \BibitemOpen
  \bibfield  {author} {\bibinfo {author} {\bibfnamefont {P.}~\bibnamefont
  {Hosur}}\ and\ \bibinfo {author} {\bibfnamefont {X.}~\bibnamefont {Qi}},\
  }\href@noop {} {\bibfield  {journal} {\bibinfo  {journal} {C.R. Phys.}\
  }\textbf {\bibinfo {volume} {14}},\ \bibinfo {pages} {857} (\bibinfo {year}
  {2013})}\BibitemShut {NoStop}%
\bibitem [{\citenamefont {Xiong}\ \emph {et~al.}(2015)\citenamefont {Xiong},
  \citenamefont {Kushwaha}, \citenamefont {Liang}, \citenamefont {Krizan},
  \citenamefont {Hirschberger}, \citenamefont {Wang}, \citenamefont {Cava},\
  and\ \citenamefont {Ong}}]{Xiong413}%
  \BibitemOpen
  \bibfield  {author} {\bibinfo {author} {\bibfnamefont {J.}~\bibnamefont
  {Xiong}}, \bibinfo {author} {\bibfnamefont {S.~K.}\ \bibnamefont {Kushwaha}},
  \bibinfo {author} {\bibfnamefont {T.}~\bibnamefont {Liang}}, \bibinfo
  {author} {\bibfnamefont {J.~W.}\ \bibnamefont {Krizan}}, \bibinfo {author}
  {\bibfnamefont {M.}~\bibnamefont {Hirschberger}}, \bibinfo {author}
  {\bibfnamefont {W.}~\bibnamefont {Wang}}, \bibinfo {author} {\bibfnamefont
  {R.~J.}\ \bibnamefont {Cava}}, \ and\ \bibinfo {author} {\bibfnamefont
  {N.~P.}\ \bibnamefont {Ong}},\ }\href {\doibase 10.1126/science.aac6089}
  {\bibfield  {journal} {\bibinfo  {journal} {Science}\ }\textbf {\bibinfo
  {volume} {350}},\ \bibinfo {pages} {413} (\bibinfo {year}
  {2015})}\BibitemShut {NoStop}%
\bibitem [{\citenamefont {Huang}\ \emph
  {et~al.}(2015{\natexlab{b}})\citenamefont {Huang}, \citenamefont {Zhao},
  \citenamefont {Long}, \citenamefont {Wang}, \citenamefont {Chen},
  \citenamefont {Yang}, \citenamefont {Liang}, \citenamefont {Xue},
  \citenamefont {Weng}, \citenamefont {Fang}, \citenamefont {Dai},\ and\
  \citenamefont {Chen}}]{Huang2015Observation}%
  \BibitemOpen
  \bibfield  {author} {\bibinfo {author} {\bibfnamefont {X.}~\bibnamefont
  {Huang}}, \bibinfo {author} {\bibfnamefont {L.}~\bibnamefont {Zhao}},
  \bibinfo {author} {\bibfnamefont {Y.}~\bibnamefont {Long}}, \bibinfo {author}
  {\bibfnamefont {P.}~\bibnamefont {Wang}}, \bibinfo {author} {\bibfnamefont
  {D.}~\bibnamefont {Chen}}, \bibinfo {author} {\bibfnamefont {Z.}~\bibnamefont
  {Yang}}, \bibinfo {author} {\bibfnamefont {H.}~\bibnamefont {Liang}},
  \bibinfo {author} {\bibfnamefont {M.}~\bibnamefont {Xue}}, \bibinfo {author}
  {\bibfnamefont {H.}~\bibnamefont {Weng}}, \bibinfo {author} {\bibfnamefont
  {Z.}~\bibnamefont {Fang}}, \bibinfo {author} {\bibfnamefont {X.}~\bibnamefont
  {Dai}}, \ and\ \bibinfo {author} {\bibfnamefont {G.}~\bibnamefont {Chen}},\
  }\href {\doibase 10.1103/PhysRevX.5.031023} {\bibfield  {journal} {\bibinfo
  {journal} {Phys. Rev. X}\ }\textbf {\bibinfo {volume} {5}},\ \bibinfo {pages}
  {031023} (\bibinfo {year} {2015}{\natexlab{b}})}\BibitemShut {NoStop}%
\bibitem [{\citenamefont {Zhang}\ \emph
  {et~al.}(2016{\natexlab{a}})\citenamefont {Zhang}, \citenamefont {Xu},
  \citenamefont {Belopolski}, \citenamefont {Yuan}, \citenamefont {Lin},
  \citenamefont {Tong}, \citenamefont {Bian}, \citenamefont {Alidoust},
  \citenamefont {Lee},\ and\ \citenamefont {Huang}}]{Zhang2016Signatures}%
  \BibitemOpen
  \bibfield  {author} {\bibinfo {author} {\bibfnamefont {C.~L.}\ \bibnamefont
  {Zhang}}, \bibinfo {author} {\bibfnamefont {S.~Y.}\ \bibnamefont {Xu}},
  \bibinfo {author} {\bibfnamefont {I.}~\bibnamefont {Belopolski}}, \bibinfo
  {author} {\bibfnamefont {Z.}~\bibnamefont {Yuan}}, \bibinfo {author}
  {\bibfnamefont {Z.}~\bibnamefont {Lin}}, \bibinfo {author} {\bibfnamefont
  {B.}~\bibnamefont {Tong}}, \bibinfo {author} {\bibfnamefont {G.}~\bibnamefont
  {Bian}}, \bibinfo {author} {\bibfnamefont {N.}~\bibnamefont {Alidoust}},
  \bibinfo {author} {\bibfnamefont {C.~C.}\ \bibnamefont {Lee}}, \ and\
  \bibinfo {author} {\bibfnamefont {S.~M.}\ \bibnamefont {Huang}},\ }\href@noop
  {} {\bibfield  {journal} {\bibinfo  {journal} {Nat. Commun.}\ }\textbf
  {\bibinfo {volume} {7}},\ \bibinfo {pages} {10735} (\bibinfo {year}
  {2016}{\natexlab{a}})}\BibitemShut {NoStop}%
\bibitem [{\citenamefont {Li}\ \emph {et~al.}(2016{\natexlab{a}})\citenamefont
  {Li}, \citenamefont {Kharzeev}, \citenamefont {Zhang}, \citenamefont {Huang},
  \citenamefont {Pletikosi{\'c}}, \citenamefont {Fedorov}, \citenamefont
  {Zhong}, \citenamefont {Schneeloch}, \citenamefont {Gu},\ and\ \citenamefont
  {Valla}}]{li2016chiral}%
  \BibitemOpen
  \bibfield  {author} {\bibinfo {author} {\bibfnamefont {Q.}~\bibnamefont
  {Li}}, \bibinfo {author} {\bibfnamefont {D.~E.}\ \bibnamefont {Kharzeev}},
  \bibinfo {author} {\bibfnamefont {C.}~\bibnamefont {Zhang}}, \bibinfo
  {author} {\bibfnamefont {Y.}~\bibnamefont {Huang}}, \bibinfo {author}
  {\bibfnamefont {I.}~\bibnamefont {Pletikosi{\'c}}}, \bibinfo {author}
  {\bibfnamefont {A.}~\bibnamefont {Fedorov}}, \bibinfo {author} {\bibfnamefont
  {R.}~\bibnamefont {Zhong}}, \bibinfo {author} {\bibfnamefont
  {J.}~\bibnamefont {Schneeloch}}, \bibinfo {author} {\bibfnamefont
  {G.}~\bibnamefont {Gu}}, \ and\ \bibinfo {author} {\bibfnamefont
  {T.}~\bibnamefont {Valla}},\ }\href@noop {} {\bibfield  {journal} {\bibinfo
  {journal} {Nat. Phys.}\ }\textbf {\bibinfo {volume} {12}},\ \bibinfo {pages}
  {550} (\bibinfo {year} {2016}{\natexlab{a}})}\BibitemShut {NoStop}%
\bibitem [{\citenamefont {Zyuzin}\ and\ \citenamefont
  {Burkov}(2012)}]{Zyuzin2012Topological}%
  \BibitemOpen
  \bibfield  {author} {\bibinfo {author} {\bibfnamefont {A.}~\bibnamefont
  {Zyuzin}}\ and\ \bibinfo {author} {\bibfnamefont {A.}~\bibnamefont
  {Burkov}},\ }\href@noop {} {\bibfield  {journal} {\bibinfo  {journal} {Phys.
  Rev. B}\ }\textbf {\bibinfo {volume} {86}},\ \bibinfo {pages} {115133}
  (\bibinfo {year} {2012})}\BibitemShut {NoStop}%
\bibitem [{\citenamefont {Chang}\ and\ \citenamefont {Yang}(2015)}]{Chang2015}%
  \BibitemOpen
  \bibfield  {author} {\bibinfo {author} {\bibfnamefont {M.-C.}\ \bibnamefont
  {Chang}}\ and\ \bibinfo {author} {\bibfnamefont {M.-F.}\ \bibnamefont
  {Yang}},\ }\href {\doibase 10.1103/PhysRevB.91.115203} {\bibfield  {journal}
  {\bibinfo  {journal} {Phys. Rev. B}\ }\textbf {\bibinfo {volume} {91}},\
  \bibinfo {pages} {115203} (\bibinfo {year} {2015})}\BibitemShut {NoStop}%
\bibitem [{\citenamefont {Vazifeh}\ and\ \citenamefont
  {Franz}(2013)}]{vazifeh2013electromagnetic}%
  \BibitemOpen
  \bibfield  {author} {\bibinfo {author} {\bibfnamefont {M.}~\bibnamefont
  {Vazifeh}}\ and\ \bibinfo {author} {\bibfnamefont {M.}~\bibnamefont
  {Franz}},\ }\href@noop {} {\bibfield  {journal} {\bibinfo  {journal} {Phys.
  Rev. Lett.}\ }\textbf {\bibinfo {volume} {111}},\ \bibinfo {pages} {027201}
  (\bibinfo {year} {2013})}\BibitemShut {NoStop}%
\bibitem [{\citenamefont {Gao}\ \emph {et~al.}(2016)\citenamefont {Gao},
  \citenamefont {Hua}, \citenamefont {Zhang},\ and\ \citenamefont
  {Zhang}}]{gao2016classification}%
  \BibitemOpen
  \bibfield  {author} {\bibinfo {author} {\bibfnamefont {Z.}~\bibnamefont
  {Gao}}, \bibinfo {author} {\bibfnamefont {M.}~\bibnamefont {Hua}}, \bibinfo
  {author} {\bibfnamefont {H.}~\bibnamefont {Zhang}}, \ and\ \bibinfo {author}
  {\bibfnamefont {X.}~\bibnamefont {Zhang}},\ }\href@noop {} {\bibfield
  {journal} {\bibinfo  {journal} {Phys. Rev. B}\ }\textbf {\bibinfo {volume}
  {93}},\ \bibinfo {pages} {205109} (\bibinfo {year} {2016})}\BibitemShut
  {NoStop}%
\bibitem [{\citenamefont {Deng}\ \emph {et~al.}(2016)\citenamefont {Deng},
  \citenamefont {Wan}, \citenamefont {Deng}, \citenamefont {Zhang},
  \citenamefont {Ding}, \citenamefont {Wang}, \citenamefont {Yan},
  \citenamefont {Huang}, \citenamefont {Zhang},\ and\ \citenamefont
  {Xu}}]{Deng2016}%
  \BibitemOpen
  \bibfield  {author} {\bibinfo {author} {\bibfnamefont {K.}~\bibnamefont
  {Deng}}, \bibinfo {author} {\bibfnamefont {G.}~\bibnamefont {Wan}}, \bibinfo
  {author} {\bibfnamefont {P.}~\bibnamefont {Deng}}, \bibinfo {author}
  {\bibfnamefont {K.}~\bibnamefont {Zhang}}, \bibinfo {author} {\bibfnamefont
  {S.}~\bibnamefont {Ding}}, \bibinfo {author} {\bibfnamefont {E.}~\bibnamefont
  {Wang}}, \bibinfo {author} {\bibfnamefont {M.}~\bibnamefont {Yan}}, \bibinfo
  {author} {\bibfnamefont {H.}~\bibnamefont {Huang}}, \bibinfo {author}
  {\bibfnamefont {H.}~\bibnamefont {Zhang}}, \ and\ \bibinfo {author}
  {\bibfnamefont {Z.}~\bibnamefont {Xu}},\ }\href@noop {} {\bibfield  {journal}
  {\bibinfo  {journal} {Nat. Phys.}\ }\textbf {\bibinfo {volume} {12}},\
  \bibinfo {pages} {1105} (\bibinfo {year} {2016})}\BibitemShut {NoStop}%
\bibitem [{\citenamefont {Wang}\ \emph
  {et~al.}(2016{\natexlab{a}})\citenamefont {Wang}, \citenamefont {Liu},
  \citenamefont {Liu}, \citenamefont {Pan}, \citenamefont {Zhang},
  \citenamefont {Zeng}, \citenamefont {Fu}, \citenamefont {Wang}, \citenamefont
  {Xu}, \citenamefont {Huang}, \citenamefont {Wang}, \citenamefont {Lu},
  \citenamefont {Xing}, \citenamefont {Wang}, \citenamefont {Wan},\ and\
  \citenamefont {Miao}}]{wang2016gate}%
  \BibitemOpen
  \bibfield  {author} {\bibinfo {author} {\bibfnamefont {Y.}~\bibnamefont
  {Wang}}, \bibinfo {author} {\bibfnamefont {E.}~\bibnamefont {Liu}}, \bibinfo
  {author} {\bibfnamefont {H.}~\bibnamefont {Liu}}, \bibinfo {author}
  {\bibfnamefont {Y.}~\bibnamefont {Pan}}, \bibinfo {author} {\bibfnamefont
  {L.}~\bibnamefont {Zhang}}, \bibinfo {author} {\bibfnamefont
  {J.}~\bibnamefont {Zeng}}, \bibinfo {author} {\bibfnamefont {Y.}~\bibnamefont
  {Fu}}, \bibinfo {author} {\bibfnamefont {M.}~\bibnamefont {Wang}}, \bibinfo
  {author} {\bibfnamefont {K.}~\bibnamefont {Xu}}, \bibinfo {author}
  {\bibfnamefont {Z.}~\bibnamefont {Huang}}, \bibinfo {author} {\bibfnamefont
  {Z.}~\bibnamefont {Wang}}, \bibinfo {author} {\bibfnamefont {H.-Z.}\
  \bibnamefont {Lu}}, \bibinfo {author} {\bibfnamefont {D.}~\bibnamefont
  {Xing}}, \bibinfo {author} {\bibfnamefont {B.}~\bibnamefont {Wang}}, \bibinfo
  {author} {\bibfnamefont {X.}~\bibnamefont {Wan}}, \ and\ \bibinfo {author}
  {\bibfnamefont {F.}~\bibnamefont {Miao}},\ }\href@noop {} {\bibfield
  {journal} {\bibinfo  {journal} {Nat. Commun.}\ }\textbf {\bibinfo {volume}
  {7}},\ \bibinfo {pages} {13142} (\bibinfo {year}
  {2016}{\natexlab{a}})}\BibitemShut {NoStop}%
\bibitem [{\citenamefont {Belopolski}\ \emph {et~al.}(2016)\citenamefont
  {Belopolski}, \citenamefont {Sanchez}, \citenamefont {Ishida}, \citenamefont
  {Pan}, \citenamefont {Yu}, \citenamefont {Xu}, \citenamefont {Chang},
  \citenamefont {Chang}, \citenamefont {Zheng}, \citenamefont {Alidoust},
  \citenamefont {Bian}, \citenamefont {Neupane}, \citenamefont {Huang},
  \citenamefont {Lee}, \citenamefont {Song}, \citenamefont {Bu}, \citenamefont
  {Wang}, \citenamefont {Li}, \citenamefont {Eda}, \citenamefont {Jeng},
  \citenamefont {Kondo}, \citenamefont {Lin}, \citenamefont {Liu},
  \citenamefont {Song}, \citenamefont {Shin},\ and\ \citenamefont
  {Hasan}}]{belopolski2016discovery}%
  \BibitemOpen
  \bibfield  {author} {\bibinfo {author} {\bibfnamefont {I.}~\bibnamefont
  {Belopolski}}, \bibinfo {author} {\bibfnamefont {D.~S.}\ \bibnamefont
  {Sanchez}}, \bibinfo {author} {\bibfnamefont {Y.}~\bibnamefont {Ishida}},
  \bibinfo {author} {\bibfnamefont {X.}~\bibnamefont {Pan}}, \bibinfo {author}
  {\bibfnamefont {P.}~\bibnamefont {Yu}}, \bibinfo {author} {\bibfnamefont
  {S.-Y.}\ \bibnamefont {Xu}}, \bibinfo {author} {\bibfnamefont
  {G.}~\bibnamefont {Chang}}, \bibinfo {author} {\bibfnamefont {T.-R.}\
  \bibnamefont {Chang}}, \bibinfo {author} {\bibfnamefont {H.}~\bibnamefont
  {Zheng}}, \bibinfo {author} {\bibfnamefont {N.}~\bibnamefont {Alidoust}},
  \bibinfo {author} {\bibfnamefont {G.}~\bibnamefont {Bian}}, \bibinfo {author}
  {\bibfnamefont {M.}~\bibnamefont {Neupane}}, \bibinfo {author} {\bibfnamefont
  {S.-M.}\ \bibnamefont {Huang}}, \bibinfo {author} {\bibfnamefont {C.-C.}\
  \bibnamefont {Lee}}, \bibinfo {author} {\bibfnamefont {Y.}~\bibnamefont
  {Song}}, \bibinfo {author} {\bibfnamefont {H.}~\bibnamefont {Bu}}, \bibinfo
  {author} {\bibfnamefont {G.}~\bibnamefont {Wang}}, \bibinfo {author}
  {\bibfnamefont {S.}~\bibnamefont {Li}}, \bibinfo {author} {\bibfnamefont
  {G.}~\bibnamefont {Eda}}, \bibinfo {author} {\bibfnamefont {H.-T.}\
  \bibnamefont {Jeng}}, \bibinfo {author} {\bibfnamefont {T.}~\bibnamefont
  {Kondo}}, \bibinfo {author} {\bibfnamefont {H.}~\bibnamefont {Lin}}, \bibinfo
  {author} {\bibfnamefont {Z.}~\bibnamefont {Liu}}, \bibinfo {author}
  {\bibfnamefont {F.}~\bibnamefont {Song}}, \bibinfo {author} {\bibfnamefont
  {S.}~\bibnamefont {Shin}}, \ and\ \bibinfo {author} {\bibfnamefont {M.~Z.}\
  \bibnamefont {Hasan}},\ }\href@noop {} {\bibfield  {journal} {\bibinfo
  {journal} {Nat. Commun.}\ }\textbf {\bibinfo {volume} {7}},\ \bibinfo {pages}
  {13643} (\bibinfo {year} {2016})}\BibitemShut {NoStop}%
\bibitem [{\citenamefont {Wang}\ \emph
  {et~al.}(2016{\natexlab{b}})\citenamefont {Wang}, \citenamefont {Zhang},
  \citenamefont {Huang}, \citenamefont {Nie}, \citenamefont {Liu},
  \citenamefont {Liang}, \citenamefont {Zhang}, \citenamefont {Shen},
  \citenamefont {Liu}, \citenamefont {Hu}, \citenamefont {Ding}, \citenamefont
  {Liu}, \citenamefont {Hu}, \citenamefont {He}, \citenamefont {Zhao},
  \citenamefont {Yu}, \citenamefont {Hu}, \citenamefont {Wei}, \citenamefont
  {Mao}, \citenamefont {Shi}, \citenamefont {Jia}, \citenamefont {Zhang},
  \citenamefont {Zhang}, \citenamefont {Yang}, \citenamefont {Wang},
  \citenamefont {Peng}, \citenamefont {Weng}, \citenamefont {Dai},
  \citenamefont {Fang}, \citenamefont {Xu}, \citenamefont {Chen},\ and\
  \citenamefont {Zhou}}]{Wang2016Observation}%
  \BibitemOpen
  \bibfield  {author} {\bibinfo {author} {\bibfnamefont {C.}~\bibnamefont
  {Wang}}, \bibinfo {author} {\bibfnamefont {Y.}~\bibnamefont {Zhang}},
  \bibinfo {author} {\bibfnamefont {J.}~\bibnamefont {Huang}}, \bibinfo
  {author} {\bibfnamefont {S.}~\bibnamefont {Nie}}, \bibinfo {author}
  {\bibfnamefont {G.}~\bibnamefont {Liu}}, \bibinfo {author} {\bibfnamefont
  {A.}~\bibnamefont {Liang}}, \bibinfo {author} {\bibfnamefont
  {Y.}~\bibnamefont {Zhang}}, \bibinfo {author} {\bibfnamefont
  {B.}~\bibnamefont {Shen}}, \bibinfo {author} {\bibfnamefont {J.}~\bibnamefont
  {Liu}}, \bibinfo {author} {\bibfnamefont {C.}~\bibnamefont {Hu}}, \bibinfo
  {author} {\bibfnamefont {Y.}~\bibnamefont {Ding}}, \bibinfo {author}
  {\bibfnamefont {D.}~\bibnamefont {Liu}}, \bibinfo {author} {\bibfnamefont
  {Y.}~\bibnamefont {Hu}}, \bibinfo {author} {\bibfnamefont {S.}~\bibnamefont
  {He}}, \bibinfo {author} {\bibfnamefont {L.}~\bibnamefont {Zhao}}, \bibinfo
  {author} {\bibfnamefont {L.}~\bibnamefont {Yu}}, \bibinfo {author}
  {\bibfnamefont {J.}~\bibnamefont {Hu}}, \bibinfo {author} {\bibfnamefont
  {J.}~\bibnamefont {Wei}}, \bibinfo {author} {\bibfnamefont {Z.}~\bibnamefont
  {Mao}}, \bibinfo {author} {\bibfnamefont {Y.}~\bibnamefont {Shi}}, \bibinfo
  {author} {\bibfnamefont {X.}~\bibnamefont {Jia}}, \bibinfo {author}
  {\bibfnamefont {F.}~\bibnamefont {Zhang}}, \bibinfo {author} {\bibfnamefont
  {S.}~\bibnamefont {Zhang}}, \bibinfo {author} {\bibfnamefont
  {F.}~\bibnamefont {Yang}}, \bibinfo {author} {\bibfnamefont {Z.}~\bibnamefont
  {Wang}}, \bibinfo {author} {\bibfnamefont {Q.}~\bibnamefont {Peng}}, \bibinfo
  {author} {\bibfnamefont {H.}~\bibnamefont {Weng}}, \bibinfo {author}
  {\bibfnamefont {X.}~\bibnamefont {Dai}}, \bibinfo {author} {\bibfnamefont
  {Z.}~\bibnamefont {Fang}}, \bibinfo {author} {\bibfnamefont {Z.}~\bibnamefont
  {Xu}}, \bibinfo {author} {\bibfnamefont {C.}~\bibnamefont {Chen}}, \ and\
  \bibinfo {author} {\bibfnamefont {X.~J.}\ \bibnamefont {Zhou}},\ }\href
  {\doibase 10.1103/PhysRevB.94.241119} {\bibfield  {journal} {\bibinfo
  {journal} {Phys. Rev. B}\ }\textbf {\bibinfo {volume} {94}},\ \bibinfo
  {pages} {241119} (\bibinfo {year} {2016}{\natexlab{b}})}\BibitemShut
  {NoStop}%
\bibitem [{\citenamefont {Jiang}\ \emph {et~al.}(2017)\citenamefont {Jiang},
  \citenamefont {Liu}, \citenamefont {Sun}, \citenamefont {Yang}, \citenamefont
  {Rajamathi}, \citenamefont {Qi}, \citenamefont {Yang}, \citenamefont {Chen},
  \citenamefont {Peng}, \citenamefont {Hwang}, \citenamefont {Sun},
  \citenamefont {Mo}, \citenamefont {Vobornik}, \citenamefont {Fujii},
  \citenamefont {Parkin}, \citenamefont {Felser}, \citenamefont {Yan},\ and\
  \citenamefont {Chen}}]{jiang2017signature}%
  \BibitemOpen
  \bibfield  {author} {\bibinfo {author} {\bibfnamefont {J.}~\bibnamefont
  {Jiang}}, \bibinfo {author} {\bibfnamefont {Z.}~\bibnamefont {Liu}}, \bibinfo
  {author} {\bibfnamefont {Y.}~\bibnamefont {Sun}}, \bibinfo {author}
  {\bibfnamefont {H.}~\bibnamefont {Yang}}, \bibinfo {author} {\bibfnamefont
  {C.}~\bibnamefont {Rajamathi}}, \bibinfo {author} {\bibfnamefont
  {Y.}~\bibnamefont {Qi}}, \bibinfo {author} {\bibfnamefont {L.}~\bibnamefont
  {Yang}}, \bibinfo {author} {\bibfnamefont {C.}~\bibnamefont {Chen}}, \bibinfo
  {author} {\bibfnamefont {H.}~\bibnamefont {Peng}}, \bibinfo {author}
  {\bibfnamefont {C.}~\bibnamefont {Hwang}}, \bibinfo {author} {\bibfnamefont
  {S.}~\bibnamefont {Sun}}, \bibinfo {author} {\bibfnamefont {S.-K.}\
  \bibnamefont {Mo}}, \bibinfo {author} {\bibfnamefont {I.}~\bibnamefont
  {Vobornik}}, \bibinfo {author} {\bibfnamefont {J.}~\bibnamefont {Fujii}},
  \bibinfo {author} {\bibfnamefont {S.}~\bibnamefont {Parkin}}, \bibinfo
  {author} {\bibfnamefont {C.}~\bibnamefont {Felser}}, \bibinfo {author}
  {\bibfnamefont {B.}~\bibnamefont {Yan}}, \ and\ \bibinfo {author}
  {\bibfnamefont {Y.}~\bibnamefont {Chen}},\ }\href@noop {} {\bibfield
  {journal} {\bibinfo  {journal} {Nat. Commun.}\ }\textbf {\bibinfo {volume}
  {8}},\ \bibinfo {pages} {13973} (\bibinfo {year} {2017})}\BibitemShut
  {NoStop}%
\bibitem [{\citenamefont {Zhang}\ \emph
  {et~al.}(2016{\natexlab{b}})\citenamefont {Zhang}, \citenamefont {Bao},
  \citenamefont {Gu}, \citenamefont {Ren}, \citenamefont {Zhang}, \citenamefont
  {Deng}, \citenamefont {Wu}, \citenamefont {Li}, \citenamefont {Feng},\ and\
  \citenamefont {Zhou}}]{zhang2016raman}%
  \BibitemOpen
  \bibfield  {author} {\bibinfo {author} {\bibfnamefont {K.}~\bibnamefont
  {Zhang}}, \bibinfo {author} {\bibfnamefont {C.}~\bibnamefont {Bao}}, \bibinfo
  {author} {\bibfnamefont {Q.}~\bibnamefont {Gu}}, \bibinfo {author}
  {\bibfnamefont {X.}~\bibnamefont {Ren}}, \bibinfo {author} {\bibfnamefont
  {H.}~\bibnamefont {Zhang}}, \bibinfo {author} {\bibfnamefont
  {K.}~\bibnamefont {Deng}}, \bibinfo {author} {\bibfnamefont {Y.}~\bibnamefont
  {Wu}}, \bibinfo {author} {\bibfnamefont {Y.}~\bibnamefont {Li}}, \bibinfo
  {author} {\bibfnamefont {J.}~\bibnamefont {Feng}}, \ and\ \bibinfo {author}
  {\bibfnamefont {S.}~\bibnamefont {Zhou}},\ }\href@noop {} {\bibfield
  {journal} {\bibinfo  {journal} {Nat. Commun.}\ }\textbf {\bibinfo {volume}
  {7}},\ \bibinfo {pages} {13552} (\bibinfo {year}
  {2016}{\natexlab{b}})}\BibitemShut {NoStop}%
\bibitem [{\citenamefont {Tamai}\ \emph {et~al.}(2016)\citenamefont {Tamai},
  \citenamefont {Wu}, \citenamefont {Cucchi}, \citenamefont {Bruno},
  \citenamefont {Ricc\`o}, \citenamefont {Kim}, \citenamefont {Hoesch},
  \citenamefont {Barreteau}, \citenamefont {Giannini}, \citenamefont {Besnard},
  \citenamefont {Soluyanov},\ and\ \citenamefont {Baumberger}}]{Tamai2016}%
  \BibitemOpen
  \bibfield  {author} {\bibinfo {author} {\bibfnamefont {A.}~\bibnamefont
  {Tamai}}, \bibinfo {author} {\bibfnamefont {Q.~S.}\ \bibnamefont {Wu}},
  \bibinfo {author} {\bibfnamefont {I.}~\bibnamefont {Cucchi}}, \bibinfo
  {author} {\bibfnamefont {F.~Y.}\ \bibnamefont {Bruno}}, \bibinfo {author}
  {\bibfnamefont {S.}~\bibnamefont {Ricc\`o}}, \bibinfo {author} {\bibfnamefont
  {T.~K.}\ \bibnamefont {Kim}}, \bibinfo {author} {\bibfnamefont
  {M.}~\bibnamefont {Hoesch}}, \bibinfo {author} {\bibfnamefont
  {C.}~\bibnamefont {Barreteau}}, \bibinfo {author} {\bibfnamefont
  {E.}~\bibnamefont {Giannini}}, \bibinfo {author} {\bibfnamefont
  {C.}~\bibnamefont {Besnard}}, \bibinfo {author} {\bibfnamefont {A.~A.}\
  \bibnamefont {Soluyanov}}, \ and\ \bibinfo {author} {\bibfnamefont
  {F.}~\bibnamefont {Baumberger}},\ }\href {\doibase 10.1103/PhysRevX.6.031021}
  {\bibfield  {journal} {\bibinfo  {journal} {Phys. Rev. X}\ }\textbf {\bibinfo
  {volume} {6}},\ \bibinfo {pages} {031021} (\bibinfo {year}
  {2016})}\BibitemShut {NoStop}%
\bibitem [{\citenamefont {Arnold}\ \emph {et~al.}(2016)\citenamefont {Arnold},
  \citenamefont {Shekhar}, \citenamefont {Wu}, \citenamefont {Sun},
  \citenamefont {Dos~Reis}, \citenamefont {Kumar}, \citenamefont {Naumann},
  \citenamefont {Ajeesh}, \citenamefont {Schmidt}, \citenamefont {Grushin},
  \citenamefont {Bardarson}, \citenamefont {Baenitz}, \citenamefont {Sokolov},
  \citenamefont {Borrmann}, \citenamefont {Nicklas}, \citenamefont {Felser},
  \citenamefont {Hassinger},\ and\ \citenamefont {Yan}}]{arnold2016negative}%
  \BibitemOpen
  \bibfield  {author} {\bibinfo {author} {\bibfnamefont {F.}~\bibnamefont
  {Arnold}}, \bibinfo {author} {\bibfnamefont {C.}~\bibnamefont {Shekhar}},
  \bibinfo {author} {\bibfnamefont {S.-C.}\ \bibnamefont {Wu}}, \bibinfo
  {author} {\bibfnamefont {Y.}~\bibnamefont {Sun}}, \bibinfo {author}
  {\bibfnamefont {R.~D.}\ \bibnamefont {Dos~Reis}}, \bibinfo {author}
  {\bibfnamefont {N.}~\bibnamefont {Kumar}}, \bibinfo {author} {\bibfnamefont
  {M.}~\bibnamefont {Naumann}}, \bibinfo {author} {\bibfnamefont {M.~O.}\
  \bibnamefont {Ajeesh}}, \bibinfo {author} {\bibfnamefont {M.}~\bibnamefont
  {Schmidt}}, \bibinfo {author} {\bibfnamefont {A.~G.}\ \bibnamefont
  {Grushin}}, \bibinfo {author} {\bibfnamefont {J.~H.}\ \bibnamefont
  {Bardarson}}, \bibinfo {author} {\bibfnamefont {M.}~\bibnamefont {Baenitz}},
  \bibinfo {author} {\bibfnamefont {D.}~\bibnamefont {Sokolov}}, \bibinfo
  {author} {\bibfnamefont {H.}~\bibnamefont {Borrmann}}, \bibinfo {author}
  {\bibfnamefont {M.}~\bibnamefont {Nicklas}}, \bibinfo {author} {\bibfnamefont
  {C.}~\bibnamefont {Felser}}, \bibinfo {author} {\bibfnamefont
  {E.}~\bibnamefont {Hassinger}}, \ and\ \bibinfo {author} {\bibfnamefont
  {B.}~\bibnamefont {Yan}},\ }\href@noop {} {\bibfield  {journal} {\bibinfo
  {journal} {Nat. Commun.}\ }\textbf {\bibinfo {volume} {7}},\ \bibinfo {pages}
  {11615} (\bibinfo {year} {2016})}\BibitemShut {NoStop}%
\bibitem [{\citenamefont {Klotz}\ \emph {et~al.}(2016)\citenamefont {Klotz},
  \citenamefont {Wu}, \citenamefont {Shekhar}, \citenamefont {Sun},
  \citenamefont {Schmidt}, \citenamefont {Nicklas}, \citenamefont {Baenitz},
  \citenamefont {Uhlarz}, \citenamefont {Wosnitza}, \citenamefont {Felser},\
  and\ \citenamefont {Yan}}]{klotz2016quantum}%
  \BibitemOpen
  \bibfield  {author} {\bibinfo {author} {\bibfnamefont {J.}~\bibnamefont
  {Klotz}}, \bibinfo {author} {\bibfnamefont {S.-C.}\ \bibnamefont {Wu}},
  \bibinfo {author} {\bibfnamefont {C.}~\bibnamefont {Shekhar}}, \bibinfo
  {author} {\bibfnamefont {Y.}~\bibnamefont {Sun}}, \bibinfo {author}
  {\bibfnamefont {M.}~\bibnamefont {Schmidt}}, \bibinfo {author} {\bibfnamefont
  {M.}~\bibnamefont {Nicklas}}, \bibinfo {author} {\bibfnamefont
  {M.}~\bibnamefont {Baenitz}}, \bibinfo {author} {\bibfnamefont
  {M.}~\bibnamefont {Uhlarz}}, \bibinfo {author} {\bibfnamefont
  {J.}~\bibnamefont {Wosnitza}}, \bibinfo {author} {\bibfnamefont
  {C.}~\bibnamefont {Felser}}, \ and\ \bibinfo {author} {\bibfnamefont
  {B.}~\bibnamefont {Yan}},\ }\href {\doibase 10.1103/PhysRevB.93.121105}
  {\bibfield  {journal} {\bibinfo  {journal} {Phys. Rev. B}\ }\textbf {\bibinfo
  {volume} {93}},\ \bibinfo {pages} {121105} (\bibinfo {year}
  {2016})}\BibitemShut {NoStop}%
\bibitem [{\citenamefont {Yan}\ and\ \citenamefont
  {Felser}(2017)}]{yan2017topological}%
  \BibitemOpen
  \bibfield  {author} {\bibinfo {author} {\bibfnamefont {B.}~\bibnamefont
  {Yan}}\ and\ \bibinfo {author} {\bibfnamefont {C.}~\bibnamefont {Felser}},\
  }\href@noop {} {\bibfield  {journal} {\bibinfo  {journal} {Annu. Rev.
  Condens. Matter Phys.}\ }\textbf {\bibinfo {volume} {8}},\ \bibinfo {pages}
  {337} (\bibinfo {year} {2017})}\BibitemShut {NoStop}%
\bibitem [{\citenamefont {Barfuss}\ \emph {et~al.}(2013)\citenamefont
  {Barfuss}, \citenamefont {Dudy}, \citenamefont {Scholz}, \citenamefont
  {Roth}, \citenamefont {H\"opfner}, \citenamefont {Blumenstein}, \citenamefont
  {Landolt}, \citenamefont {Dil}, \citenamefont {Plumb}, \citenamefont
  {Radovic}, \citenamefont {Bostwick}, \citenamefont {Rotenberg}, \citenamefont
  {Fleszar}, \citenamefont {Bihlmayer}, \citenamefont {Wortmann}, \citenamefont
  {Li}, \citenamefont {Hanke}, \citenamefont {Claessen},\ and\ \citenamefont
  {Sch\"afer}}]{Barfuss2013}%
  \BibitemOpen
  \bibfield  {author} {\bibinfo {author} {\bibfnamefont {A.}~\bibnamefont
  {Barfuss}}, \bibinfo {author} {\bibfnamefont {L.}~\bibnamefont {Dudy}},
  \bibinfo {author} {\bibfnamefont {M.~R.}\ \bibnamefont {Scholz}}, \bibinfo
  {author} {\bibfnamefont {H.}~\bibnamefont {Roth}}, \bibinfo {author}
  {\bibfnamefont {P.}~\bibnamefont {H\"opfner}}, \bibinfo {author}
  {\bibfnamefont {C.}~\bibnamefont {Blumenstein}}, \bibinfo {author}
  {\bibfnamefont {G.}~\bibnamefont {Landolt}}, \bibinfo {author} {\bibfnamefont
  {J.~H.}\ \bibnamefont {Dil}}, \bibinfo {author} {\bibfnamefont {N.~C.}\
  \bibnamefont {Plumb}}, \bibinfo {author} {\bibfnamefont {M.}~\bibnamefont
  {Radovic}}, \bibinfo {author} {\bibfnamefont {A.}~\bibnamefont {Bostwick}},
  \bibinfo {author} {\bibfnamefont {E.}~\bibnamefont {Rotenberg}}, \bibinfo
  {author} {\bibfnamefont {A.}~\bibnamefont {Fleszar}}, \bibinfo {author}
  {\bibfnamefont {G.}~\bibnamefont {Bihlmayer}}, \bibinfo {author}
  {\bibfnamefont {D.}~\bibnamefont {Wortmann}}, \bibinfo {author}
  {\bibfnamefont {G.}~\bibnamefont {Li}}, \bibinfo {author} {\bibfnamefont
  {W.}~\bibnamefont {Hanke}}, \bibinfo {author} {\bibfnamefont
  {R.}~\bibnamefont {Claessen}}, \ and\ \bibinfo {author} {\bibfnamefont
  {J.}~\bibnamefont {Sch\"afer}},\ }\href {\doibase
  10.1103/PhysRevLett.111.157205} {\bibfield  {journal} {\bibinfo  {journal}
  {Phys. Rev. Lett.}\ }\textbf {\bibinfo {volume} {111}},\ \bibinfo {pages}
  {157205} (\bibinfo {year} {2013})}\BibitemShut {NoStop}%
\bibitem [{\citenamefont {Ohtsubo}\ \emph {et~al.}(2013)\citenamefont
  {Ohtsubo}, \citenamefont {Le~F\`evre}, \citenamefont {Bertran},\ and\
  \citenamefont {Taleb-Ibrahimi}}]{Ohtsubo2013}%
  \BibitemOpen
  \bibfield  {author} {\bibinfo {author} {\bibfnamefont {Y.}~\bibnamefont
  {Ohtsubo}}, \bibinfo {author} {\bibfnamefont {P.}~\bibnamefont {Le~F\`evre}},
  \bibinfo {author} {\bibfnamefont {F.~m.~c.}\ \bibnamefont {Bertran}}, \ and\
  \bibinfo {author} {\bibfnamefont {A.}~\bibnamefont {Taleb-Ibrahimi}},\ }\href
  {\doibase 10.1103/PhysRevLett.111.216401} {\bibfield  {journal} {\bibinfo
  {journal} {Phys. Rev. Lett.}\ }\textbf {\bibinfo {volume} {111}},\ \bibinfo
  {pages} {216401} (\bibinfo {year} {2013})}\BibitemShut {NoStop}%
\bibitem [{\citenamefont {Scholz}\ \emph {et~al.}(2018)\citenamefont {Scholz},
  \citenamefont {Rogalev}, \citenamefont {Dudy}, \citenamefont {Reis},
  \citenamefont {Adler}, \citenamefont {Aulbach}, \citenamefont
  {Collins-McIntyre}, \citenamefont {Duffy}, \citenamefont {Yang},
  \citenamefont {Chen}, \citenamefont {Hesjedal}, \citenamefont {Liu},
  \citenamefont {Hoesch}, \citenamefont {Muff}, \citenamefont {Dil},
  \citenamefont {Sch\"afer},\ and\ \citenamefont {Claessen}}]{scholz2018}%
  \BibitemOpen
  \bibfield  {author} {\bibinfo {author} {\bibfnamefont {M.~R.}\ \bibnamefont
  {Scholz}}, \bibinfo {author} {\bibfnamefont {V.~A.}\ \bibnamefont {Rogalev}},
  \bibinfo {author} {\bibfnamefont {L.}~\bibnamefont {Dudy}}, \bibinfo {author}
  {\bibfnamefont {F.}~\bibnamefont {Reis}}, \bibinfo {author} {\bibfnamefont
  {F.}~\bibnamefont {Adler}}, \bibinfo {author} {\bibfnamefont
  {J.}~\bibnamefont {Aulbach}}, \bibinfo {author} {\bibfnamefont {L.~J.}\
  \bibnamefont {Collins-McIntyre}}, \bibinfo {author} {\bibfnamefont {L.~B.}\
  \bibnamefont {Duffy}}, \bibinfo {author} {\bibfnamefont {H.~F.}\ \bibnamefont
  {Yang}}, \bibinfo {author} {\bibfnamefont {Y.~L.}\ \bibnamefont {Chen}},
  \bibinfo {author} {\bibfnamefont {T.}~\bibnamefont {Hesjedal}}, \bibinfo
  {author} {\bibfnamefont {Z.~K.}\ \bibnamefont {Liu}}, \bibinfo {author}
  {\bibfnamefont {M.}~\bibnamefont {Hoesch}}, \bibinfo {author} {\bibfnamefont
  {S.}~\bibnamefont {Muff}}, \bibinfo {author} {\bibfnamefont {J.~H.}\
  \bibnamefont {Dil}}, \bibinfo {author} {\bibfnamefont {J.}~\bibnamefont
  {Sch\"afer}}, \ and\ \bibinfo {author} {\bibfnamefont {R.}~\bibnamefont
  {Claessen}},\ }\href {\doibase 10.1103/PhysRevB.97.075101} {\bibfield
  {journal} {\bibinfo  {journal} {Phys. Rev. B}\ }\textbf {\bibinfo {volume}
  {97}},\ \bibinfo {pages} {075101} (\bibinfo {year} {2018})}\BibitemShut
  {NoStop}%
\bibitem [{\citenamefont {Zhu}\ \emph {et~al.}(2015)\citenamefont {Zhu},
  \citenamefont {Chen}, \citenamefont {Xu}, \citenamefont {Gao}, \citenamefont
  {Guan}, \citenamefont {Liu}, \citenamefont {Qian}, \citenamefont {Zhang},\
  and\ \citenamefont {Jia}}]{zhu2015epitaxial}%
  \BibitemOpen
  \bibfield  {author} {\bibinfo {author} {\bibfnamefont {F.-F.}\ \bibnamefont
  {Zhu}}, \bibinfo {author} {\bibfnamefont {W.-J.}\ \bibnamefont {Chen}},
  \bibinfo {author} {\bibfnamefont {Y.}~\bibnamefont {Xu}}, \bibinfo {author}
  {\bibfnamefont {C.-L.}\ \bibnamefont {Gao}}, \bibinfo {author} {\bibfnamefont
  {D.-D.}\ \bibnamefont {Guan}}, \bibinfo {author} {\bibfnamefont {C.-H.}\
  \bibnamefont {Liu}}, \bibinfo {author} {\bibfnamefont {D.}~\bibnamefont
  {Qian}}, \bibinfo {author} {\bibfnamefont {S.-C.}\ \bibnamefont {Zhang}}, \
  and\ \bibinfo {author} {\bibfnamefont {J.-F.}\ \bibnamefont {Jia}},\
  }\href@noop {} {\bibfield  {journal} {\bibinfo  {journal} {Nat. Mater.}\
  }\textbf {\bibinfo {volume} {14}},\ \bibinfo {pages} {1020} (\bibinfo {year}
  {2015})}\BibitemShut {NoStop}%
\bibitem [{\citenamefont {Liao}\ \emph {et~al.}(2018)\citenamefont {Liao},
  \citenamefont {Zang}, \citenamefont {Guan}, \citenamefont {Li}, \citenamefont
  {Gong}, \citenamefont {Zhu}, \citenamefont {Hu}, \citenamefont {Zhang},
  \citenamefont {Wang}, \citenamefont {He}, \citenamefont {Ma}, \citenamefont
  {Zhang},\ and\ \citenamefont {Qi-Kun}}]{liao2017superconductivity}%
  \BibitemOpen
  \bibfield  {author} {\bibinfo {author} {\bibfnamefont {M.}~\bibnamefont
  {Liao}}, \bibinfo {author} {\bibfnamefont {Y.}~\bibnamefont {Zang}}, \bibinfo
  {author} {\bibfnamefont {Z.}~\bibnamefont {Guan}}, \bibinfo {author}
  {\bibfnamefont {H.}~\bibnamefont {Li}}, \bibinfo {author} {\bibfnamefont
  {Y.}~\bibnamefont {Gong}}, \bibinfo {author} {\bibfnamefont {K.}~\bibnamefont
  {Zhu}}, \bibinfo {author} {\bibfnamefont {X.-P.}\ \bibnamefont {Hu}},
  \bibinfo {author} {\bibfnamefont {D.}~\bibnamefont {Zhang}}, \bibinfo
  {author} {\bibfnamefont {Y.-Y.}\ \bibnamefont {Wang}}, \bibinfo {author}
  {\bibfnamefont {K.}~\bibnamefont {He}}, \bibinfo {author} {\bibfnamefont
  {X.-C.}\ \bibnamefont {Ma}}, \bibinfo {author} {\bibfnamefont {S.-C.}\
  \bibnamefont {Zhang}}, \ and\ \bibinfo {author} {\bibfnamefont
  {X.}~\bibnamefont {Qi-Kun}},\ }\href@noop {} {\bibfield  {journal} {\bibinfo
  {journal} {Nat. Phys.}\ }\textbf {\bibinfo {volume} {14}},\ \bibinfo {pages}
  {344} (\bibinfo {year} {2018})}\BibitemShut {NoStop}%
\bibitem [{\citenamefont {Li}\ \emph {et~al.}(2016{\natexlab{b}})\citenamefont
  {Li}, \citenamefont {He}, \citenamefont {Lu}, \citenamefont {Zhang},
  \citenamefont {Liu}, \citenamefont {Ma}, \citenamefont {Fan}, \citenamefont
  {Shen},\ and\ \citenamefont {Wang}}]{Hui2015Negative}%
  \BibitemOpen
  \bibfield  {author} {\bibinfo {author} {\bibfnamefont {H.}~\bibnamefont
  {Li}}, \bibinfo {author} {\bibfnamefont {H.}~\bibnamefont {He}}, \bibinfo
  {author} {\bibfnamefont {H.~Z.}\ \bibnamefont {Lu}}, \bibinfo {author}
  {\bibfnamefont {H.}~\bibnamefont {Zhang}}, \bibinfo {author} {\bibfnamefont
  {H.}~\bibnamefont {Liu}}, \bibinfo {author} {\bibfnamefont {R.}~\bibnamefont
  {Ma}}, \bibinfo {author} {\bibfnamefont {Z.}~\bibnamefont {Fan}}, \bibinfo
  {author} {\bibfnamefont {S.~Q.}\ \bibnamefont {Shen}}, \ and\ \bibinfo
  {author} {\bibfnamefont {J.}~\bibnamefont {Wang}},\ }\href@noop {} {\bibfield
   {journal} {\bibinfo  {journal} {Nat. Commun.}\ }\textbf {\bibinfo {volume}
  {7}},\ \bibinfo {pages} {10301} (\bibinfo {year}
  {2016}{\natexlab{b}})}\BibitemShut {NoStop}%
\bibitem [{\citenamefont {Li}\ \emph {et~al.}(2015)\citenamefont {Li},
  \citenamefont {Wang}, \citenamefont {Liu}, \citenamefont {Jian},
  \citenamefont {Liao},\ and\ \citenamefont {Yu}}]{Li2015Giant}%
  \BibitemOpen
  \bibfield  {author} {\bibinfo {author} {\bibfnamefont {C.~Z.}\ \bibnamefont
  {Li}}, \bibinfo {author} {\bibfnamefont {L.~X.}\ \bibnamefont {Wang}},
  \bibinfo {author} {\bibfnamefont {H.}~\bibnamefont {Liu}}, \bibinfo {author}
  {\bibfnamefont {W.}~\bibnamefont {Jian}}, \bibinfo {author} {\bibfnamefont
  {Z.~M.}\ \bibnamefont {Liao}}, \ and\ \bibinfo {author} {\bibfnamefont
  {D.~P.}\ \bibnamefont {Yu}},\ }\href@noop {} {\bibfield  {journal} {\bibinfo
  {journal} {Nat. Commun.}\ }\textbf {\bibinfo {volume} {6}},\ \bibinfo {pages}
  {10137} (\bibinfo {year} {2015})}\BibitemShut {NoStop}%
\bibitem [{\citenamefont {Liang}\ \emph {et~al.}(2015)\citenamefont {Liang},
  \citenamefont {Gibson}, \citenamefont {Ali}, \citenamefont {Liu},
  \citenamefont {Cava},\ and\ \citenamefont {Ong}}]{Liang2015Ultrahigh}%
  \BibitemOpen
  \bibfield  {author} {\bibinfo {author} {\bibfnamefont {T.}~\bibnamefont
  {Liang}}, \bibinfo {author} {\bibfnamefont {Q.}~\bibnamefont {Gibson}},
  \bibinfo {author} {\bibfnamefont {M.~N.}\ \bibnamefont {Ali}}, \bibinfo
  {author} {\bibfnamefont {M.}~\bibnamefont {Liu}}, \bibinfo {author}
  {\bibfnamefont {R.~J.}\ \bibnamefont {Cava}}, \ and\ \bibinfo {author}
  {\bibfnamefont {N.~P.}\ \bibnamefont {Ong}},\ }\href@noop {} {\bibfield
  {journal} {\bibinfo  {journal} {Nat. Mater.}\ }\textbf {\bibinfo {volume}
  {14}},\ \bibinfo {pages} {280} (\bibinfo {year} {2015})}\BibitemShut
  {NoStop}%
\bibitem [{\citenamefont {Hirschberger}\ \emph {et~al.}(2016)\citenamefont
  {Hirschberger}, \citenamefont {Kushwaha}, \citenamefont {Wang}, \citenamefont
  {Gibson}, \citenamefont {Liang}, \citenamefont {Belvin}, \citenamefont
  {Bernevig}, \citenamefont {Cava},\ and\ \citenamefont
  {Ong}}]{Hirschberger2016}%
  \BibitemOpen
  \bibfield  {author} {\bibinfo {author} {\bibfnamefont {M.}~\bibnamefont
  {Hirschberger}}, \bibinfo {author} {\bibfnamefont {S.}~\bibnamefont
  {Kushwaha}}, \bibinfo {author} {\bibfnamefont {Z.}~\bibnamefont {Wang}},
  \bibinfo {author} {\bibfnamefont {Q.}~\bibnamefont {Gibson}}, \bibinfo
  {author} {\bibfnamefont {S.}~\bibnamefont {Liang}}, \bibinfo {author}
  {\bibfnamefont {C.~A.}\ \bibnamefont {Belvin}}, \bibinfo {author}
  {\bibfnamefont {B.~A.}\ \bibnamefont {Bernevig}}, \bibinfo {author}
  {\bibfnamefont {R.~J.}\ \bibnamefont {Cava}}, \ and\ \bibinfo {author}
  {\bibfnamefont {N.~P.}\ \bibnamefont {Ong}},\ }\href@noop {} {\bibfield
  {journal} {\bibinfo  {journal} {Nat. Mater.}\ }\textbf {\bibinfo {volume}
  {15}},\ \bibinfo {pages} {1161} (\bibinfo {year} {2016})}\BibitemShut
  {NoStop}%
\bibitem [{\citenamefont {Shekhar}\ \emph {et~al.}()\citenamefont {Shekhar},
  \citenamefont {Nayak}, \citenamefont {Singh}, \citenamefont {Kumar},
  \citenamefont {Wu}, \citenamefont {Zhang}, \citenamefont {Komarek},
  \citenamefont {Kampert}, \citenamefont {Skourski},\ and\ \citenamefont
  {Wosnitza}}]{Shekhar2016Observation}%
  \BibitemOpen
  \bibfield  {author} {\bibinfo {author} {\bibfnamefont {C.}~\bibnamefont
  {Shekhar}}, \bibinfo {author} {\bibfnamefont {A.~K.}\ \bibnamefont {Nayak}},
  \bibinfo {author} {\bibfnamefont {S.}~\bibnamefont {Singh}}, \bibinfo
  {author} {\bibfnamefont {N.}~\bibnamefont {Kumar}}, \bibinfo {author}
  {\bibfnamefont {S.~C.}\ \bibnamefont {Wu}}, \bibinfo {author} {\bibfnamefont
  {Y.}~\bibnamefont {Zhang}}, \bibinfo {author} {\bibfnamefont {A.~C.}\
  \bibnamefont {Komarek}}, \bibinfo {author} {\bibfnamefont {E.}~\bibnamefont
  {Kampert}}, \bibinfo {author} {\bibfnamefont {Y.}~\bibnamefont {Skourski}}, \
  and\ \bibinfo {author} {\bibfnamefont {J.}~\bibnamefont {Wosnitza}},\
  }\href@noop {} {\bibinfo  {journal} {arXiv:1604.01641}\ }\BibitemShut
  {NoStop}%
\bibitem [{\citenamefont {Cano}\ \emph {et~al.}(2017)\citenamefont {Cano},
  \citenamefont {Bradlyn}, \citenamefont {Wang}, \citenamefont {Hirschberger},
  \citenamefont {Ong},\ and\ \citenamefont {Bernevig}}]{Cano}%
  \BibitemOpen
\bibfield  {journal} {  }\bibfield  {author} {\bibinfo {author} {\bibfnamefont
  {J.}~\bibnamefont {Cano}}, \bibinfo {author} {\bibfnamefont {B.}~\bibnamefont
  {Bradlyn}}, \bibinfo {author} {\bibfnamefont {Z.}~\bibnamefont {Wang}},
  \bibinfo {author} {\bibfnamefont {M.}~\bibnamefont {Hirschberger}}, \bibinfo
  {author} {\bibfnamefont {N.~P.}\ \bibnamefont {Ong}}, \ and\ \bibinfo
  {author} {\bibfnamefont {B.~A.}\ \bibnamefont {Bernevig}},\ }\href {\doibase
  10.1103/PhysRevB.95.161306} {\bibfield  {journal} {\bibinfo  {journal} {Phys.
  Rev. B}\ }\textbf {\bibinfo {volume} {95}},\ \bibinfo {pages} {161306}
  (\bibinfo {year} {2017})}\BibitemShut {NoStop}%
\bibitem [{\citenamefont {Oh}\ and\ \citenamefont
  {Yang}(2017)}]{oh2017topological}%
  \BibitemOpen
  \bibfield  {author} {\bibinfo {author} {\bibfnamefont {T.}~\bibnamefont
  {Oh}}\ and\ \bibinfo {author} {\bibfnamefont {B.-J.}\ \bibnamefont {Yang}},\
  }\href@noop {} {\bibfield  {journal} {\bibinfo  {journal} {arXiv:1709.06796}\
  } (\bibinfo {year} {2017})}\BibitemShut {NoStop}%
\bibitem [{\citenamefont {Lindner}\ \emph {et~al.}(2011)\citenamefont
  {Lindner}, \citenamefont {Refael},\ and\ \citenamefont
  {Galitski}}]{lindner2011floquet}%
  \BibitemOpen
  \bibfield  {author} {\bibinfo {author} {\bibfnamefont {N.~H.}\ \bibnamefont
  {Lindner}}, \bibinfo {author} {\bibfnamefont {G.}~\bibnamefont {Refael}}, \
  and\ \bibinfo {author} {\bibfnamefont {V.}~\bibnamefont {Galitski}},\
  }\href@noop {} {\bibfield  {journal} {\bibinfo  {journal} {Nat. Phys.}\
  }\textbf {\bibinfo {volume} {7}},\ \bibinfo {pages} {490} (\bibinfo {year}
  {2011})}\BibitemShut {NoStop}%
\bibitem [{\citenamefont {Cayssol}\ \emph {et~al.}(2013)\citenamefont
  {Cayssol}, \citenamefont {D{\'o}ra}, \citenamefont {Simon},\ and\
  \citenamefont {Moessner}}]{cayssol2013floquet}%
  \BibitemOpen
  \bibfield  {author} {\bibinfo {author} {\bibfnamefont {J.}~\bibnamefont
  {Cayssol}}, \bibinfo {author} {\bibfnamefont {B.}~\bibnamefont {D{\'o}ra}},
  \bibinfo {author} {\bibfnamefont {F.}~\bibnamefont {Simon}}, \ and\ \bibinfo
  {author} {\bibfnamefont {R.}~\bibnamefont {Moessner}},\ }\href@noop {}
  {\bibfield  {journal} {\bibinfo  {journal} {Phys. Status Solidi RRL}\
  }\textbf {\bibinfo {volume} {7}},\ \bibinfo {pages} {101} (\bibinfo {year}
  {2013})}\BibitemShut {NoStop}%
\bibitem [{\citenamefont {Narayan}(2015)}]{Narayan2015}%
  \BibitemOpen
  \bibfield  {author} {\bibinfo {author} {\bibfnamefont {A.}~\bibnamefont
  {Narayan}},\ }\href {\doibase 10.1103/PhysRevB.91.205445} {\bibfield
  {journal} {\bibinfo  {journal} {Phys. Rev. B}\ }\textbf {\bibinfo {volume}
  {91}},\ \bibinfo {pages} {205445} (\bibinfo {year} {2015})}\BibitemShut
  {NoStop}%
\bibitem [{\citenamefont {Hubener}\ \emph {et~al.}(2017)\citenamefont
  {Hubener}, \citenamefont {Sentef}, \citenamefont {De~Giovannini},
  \citenamefont {Kemper},\ and\ \citenamefont {Rubio}}]{Hubener2017}%
  \BibitemOpen
  \bibfield  {author} {\bibinfo {author} {\bibfnamefont {H.}~\bibnamefont
  {Hubener}}, \bibinfo {author} {\bibfnamefont {M.~A.}\ \bibnamefont {Sentef}},
  \bibinfo {author} {\bibfnamefont {U.}~\bibnamefont {De~Giovannini}}, \bibinfo
  {author} {\bibfnamefont {A.~F.}\ \bibnamefont {Kemper}}, \ and\ \bibinfo
  {author} {\bibfnamefont {A.}~\bibnamefont {Rubio}},\ }\href@noop {}
  {\bibfield  {journal} {\bibinfo  {journal} {Nat. Commun.}\ }\textbf {\bibinfo
  {volume} {8}},\ \bibinfo {pages} {13940} (\bibinfo {year}
  {2017})}\BibitemShut {NoStop}%
\bibitem [{\citenamefont {Chan}\ \emph
  {et~al.}(2016{\natexlab{a}})\citenamefont {Chan}, \citenamefont {Lee},
  \citenamefont {Burch}, \citenamefont {Han},\ and\ \citenamefont
  {Ran}}]{Chan2016a}%
  \BibitemOpen
  \bibfield  {author} {\bibinfo {author} {\bibfnamefont {C.-K.}\ \bibnamefont
  {Chan}}, \bibinfo {author} {\bibfnamefont {P.~A.}\ \bibnamefont {Lee}},
  \bibinfo {author} {\bibfnamefont {K.~S.}\ \bibnamefont {Burch}}, \bibinfo
  {author} {\bibfnamefont {J.~H.}\ \bibnamefont {Han}}, \ and\ \bibinfo
  {author} {\bibfnamefont {Y.}~\bibnamefont {Ran}},\ }\href {\doibase
  10.1103/PhysRevLett.116.026805} {\bibfield  {journal} {\bibinfo  {journal}
  {Phys. Rev. Lett.}\ }\textbf {\bibinfo {volume} {116}},\ \bibinfo {pages}
  {026805} (\bibinfo {year} {2016}{\natexlab{a}})}\BibitemShut {NoStop}%
\bibitem [{\citenamefont {Chan}\ \emph
  {et~al.}(2016{\natexlab{b}})\citenamefont {Chan}, \citenamefont {Oh},
  \citenamefont {Han},\ and\ \citenamefont {Lee}}]{Chan2016b}%
  \BibitemOpen
  \bibfield  {author} {\bibinfo {author} {\bibfnamefont {C.-K.}\ \bibnamefont
  {Chan}}, \bibinfo {author} {\bibfnamefont {Y.-T.}\ \bibnamefont {Oh}},
  \bibinfo {author} {\bibfnamefont {J.~H.}\ \bibnamefont {Han}}, \ and\
  \bibinfo {author} {\bibfnamefont {P.~A.}\ \bibnamefont {Lee}},\ }\href
  {\doibase 10.1103/PhysRevB.94.121106} {\bibfield  {journal} {\bibinfo
  {journal} {Phys. Rev. B}\ }\textbf {\bibinfo {volume} {94}},\ \bibinfo
  {pages} {121106} (\bibinfo {year} {2016}{\natexlab{b}})}\BibitemShut
  {NoStop}%
\bibitem [{\citenamefont {Yan}\ and\ \citenamefont {Wang}(2016)}]{Yan2016}%
  \BibitemOpen
  \bibfield  {author} {\bibinfo {author} {\bibfnamefont {Z.}~\bibnamefont
  {Yan}}\ and\ \bibinfo {author} {\bibfnamefont {Z.}~\bibnamefont {Wang}},\
  }\href {\doibase 10.1103/PhysRevLett.117.087402} {\bibfield  {journal}
  {\bibinfo  {journal} {Phys. Rev. Lett.}\ }\textbf {\bibinfo {volume} {117}},\
  \bibinfo {pages} {087402} (\bibinfo {year} {2016})}\BibitemShut {NoStop}%
\bibitem [{\citenamefont {Taguchi}\ \emph {et~al.}(2016)\citenamefont
  {Taguchi}, \citenamefont {Xu}, \citenamefont {Yamakage},\ and\ \citenamefont
  {Law}}]{Taguchi2016}%
  \BibitemOpen
  \bibfield  {author} {\bibinfo {author} {\bibfnamefont {K.}~\bibnamefont
  {Taguchi}}, \bibinfo {author} {\bibfnamefont {D.-H.}\ \bibnamefont {Xu}},
  \bibinfo {author} {\bibfnamefont {A.}~\bibnamefont {Yamakage}}, \ and\
  \bibinfo {author} {\bibfnamefont {K.~T.}\ \bibnamefont {Law}},\ }\href
  {\doibase 10.1103/PhysRevB.94.155206} {\bibfield  {journal} {\bibinfo
  {journal} {Phys. Rev. B}\ }\textbf {\bibinfo {volume} {94}},\ \bibinfo
  {pages} {155206} (\bibinfo {year} {2016})}\BibitemShut {NoStop}%
\bibitem [{\citenamefont {Bomantara}\ \emph {et~al.}(2016)\citenamefont
  {Bomantara}, \citenamefont {Raghava}, \citenamefont {Zhou},\ and\
  \citenamefont {Gong}}]{Bomantara2016}%
  \BibitemOpen
  \bibfield  {author} {\bibinfo {author} {\bibfnamefont {R.~W.}\ \bibnamefont
  {Bomantara}}, \bibinfo {author} {\bibfnamefont {G.~N.}\ \bibnamefont
  {Raghava}}, \bibinfo {author} {\bibfnamefont {L.}~\bibnamefont {Zhou}}, \
  and\ \bibinfo {author} {\bibfnamefont {J.}~\bibnamefont {Gong}},\ }\href
  {\doibase 10.1103/PhysRevE.93.022209} {\bibfield  {journal} {\bibinfo
  {journal} {Phys. Rev. E}\ }\textbf {\bibinfo {volume} {93}},\ \bibinfo
  {pages} {022209} (\bibinfo {year} {2016})}\BibitemShut {NoStop}%
\bibitem [{\citenamefont {Wang}\ \emph {et~al.}(2014)\citenamefont {Wang},
  \citenamefont {Wang}, \citenamefont {Shen}, \citenamefont {Sheng},\ and\
  \citenamefont {Xing}}]{Rui2014}%
  \BibitemOpen
  \bibfield  {author} {\bibinfo {author} {\bibfnamefont {R.}~\bibnamefont
  {Wang}}, \bibinfo {author} {\bibfnamefont {B.}~\bibnamefont {Wang}}, \bibinfo
  {author} {\bibfnamefont {R.}~\bibnamefont {Shen}}, \bibinfo {author}
  {\bibfnamefont {L.}~\bibnamefont {Sheng}}, \ and\ \bibinfo {author}
  {\bibfnamefont {D.~Y.}\ \bibnamefont {Xing}},\ }\href
  {http://stacks.iop.org/0295-5075/105/i=1/a=17004} {\bibfield  {journal}
  {\bibinfo  {journal} {Europhys. Lett.}\ }\textbf {\bibinfo {volume} {105}},\
  \bibinfo {pages} {17004} (\bibinfo {year} {2014})}\BibitemShut {NoStop}%
\bibitem [{\citenamefont {Chen}\ \emph {et~al.}(2016)\citenamefont {Chen},
  \citenamefont {Su}, \citenamefont {Deng}, \citenamefont {Ruan}, \citenamefont
  {Luo}, \citenamefont {Shao}, \citenamefont {Sheng},\ and\ \citenamefont
  {Xing}}]{Chen2016}%
  \BibitemOpen
  \bibfield  {author} {\bibinfo {author} {\bibfnamefont {M.~N.}\ \bibnamefont
  {Chen}}, \bibinfo {author} {\bibfnamefont {W.}~\bibnamefont {Su}}, \bibinfo
  {author} {\bibfnamefont {M.~X.}\ \bibnamefont {Deng}}, \bibinfo {author}
  {\bibfnamefont {J.}~\bibnamefont {Ruan}}, \bibinfo {author} {\bibfnamefont
  {W.}~\bibnamefont {Luo}}, \bibinfo {author} {\bibfnamefont {D.~X.}\
  \bibnamefont {Shao}}, \bibinfo {author} {\bibfnamefont {L.}~\bibnamefont
  {Sheng}}, \ and\ \bibinfo {author} {\bibfnamefont {D.~Y.}\ \bibnamefont
  {Xing}},\ }\href {\doibase 10.1103/PhysRevB.94.205429} {\bibfield  {journal}
  {\bibinfo  {journal} {Phys. Rev. B}\ }\textbf {\bibinfo {volume} {94}},\
  \bibinfo {pages} {205429} (\bibinfo {year} {2016})}\BibitemShut {NoStop}%
\bibitem [{\citenamefont {Zhang}\ \emph
  {et~al.}(2016{\natexlab{c}})\citenamefont {Zhang}, \citenamefont {Ong},\ and\
  \citenamefont {Nagaosa}}]{Zhang2016}%
  \BibitemOpen
  \bibfield  {author} {\bibinfo {author} {\bibfnamefont {X.-X.}\ \bibnamefont
  {Zhang}}, \bibinfo {author} {\bibfnamefont {T.~T.}\ \bibnamefont {Ong}}, \
  and\ \bibinfo {author} {\bibfnamefont {N.}~\bibnamefont {Nagaosa}},\ }\href
  {\doibase 10.1103/PhysRevB.94.235137} {\bibfield  {journal} {\bibinfo
  {journal} {Phys. Rev. B}\ }\textbf {\bibinfo {volume} {94}},\ \bibinfo
  {pages} {235137} (\bibinfo {year} {2016}{\natexlab{c}})}\BibitemShut
  {NoStop}%
\bibitem [{\citenamefont {Bomantara}\ and\ \citenamefont
  {Gong}(2016)}]{Bomantara2016generating}%
  \BibitemOpen
  \bibfield  {author} {\bibinfo {author} {\bibfnamefont {R.~W.}\ \bibnamefont
  {Bomantara}}\ and\ \bibinfo {author} {\bibfnamefont {J.}~\bibnamefont
  {Gong}},\ }\href {\doibase 10.1103/PhysRevB.94.235447} {\bibfield  {journal}
  {\bibinfo  {journal} {Phys. Rev. B}\ }\textbf {\bibinfo {volume} {94}},\
  \bibinfo {pages} {235447} (\bibinfo {year} {2016})}\BibitemShut {NoStop}%
\bibitem [{\citenamefont {Zhou}\ \emph {et~al.}(2016)\citenamefont {Zhou},
  \citenamefont {Chen},\ and\ \citenamefont {Gong}}]{Zhou2016Floquet}%
  \BibitemOpen
  \bibfield  {author} {\bibinfo {author} {\bibfnamefont {L.}~\bibnamefont
  {Zhou}}, \bibinfo {author} {\bibfnamefont {C.}~\bibnamefont {Chen}}, \ and\
  \bibinfo {author} {\bibfnamefont {J.}~\bibnamefont {Gong}},\ }\href {\doibase
  10.1103/PhysRevB.94.075443} {\bibfield  {journal} {\bibinfo  {journal} {Phys.
  Rev. B}\ }\textbf {\bibinfo {volume} {94}},\ \bibinfo {pages} {075443}
  (\bibinfo {year} {2016})}\BibitemShut {NoStop}%
\bibitem [{\citenamefont {Yan}\ and\ \citenamefont {Wang}(2017)}]{Yan2017}%
  \BibitemOpen
  \bibfield  {author} {\bibinfo {author} {\bibfnamefont {Z.}~\bibnamefont
  {Yan}}\ and\ \bibinfo {author} {\bibfnamefont {Z.}~\bibnamefont {Wang}},\
  }\href {\doibase 10.1103/PhysRevB.96.041206} {\bibfield  {journal} {\bibinfo
  {journal} {Phys. Rev. B}\ }\textbf {\bibinfo {volume} {96}},\ \bibinfo
  {pages} {041206} (\bibinfo {year} {2017})}\BibitemShut {NoStop}%
\bibitem [{\citenamefont {Ezawa}(2017)}]{Ezawa2017}%
  \BibitemOpen
  \bibfield  {author} {\bibinfo {author} {\bibfnamefont {M.}~\bibnamefont
  {Ezawa}},\ }\href {\doibase 10.1103/PhysRevB.96.041205} {\bibfield  {journal}
  {\bibinfo  {journal} {Phys. Rev. B}\ }\textbf {\bibinfo {volume} {96}},\
  \bibinfo {pages} {041205} (\bibinfo {year} {2017})}\BibitemShut {NoStop}%
\bibitem [{\citenamefont {Gupta}(2017)}]{Gupta2017}%
  \BibitemOpen
  \bibfield  {author} {\bibinfo {author} {\bibfnamefont {A.}~\bibnamefont
  {Gupta}},\ }\href@noop {} {\bibfield  {journal} {\bibinfo  {journal}
  {arXiv:1703.07271}\ } (\bibinfo {year} {2017})}\BibitemShut {NoStop}%
\bibitem [{\citenamefont {Wang}\ \emph {et~al.}(2017)\citenamefont {Wang},
  \citenamefont {Chen}, \citenamefont {Bomantara}, \citenamefont {Gong},\ and\
  \citenamefont {Xing}}]{Wang2017Line}%
  \BibitemOpen
  \bibfield  {author} {\bibinfo {author} {\bibfnamefont {H.-Q.}\ \bibnamefont
  {Wang}}, \bibinfo {author} {\bibfnamefont {M.~N.}\ \bibnamefont {Chen}},
  \bibinfo {author} {\bibfnamefont {R.~W.}\ \bibnamefont {Bomantara}}, \bibinfo
  {author} {\bibfnamefont {J.}~\bibnamefont {Gong}}, \ and\ \bibinfo {author}
  {\bibfnamefont {D.~Y.}\ \bibnamefont {Xing}},\ }\href {\doibase
  10.1103/PhysRevB.95.075136} {\bibfield  {journal} {\bibinfo  {journal} {Phys.
  Rev. B}\ }\textbf {\bibinfo {volume} {95}},\ \bibinfo {pages} {075136}
  (\bibinfo {year} {2017})}\BibitemShut {NoStop}%
\bibitem [{\citenamefont {Soluyanov}\ \emph {et~al.}(2015)\citenamefont
  {Soluyanov}, \citenamefont {Gresch}, \citenamefont {Wang}, \citenamefont
  {Wu}, \citenamefont {Troyer}, \citenamefont {Dai},\ and\ \citenamefont
  {Bernevig}}]{Soluyanov2015}%
  \BibitemOpen
  \bibfield  {author} {\bibinfo {author} {\bibfnamefont {A.~A.}\ \bibnamefont
  {Soluyanov}}, \bibinfo {author} {\bibfnamefont {D.}~\bibnamefont {Gresch}},
  \bibinfo {author} {\bibfnamefont {Z.}~\bibnamefont {Wang}}, \bibinfo {author}
  {\bibfnamefont {Q.~S.}\ \bibnamefont {Wu}}, \bibinfo {author} {\bibfnamefont
  {M.}~\bibnamefont {Troyer}}, \bibinfo {author} {\bibfnamefont
  {X.}~\bibnamefont {Dai}}, \ and\ \bibinfo {author} {\bibfnamefont {B.~A.}\
  \bibnamefont {Bernevig}},\ }\href@noop {} {\bibfield  {journal} {\bibinfo
  {journal} {Nature}\ }\textbf {\bibinfo {volume} {527}},\ \bibinfo {pages}
  {495} (\bibinfo {year} {2015})}\BibitemShut {NoStop}%
\bibitem [{\citenamefont {Wang}\ \emph
  {et~al.}(2016{\natexlab{c}})\citenamefont {Wang}, \citenamefont {Gresch},
  \citenamefont {Soluyanov}, \citenamefont {Xie}, \citenamefont {Kushwaha},
  \citenamefont {Dai}, \citenamefont {Troyer}, \citenamefont {Cava},\ and\
  \citenamefont {Bernevig}}]{Wang2016}%
  \BibitemOpen
  \bibfield  {author} {\bibinfo {author} {\bibfnamefont {Z.}~\bibnamefont
  {Wang}}, \bibinfo {author} {\bibfnamefont {D.}~\bibnamefont {Gresch}},
  \bibinfo {author} {\bibfnamefont {A.~A.}\ \bibnamefont {Soluyanov}}, \bibinfo
  {author} {\bibfnamefont {W.}~\bibnamefont {Xie}}, \bibinfo {author}
  {\bibfnamefont {S.}~\bibnamefont {Kushwaha}}, \bibinfo {author}
  {\bibfnamefont {X.}~\bibnamefont {Dai}}, \bibinfo {author} {\bibfnamefont
  {M.}~\bibnamefont {Troyer}}, \bibinfo {author} {\bibfnamefont {R.~J.}\
  \bibnamefont {Cava}}, \ and\ \bibinfo {author} {\bibfnamefont {B.~A.}\
  \bibnamefont {Bernevig}},\ }\href {\doibase 10.1103/PhysRevLett.117.056805}
  {\bibfield  {journal} {\bibinfo  {journal} {Phys. Rev. Lett.}\ }\textbf
  {\bibinfo {volume} {117}},\ \bibinfo {pages} {056805} (\bibinfo {year}
  {2016}{\natexlab{c}})}\BibitemShut {NoStop}%
\bibitem [{\citenamefont {Aut\`es}\ \emph {et~al.}(2016)\citenamefont
  {Aut\`es}, \citenamefont {Gresch}, \citenamefont {Troyer}, \citenamefont
  {Soluyanov},\ and\ \citenamefont {Yazyev}}]{Aut2016}%
  \BibitemOpen
  \bibfield  {author} {\bibinfo {author} {\bibfnamefont {G.}~\bibnamefont
  {Aut\`es}}, \bibinfo {author} {\bibfnamefont {D.}~\bibnamefont {Gresch}},
  \bibinfo {author} {\bibfnamefont {M.}~\bibnamefont {Troyer}}, \bibinfo
  {author} {\bibfnamefont {A.~A.}\ \bibnamefont {Soluyanov}}, \ and\ \bibinfo
  {author} {\bibfnamefont {O.~V.}\ \bibnamefont {Yazyev}},\ }\href {\doibase
  10.1103/PhysRevLett.117.066402} {\bibfield  {journal} {\bibinfo  {journal}
  {Phys. Rev. Lett.}\ }\textbf {\bibinfo {volume} {117}},\ \bibinfo {pages}
  {066402} (\bibinfo {year} {2016})}\BibitemShut {NoStop}%
\bibitem [{\citenamefont {Sun}\ \emph {et~al.}(2015)\citenamefont {Sun},
  \citenamefont {Wu}, \citenamefont {Ali}, \citenamefont {Felser},\ and\
  \citenamefont {Yan}}]{Sun2015}%
  \BibitemOpen
  \bibfield  {author} {\bibinfo {author} {\bibfnamefont {Y.}~\bibnamefont
  {Sun}}, \bibinfo {author} {\bibfnamefont {S.-C.}\ \bibnamefont {Wu}},
  \bibinfo {author} {\bibfnamefont {M.~N.}\ \bibnamefont {Ali}}, \bibinfo
  {author} {\bibfnamefont {C.}~\bibnamefont {Felser}}, \ and\ \bibinfo {author}
  {\bibfnamefont {B.}~\bibnamefont {Yan}},\ }\href {\doibase
  10.1103/PhysRevB.92.161107} {\bibfield  {journal} {\bibinfo  {journal} {Phys.
  Rev. B}\ }\textbf {\bibinfo {volume} {92}},\ \bibinfo {pages} {161107}
  (\bibinfo {year} {2015})}\BibitemShut {NoStop}%
\bibitem [{\citenamefont {Fang}\ \emph {et~al.}(2012)\citenamefont {Fang},
  \citenamefont {Gilbert}, \citenamefont {Dai},\ and\ \citenamefont
  {Bernevig}}]{Fang2012}%
  \BibitemOpen
  \bibfield  {author} {\bibinfo {author} {\bibfnamefont {C.}~\bibnamefont
  {Fang}}, \bibinfo {author} {\bibfnamefont {M.~J.}\ \bibnamefont {Gilbert}},
  \bibinfo {author} {\bibfnamefont {X.}~\bibnamefont {Dai}}, \ and\ \bibinfo
  {author} {\bibfnamefont {B.~A.}\ \bibnamefont {Bernevig}},\ }\href {\doibase
  10.1103/PhysRevLett.108.266802} {\bibfield  {journal} {\bibinfo  {journal}
  {Phys. Rev. Lett.}\ }\textbf {\bibinfo {volume} {108}},\ \bibinfo {pages}
  {266802} (\bibinfo {year} {2012})}\BibitemShut {NoStop}%
\bibitem [{\citenamefont {Huang}\ \emph {et~al.}(2016)\citenamefont {Huang},
  \citenamefont {Xu}, \citenamefont {Belopolski}, \citenamefont {Lee},
  \citenamefont {Chang}, \citenamefont {Chang}, \citenamefont {Wang},
  \citenamefont {Alidoust}, \citenamefont {Bian}, \citenamefont {Neupane},
  \citenamefont {Sanchez}, \citenamefont {Zheng}, \citenamefont {Jeng},
  \citenamefont {Bansil}, \citenamefont {Neupert}, \citenamefont {Lin},\ and\
  \citenamefont {Hasan}}]{Huang2015New}%
  \BibitemOpen
  \bibfield  {author} {\bibinfo {author} {\bibfnamefont {S.-M.}\ \bibnamefont
  {Huang}}, \bibinfo {author} {\bibfnamefont {S.-Y.}\ \bibnamefont {Xu}},
  \bibinfo {author} {\bibfnamefont {I.}~\bibnamefont {Belopolski}}, \bibinfo
  {author} {\bibfnamefont {C.-C.}\ \bibnamefont {Lee}}, \bibinfo {author}
  {\bibfnamefont {G.}~\bibnamefont {Chang}}, \bibinfo {author} {\bibfnamefont
  {T.-R.}\ \bibnamefont {Chang}}, \bibinfo {author} {\bibfnamefont
  {B.}~\bibnamefont {Wang}}, \bibinfo {author} {\bibfnamefont {N.}~\bibnamefont
  {Alidoust}}, \bibinfo {author} {\bibfnamefont {G.}~\bibnamefont {Bian}},
  \bibinfo {author} {\bibfnamefont {M.}~\bibnamefont {Neupane}}, \bibinfo
  {author} {\bibfnamefont {D.}~\bibnamefont {Sanchez}}, \bibinfo {author}
  {\bibfnamefont {H.}~\bibnamefont {Zheng}}, \bibinfo {author} {\bibfnamefont
  {H.-T.}\ \bibnamefont {Jeng}}, \bibinfo {author} {\bibfnamefont
  {A.}~\bibnamefont {Bansil}}, \bibinfo {author} {\bibfnamefont
  {T.}~\bibnamefont {Neupert}}, \bibinfo {author} {\bibfnamefont
  {H.}~\bibnamefont {Lin}}, \ and\ \bibinfo {author} {\bibfnamefont {M.~Z.}\
  \bibnamefont {Hasan}},\ }\href {\doibase 10.1073/pnas.1514581113} {\bibfield
  {journal} {\bibinfo  {journal} {Proc. Natl. Acad. Sci.}\ }\textbf {\bibinfo
  {volume} {113}},\ \bibinfo {pages} {1180} (\bibinfo {year}
  {2016})}\BibitemShut {NoStop}%
\bibitem [{\citenamefont {Yao}\ and\ \citenamefont {Chen}(2017)}]{yao2017pr}%
  \BibitemOpen
  \bibfield  {author} {\bibinfo {author} {\bibfnamefont {X.~P.}\ \bibnamefont
  {Yao}}\ and\ \bibinfo {author} {\bibfnamefont {G.}~\bibnamefont {Chen}},\
  }\href@noop {} {\bibfield  {journal} {\bibinfo  {journal} {arXiv:1712.06534}\
  } (\bibinfo {year} {2017})}\BibitemShut {NoStop}%
\bibitem [{\citenamefont {Ghorashi}\ \emph {et~al.}(2018)\citenamefont
  {Ghorashi}, \citenamefont {Hosur},\ and\ \citenamefont
  {Ting}}]{Ghorashi2018}%
  \BibitemOpen
  \bibfield  {author} {\bibinfo {author} {\bibfnamefont {S.~A.~A.}\
  \bibnamefont {Ghorashi}}, \bibinfo {author} {\bibfnamefont {P.}~\bibnamefont
  {Hosur}}, \ and\ \bibinfo {author} {\bibfnamefont {C.-S.}\ \bibnamefont
  {Ting}},\ }\href {\doibase 10.1103/PhysRevB.97.205402} {\bibfield  {journal}
  {\bibinfo  {journal} {Phys. Rev. B}\ }\textbf {\bibinfo {volume} {97}},\
  \bibinfo {pages} {205402} (\bibinfo {year} {2018})}\BibitemShut {NoStop}%
\bibitem [{\citenamefont {Kresse}\ and\ \citenamefont
  {Joubert}(1999)}]{kresse1999}%
  \BibitemOpen
  \bibfield  {author} {\bibinfo {author} {\bibfnamefont {G.}~\bibnamefont
  {Kresse}}\ and\ \bibinfo {author} {\bibfnamefont {D.}~\bibnamefont
  {Joubert}},\ }\href {\doibase 10.1103/PhysRevB.59.1758} {\bibfield  {journal}
  {\bibinfo  {journal} {Phys. Rev. B}\ }\textbf {\bibinfo {volume} {59}},\
  \bibinfo {pages} {1758} (\bibinfo {year} {1999})}\BibitemShut {NoStop}%
\bibitem [{\citenamefont {Heyd}\ \emph {et~al.}(2003)\citenamefont {Heyd},
  \citenamefont {Scuseria},\ and\ \citenamefont {Ernzerhof}}]{Heyd2003}%
  \BibitemOpen
  \bibfield  {author} {\bibinfo {author} {\bibfnamefont {J.}~\bibnamefont
  {Heyd}}, \bibinfo {author} {\bibfnamefont {G.~E.}\ \bibnamefont {Scuseria}},
  \ and\ \bibinfo {author} {\bibfnamefont {M.}~\bibnamefont {Ernzerhof}},\
  }\href {\doibase 10.1063/1.1564060} {\bibfield  {journal} {\bibinfo
  {journal} {J. Chem. Phys.}\ }\textbf {\bibinfo {volume} {118}},\ \bibinfo
  {pages} {8207} (\bibinfo {year} {2003})}\BibitemShut {NoStop}%
\bibitem [{\citenamefont {Xu}\ \emph {et~al.}(2013)\citenamefont {Xu},
  \citenamefont {Yan}, \citenamefont {Zhang}, \citenamefont {Wang},
  \citenamefont {Xu}, \citenamefont {Tang}, \citenamefont {Duan},\ and\
  \citenamefont {Zhang}}]{Xu2013Large}%
  \BibitemOpen
  \bibfield  {author} {\bibinfo {author} {\bibfnamefont {Y.}~\bibnamefont
  {Xu}}, \bibinfo {author} {\bibfnamefont {B.}~\bibnamefont {Yan}}, \bibinfo
  {author} {\bibfnamefont {H.-J.}\ \bibnamefont {Zhang}}, \bibinfo {author}
  {\bibfnamefont {J.}~\bibnamefont {Wang}}, \bibinfo {author} {\bibfnamefont
  {G.}~\bibnamefont {Xu}}, \bibinfo {author} {\bibfnamefont {P.}~\bibnamefont
  {Tang}}, \bibinfo {author} {\bibfnamefont {W.}~\bibnamefont {Duan}}, \ and\
  \bibinfo {author} {\bibfnamefont {S.-C.}\ \bibnamefont {Zhang}},\ }\href
  {\doibase 10.1103/PhysRevLett.111.136804} {\bibfield  {journal} {\bibinfo
  {journal} {Phys. Rev. Lett.}\ }\textbf {\bibinfo {volume} {111}},\ \bibinfo
  {pages} {136804} (\bibinfo {year} {2013})}\BibitemShut {NoStop}%
\bibitem [{\citenamefont {Luttinger}(1956)}]{Luttinger1956}%
  \BibitemOpen
  \bibfield  {author} {\bibinfo {author} {\bibfnamefont {J.~M.}\ \bibnamefont
  {Luttinger}},\ }\href {\doibase 10.1103/PhysRev.102.1030} {\bibfield
  {journal} {\bibinfo  {journal} {Phys. Rev.}\ }\textbf {\bibinfo {volume}
  {102}},\ \bibinfo {pages} {1030} (\bibinfo {year} {1956})}\BibitemShut
  {NoStop}%
\bibitem [{\citenamefont {Lawaetz}(1971)}]{Lawaetz1971}%
  \BibitemOpen
  \bibfield  {author} {\bibinfo {author} {\bibfnamefont {P.}~\bibnamefont
  {Lawaetz}},\ }\href {\doibase 10.1103/PhysRevB.4.3460} {\bibfield  {journal}
  {\bibinfo  {journal} {Phys. Rev. B}\ }\textbf {\bibinfo {volume} {4}},\
  \bibinfo {pages} {3460} (\bibinfo {year} {1971})}\BibitemShut {NoStop}%
\bibitem [{\citenamefont {Fu}\ and\ \citenamefont {Kane}(2007)}]{Fu2007}%
  \BibitemOpen
  \bibfield  {author} {\bibinfo {author} {\bibfnamefont {L.}~\bibnamefont
  {Fu}}\ and\ \bibinfo {author} {\bibfnamefont {C.~L.}\ \bibnamefont {Kane}},\
  }\href {\doibase 10.1103/PhysRevB.76.045302} {\bibfield  {journal} {\bibinfo
  {journal} {Phys. Rev. B}\ }\textbf {\bibinfo {volume} {76}},\ \bibinfo
  {pages} {045302} (\bibinfo {year} {2007})}\BibitemShut {NoStop}%
\bibitem [{\citenamefont {Marzari}\ and\ \citenamefont
  {Vanderbilt}(1997)}]{Marzari1997}%
  \BibitemOpen
  \bibfield  {author} {\bibinfo {author} {\bibfnamefont {N.}~\bibnamefont
  {Marzari}}\ and\ \bibinfo {author} {\bibfnamefont {D.}~\bibnamefont
  {Vanderbilt}},\ }\href {\doibase 10.1103/PhysRevB.56.12847} {\bibfield
  {journal} {\bibinfo  {journal} {Phys. Rev. B}\ }\textbf {\bibinfo {volume}
  {56}},\ \bibinfo {pages} {12847} (\bibinfo {year} {1997})}\BibitemShut
  {NoStop}%
\bibitem [{\citenamefont {Souza}\ \emph {et~al.}(2001)\citenamefont {Souza},
  \citenamefont {Marzari},\ and\ \citenamefont {Vanderbilt}}]{Souza2001}%
  \BibitemOpen
  \bibfield  {author} {\bibinfo {author} {\bibfnamefont {I.}~\bibnamefont
  {Souza}}, \bibinfo {author} {\bibfnamefont {N.}~\bibnamefont {Marzari}}, \
  and\ \bibinfo {author} {\bibfnamefont {D.}~\bibnamefont {Vanderbilt}},\
  }\href {\doibase 10.1103/PhysRevB.65.035109} {\bibfield  {journal} {\bibinfo
  {journal} {Phys. Rev. B}\ }\textbf {\bibinfo {volume} {65}},\ \bibinfo
  {pages} {035109} (\bibinfo {year} {2001})}\BibitemShut {NoStop}%
\bibitem [{\citenamefont {Zhang}\ \emph
  {et~al.}(2009{\natexlab{b}})\citenamefont {Zhang}, \citenamefont {Liu},
  \citenamefont {Qi}, \citenamefont {Deng}, \citenamefont {Dai}, \citenamefont
  {Zhang},\ and\ \citenamefont {Fang}}]{Zhang2009a}%
  \BibitemOpen
  \bibfield  {author} {\bibinfo {author} {\bibfnamefont {H.-J.}\ \bibnamefont
  {Zhang}}, \bibinfo {author} {\bibfnamefont {C.-X.}\ \bibnamefont {Liu}},
  \bibinfo {author} {\bibfnamefont {X.-L.}\ \bibnamefont {Qi}}, \bibinfo
  {author} {\bibfnamefont {X.-Y.}\ \bibnamefont {Deng}}, \bibinfo {author}
  {\bibfnamefont {X.}~\bibnamefont {Dai}}, \bibinfo {author} {\bibfnamefont
  {S.-C.}\ \bibnamefont {Zhang}}, \ and\ \bibinfo {author} {\bibfnamefont
  {Z.}~\bibnamefont {Fang}},\ }\href {\doibase 10.1103/PhysRevB.80.085307}
  {\bibfield  {journal} {\bibinfo  {journal} {Phys. Rev. B}\ }\textbf {\bibinfo
  {volume} {80}},\ \bibinfo {pages} {085307} (\bibinfo {year}
  {2009}{\natexlab{b}})}\BibitemShut {NoStop}%
\bibitem [{\citenamefont {Datta}(1995)}]{Datta}%
  \BibitemOpen
  \bibfield  {author} {\bibinfo {author} {\bibfnamefont {S.}~\bibnamefont
  {Datta}},\ }\href@noop {} {\emph {\bibinfo {title} {Electronic transport in
  mesoscopic systems}}}\ (\bibinfo  {publisher} {Cambridge University Press,
  Cambridge},\ \bibinfo {year} {1995})\BibitemShut {NoStop}%
\bibitem [{\citenamefont {Ewald}\ and\ \citenamefont
  {Tufte}(1959)}]{Ewald1959Electronic}%
  \BibitemOpen
  \bibfield  {author} {\bibinfo {author} {\bibfnamefont {A.~W.}\ \bibnamefont
  {Ewald}}\ and\ \bibinfo {author} {\bibfnamefont {O.~N.}\ \bibnamefont
  {Tufte}},\ }\href@noop {} {\bibfield  {journal} {\bibinfo  {journal} {J.
  Phys. Chem. Solids}\ }\textbf {\bibinfo {volume} {8}},\ \bibinfo {pages}
  {523} (\bibinfo {year} {1959})}\BibitemShut {NoStop}%
\bibitem [{\citenamefont {Kim}\ \emph {et~al.}(2017)\citenamefont {Kim},
  \citenamefont {Ryoo},\ and\ \citenamefont {Park}}]{Kim2017}%
  \BibitemOpen
  \bibfield  {author} {\bibinfo {author} {\bibfnamefont {P.}~\bibnamefont
  {Kim}}, \bibinfo {author} {\bibfnamefont {J.~H.}\ \bibnamefont {Ryoo}}, \
  and\ \bibinfo {author} {\bibfnamefont {C.-H.}\ \bibnamefont {Park}},\ }\href
  {\doibase 10.1103/PhysRevLett.119.266401} {\bibfield  {journal} {\bibinfo
  {journal} {Phys. Rev. Lett.}\ }\textbf {\bibinfo {volume} {119}},\ \bibinfo
  {pages} {266401} (\bibinfo {year} {2017})}\BibitemShut {NoStop}%
\bibitem [{\citenamefont {Zhang}\ \emph {et~al.}(2017)\citenamefont {Zhang},
  \citenamefont {Xu}, \citenamefont {Lin}, \citenamefont {Du}, \citenamefont
  {Guo}, \citenamefont {Lee}, \citenamefont {Lu}, \citenamefont {Feng},
  \citenamefont {Huang}, \citenamefont {Chang}, \citenamefont {Hsu},
  \citenamefont {Liu}, \citenamefont {Lin}, \citenamefont {Li}, \citenamefont
  {Zhang}, \citenamefont {Zhang}, \citenamefont {Xie}, \citenamefont {Neupert},
  \citenamefont {Hasan}, \citenamefont {Lu}, \citenamefont {Wang},\ and\
  \citenamefont {Jia}}]{Zhang2017Magnetic}%
  \BibitemOpen
  \bibfield  {author} {\bibinfo {author} {\bibfnamefont {C.-L.}\ \bibnamefont
  {Zhang}}, \bibinfo {author} {\bibfnamefont {S.-Y.}\ \bibnamefont {Xu}},
  \bibinfo {author} {\bibfnamefont {Z.}~\bibnamefont {Lin}}, \bibinfo {author}
  {\bibfnamefont {Z.~Z.}\ \bibnamefont {Du}}, \bibinfo {author} {\bibfnamefont
  {C.}~\bibnamefont {Guo}}, \bibinfo {author} {\bibfnamefont {C.-C.}\
  \bibnamefont {Lee}}, \bibinfo {author} {\bibfnamefont {H.}~\bibnamefont
  {Lu}}, \bibinfo {author} {\bibfnamefont {Y.}~\bibnamefont {Feng}}, \bibinfo
  {author} {\bibfnamefont {S.-M.}\ \bibnamefont {Huang}}, \bibinfo {author}
  {\bibfnamefont {G.}~\bibnamefont {Chang}}, \bibinfo {author} {\bibfnamefont
  {C.-H.}\ \bibnamefont {Hsu}}, \bibinfo {author} {\bibfnamefont
  {H.}~\bibnamefont {Liu}}, \bibinfo {author} {\bibfnamefont {H.}~\bibnamefont
  {Lin}}, \bibinfo {author} {\bibfnamefont {L.}~\bibnamefont {Li}}, \bibinfo
  {author} {\bibfnamefont {C.}~\bibnamefont {Zhang}}, \bibinfo {author}
  {\bibfnamefont {J.}~\bibnamefont {Zhang}}, \bibinfo {author} {\bibfnamefont
  {X.-C.}\ \bibnamefont {Xie}}, \bibinfo {author} {\bibfnamefont
  {T.}~\bibnamefont {Neupert}}, \bibinfo {author} {\bibfnamefont {M.~Z.}\
  \bibnamefont {Hasan}}, \bibinfo {author} {\bibfnamefont {H.-Z.}\ \bibnamefont
  {Lu}}, \bibinfo {author} {\bibfnamefont {J.}~\bibnamefont {Wang}}, \ and\
  \bibinfo {author} {\bibfnamefont {S.}~\bibnamefont {Jia}},\ }\href {\doibase
  10.1038/nphys4183} {\bibfield  {journal} {\bibinfo  {journal} {Nat. Phys.}\
  }\textbf {\bibinfo {volume} {13}},\ \bibinfo {pages} {979} (\bibinfo {year}
  {2017})}\BibitemShut {NoStop}%
\bibitem [{\citenamefont {Kitagawa}\ \emph {et~al.}(2011)\citenamefont
  {Kitagawa}, \citenamefont {Oka}, \citenamefont {Brataas}, \citenamefont
  {Fu},\ and\ \citenamefont {Demler}}]{Kitagawa2011}%
  \BibitemOpen
  \bibfield  {author} {\bibinfo {author} {\bibfnamefont {T.}~\bibnamefont
  {Kitagawa}}, \bibinfo {author} {\bibfnamefont {T.}~\bibnamefont {Oka}},
  \bibinfo {author} {\bibfnamefont {A.}~\bibnamefont {Brataas}}, \bibinfo
  {author} {\bibfnamefont {L.}~\bibnamefont {Fu}}, \ and\ \bibinfo {author}
  {\bibfnamefont {E.}~\bibnamefont {Demler}},\ }\href {\doibase
  10.1103/PhysRevB.84.235108} {\bibfield  {journal} {\bibinfo  {journal} {Phys.
  Rev. B}\ }\textbf {\bibinfo {volume} {84}},\ \bibinfo {pages} {235108}
  (\bibinfo {year} {2011})}\BibitemShut {NoStop}%
\bibitem [{\citenamefont {Bernevig}\ and\ \citenamefont
  {Hughes}(2013)}]{Bernevig2013Topological}%
  \BibitemOpen
  \bibfield  {author} {\bibinfo {author} {\bibfnamefont {B.~A.}\ \bibnamefont
  {Bernevig}}\ and\ \bibinfo {author} {\bibfnamefont {T.~L.}\ \bibnamefont
  {Hughes}},\ }\href@noop {} {\emph {\bibinfo {title} {Topological insulators
  and topological superconductors}}}\ (\bibinfo  {publisher} {Princeton
  University Press, Princeton},\ \bibinfo {year} {2013})\BibitemShut {NoStop}%
\bibitem [{\citenamefont {Lu}\ and\ \citenamefont
  {Shen}(2017)}]{lu2017quantum}%
  \BibitemOpen
  \bibfield  {author} {\bibinfo {author} {\bibfnamefont {H.-Z.}\ \bibnamefont
  {Lu}}\ and\ \bibinfo {author} {\bibfnamefont {S.-Q.}\ \bibnamefont {Shen}},\
  }\href@noop {} {\bibfield  {journal} {\bibinfo  {journal} {Front. Phys}\
  }\textbf {\bibinfo {volume} {12}},\ \bibinfo {pages} {127201} (\bibinfo
  {year} {2017})}\BibitemShut {NoStop}%
\bibitem [{\citenamefont {Xiao}\ \emph {et~al.}(2010)\citenamefont {Xiao},
  \citenamefont {Chang},\ and\ \citenamefont {Niu}}]{Xiao2010}%
  \BibitemOpen
  \bibfield  {author} {\bibinfo {author} {\bibfnamefont {D.}~\bibnamefont
  {Xiao}}, \bibinfo {author} {\bibfnamefont {M.-C.}\ \bibnamefont {Chang}}, \
  and\ \bibinfo {author} {\bibfnamefont {Q.}~\bibnamefont {Niu}},\ }\href
  {\doibase 10.1103/RevModPhys.82.1959} {\bibfield  {journal} {\bibinfo
  {journal} {Rev. Mod. Phys.}\ }\textbf {\bibinfo {volume} {82}},\ \bibinfo
  {pages} {1959} (\bibinfo {year} {2010})}\BibitemShut {NoStop}%
\end{thebibliography}%

\end{document}